\documentclass[a4paper,11pt]{article}
\pdfoutput=1

\usepackage[utf8]{inputenc}
\usepackage{jheppub_mod}
\usepackage{comment}
\usepackage{bm}
\usepackage[dvipsnames]{xcolor}
\usepackage{tikz}
\usepackage{alltt}
\usepackage{color,xcolor}
\usepackage{ifthen}
\usetikzlibrary{calc}
\usepackage{pgfplots}
\usepackage{color,xcolor}
\usepgfplotslibrary{fillbetween}
\usetikzlibrary{positioning}
\usetikzlibrary{shapes.geometric}
\usetikzlibrary{ decorations.markings}
\usetikzlibrary{graphs,graphs.standard,quotes}
\usepackage{ifthen}
\usepackage {empheq}
\pgfplotsset{holdot/.style={color=black,only marks,mark=*},compat=newest}

\newcommand{\lcb}{{\tt {\char '173}}}
\newcommand{\rcb}{{\tt {\char '175}}}

\newcommand*{\TickSize}{3pt}

\newcommand{\D}{\mathrm{d}}
\newcommand{\A}{\mathcal{A}}

\renewcommand{\O}{\mathcal{O}}
\newcommand{\x}{\bs{x}}
\newcommand{\dd}{\mathrm{d}}

\DeclareMathOperator{\Li}{Li}

\newcommand{\bs}[1]{\boldsymbol{#1}}

\newcommand{\nn}{\nonumber}

\newcommand{\g}{\gamma}
\newcommand{\ep}{\epsilon}
\newcommand{\z}{\zeta}

\renewcommand{\d}{\delta}
\renewcommand{\t}{\theta}
\newcommand{\tb}{\bar{\theta}}
\newcommand{\ino}{i}

\newcommand{\p}{\partial}
\newcommand{\vphi}{\varphi}
\renewcommand{\[}{\begin{equation}}
\renewcommand{\]}{\end{equation}}

\begin{document}

\title{$\bs{\mathcal{A}}$-hypergeometric functions 
and creation operators for Feynman and Witten diagrams}

\author{Francesca Caloro and Paul McFadden}
\affiliation{School of Mathematics, 
	Statistics \& Physics, Newcastle University, 
	Newcastle NE1 7RU, U.K.}

\emailAdd{f.caloro2@ncl.ac.uk, paul.l.mcfadden@ncl.ac.uk}

\abstract{ 
Both Feynman integrals and holographic Witten diagrams can be represented as  multivariable hypergeometric functions of a class studied by 
Gel'fand, Kapranov \& Zelevinsky known as GKZ or $\mathcal{A}$-hypergeometric functions.  
Among other advantages, this formalism enables the systematic construction of highly non-trivial weight-shifting operators known as `creation' operators.  
We derive these operators from a physics perspective, highlighting their close relation  to the spectral singularities of the integral as 
encoded by the facets of the Newton polytope.   
Many examples for Feynman and Witten diagrams are given, including novel weight-shifting operators for holographic contact diagrams. 
These in turn allow momentum-space  exchange diagrams of different operator dimensions to be related while keeping the spacetime dimension fixed. 
In contrast to previous constructions, only non-derivative vertices are involved.
}

\maketitle

\section{Introduction}

It has long been suspected that Feynman integrals represent 
a multi-variable generalisation of 
hypergeometric functions \cite{Regge, Kashiwara:1977nf}.  
Recently \cite{Vanhove:2018mto, de_la_Cruz_2019, Klausen:2019hrg,Klausen:2021yrt,  Klausen:2023gui,Klemm_2020,Feng_2020,Chestnov:2022alh,Ananthanarayan:2022ntm,Zhang:2023fil}, this connection has been 
sharpened 
by writing Feynman integrals as Gel'fand-Kapranov-Zelevinksy (GKZ) or $\mathcal{A}$-hypergeometric functions  \cite{GKZ_1, GKZ_2,GKZ_3, GKZ_book}.   
As shown in \cite{de_la_Cruz_2019, Klausen:2019hrg}, this can be achieved simply
 by expressing Feynman integrals in Lee-Pomeransky form \cite{Lee:2013hzt}, where only a single denominator polynomial appears, followed by uplifting to a higher-dimensional space of generalised momenta. 
These $\mathcal{A}$-hypergeometric functions are well-studied 
in the mathematics literature  \cite{Stienstra:2005nr, BeukersNotes, cattani2006three, takayama2020hypergeometric, saito2013grobner, reichelt2021algebraic} and satisfy various linear partial differential equations whose form can be read off in systematic fashion from a certain matrix -- the $\mathcal{A}$-matrix -- which encodes both the structure of the integral 
as well as all kinematic and spectral singularities.

A task of great practical interest is then 
to construct hypergeometric 
{\it shift operators} connecting integrals of different parameter values. 
These operators enable a known `seed' integral to be converted, by simple differentiation, into an entire series of new integrals.
For Feynman integrals, the parameters are typically  the powers of various propagators and the spacetime dimension.  Here we will also study Witten diagrams in anti-de Sitter spacetime
for which the relevant parameters, besides the spacetime dimension, are 
the scaling dimensions of operators in the holographically dual conformal field theory. 

While various techniques for constructing 
shift operators for Feynman integrals \cite{Tarasov:1996br, Lee:2010wea,  Bytev:2009kb, Bytev:2013gva, Bitoun:2017nre} and Witten diagrams \cite{Dolan:2000ut,Bzowski:2015yxv,Karateev:2017jgd, Costa:2018mcg, Baumann:2019oyu,Rigatos:2022eos, Bzowski:2022rlz} are known, the GKZ formalism offers a more powerful and unified approach.  Besides the elementary shift operators,  known as `annihilation' operators in the mathematics literature, their {\it inverses} -- a highly non-trivial class of operators known as `creation' operators -- can be systematically constructed \cite{Saito_param_shift, Saito_restrictions, Saito_Lauricella, saito_sturmfels_takayama_1999}.
Together, these creation and annihilation operators 
form a full set of shift  operators connecting $\mathcal{A}$-hypergeometric functions of different parameter values,  
just as the ordinary creation and annihilation (or ladder) operators 
connect different eigenstates of the quantum harmonic oscillator.

\begin{figure}
\centering
\begin{tikzpicture}[scale=2]
   \draw[thick,black] (1,0,0) -- (0,1,0) -- (0,0,1); 
  \draw[thick,black]  (-1,0,0) -- (0,-1,0) ;
   \draw[thick,black] (-1,0,0) -- (0,1,0) ;
    \draw[thick,black] (-1,0,0) -- (0,0,1)--(1,0,0) ;
    \draw[thick,black] (0,0,1) -- (0,-1,0)--(1,0,0) ;
   \draw[style=dashed, color=black] (0,-1,0)-- (0,0,-1)--(-1,0,0);
   \draw[style=dashed, color=black] (1,0,0)-- (0,0,-1)--(0,1,0); 
 
   \draw[thick,black,fill=YellowGreen, fill opacity=0.2] (-1,0,0)--(0,-1,0) -- (1,0,0) -- (0,1,0) --cycle;

   \end{tikzpicture}
   \caption{The Newton polytope for a 3-point contact Witten diagram in momentum space 
   is an octahedron as shown.  At $n$-points, we obtain an $n$-dimensional cross-polytope.  The spectral singularities consist of an infinite series of hyperplanes parallel to the facets of this Newton polytope, while the integral is convergent for parameter values lying inside the polytope.  Identification of the singularities enables a systematic construction of all creation-type shift operators. \label{3Kpolytope}}
\label{fig:octa}
\end{figure}
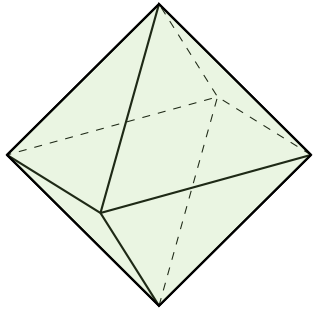

A key aim of this paper is to 
show that  creation operators can be constructed 
directly 
from knowledge of the spectral singularities of an $\mathcal{A}$-hypergeometric function, namely, 
the special set of parameter values for which the corresponding GKZ integral representation diverges.
These singularities can be computed directly from the $\mathcal{A}$-matrix of the integral.  Remarkably, they correspond geometrically to an infinite series of hyperplanes parallel to the 
co-dimension one facets of the Newton polytope  associated with the integral's denominator \cite{nilsson2010mellin,berkesch2013eulermellin}, as shown in figure \ref{3Kpolytope}.
Standard convex hulling algorithms exist for computing such facets 
allowing a simple identification of all  singularities.  

To construct creation operators, we start with a pair of integrals 
 connected by an annihilation operator.  As we will review, this annihilator consists of   a single derivative with respect to one of the GKZ generalised momenta. 
Specifically, we are interested in cases with parameters such that the starting integral is {\it divergent} while the resulting integral is  {\it finite}.  
(To regulate divergences, we assume a dimensional scheme where  parameters are infinitesimally shifted away from their singular values.)
The divergences are thus projected out by the action of the annihilator.  
As the inverse of the annihilator, the creation operator must then produce the reverse shift, from the finite integral to the divergent one. 
Clearly, however, this cannot be achieved directly: the result of acting with a finite differential operator on a finite integral must necessarily be finite.  
Instead, the outcome must be a finite product of 
 the divergent integral multiplied by a vanishing function of the parameters. 
This  function,  whose zeros serve to cancel out the divergence, is known as the {\it $b$-function} and holds the key to the construction of creation operators.  

From a knowledge of the singular parameter values,  
we can predict the necessary zeros of the $b$-function and hence its minimal form as a polynomial.  
Then, acting on an integral with both the annihilator and the (as yet unknown) creation operator, we must recover the original integral multiplied by the $b$-function.
In the GKZ formalism, however,  any polynomial in the parameters can  be  traded for an equivalent polynomial in Euler operators acting on the generalised  momenta. 
Applying this procedure to the $b$-function, the resulting differential operator must thus be factorisable into a product of the annihilation and the creation operator.  As the annihilator is just a single derivative, this factorisation is easily performed (with the aid of a further set of PDEs known as the toric equations) revealing the identity of the creation operator. 
As a final step, one then projects back from the higher-dimensional GKZ space of generalised momenta to that of the physical variables (the external momenta and masses), with the aid of an auxiliary set of Euler equations.

We hope this simple physical approach, based on the  spectral singularities of the GKZ integral, 
will facilitate
the application 
of creation operators to a range of  physical
systems. 
As an initial demonstration of the possibilities, we have used the formalism to construct new shift operators for a range of simple Feynman integrals, as well as Witten diagrams encoding momentum-space correlators in holographic conformal field theories.  These latter objects are intimately related to cosmological correlators in de Sitter spacetime, and the new shift operators we construct can also be applied in this context.     
In particular, we have found new shift operators connecting both exchange and contact 4-point Witten diagrams, with arbitrary external scaling dimensions, to corresponding diagrams with shifted scaling dimensions but the same spacetime dimension.
Until now, such operators were only available in the case where diagrams with {\it non-derivative} vertices are mapped to those with {\it derivative} vertices, and for a restricted set of scaling dimensions at that \cite{Baumann:2019oyu,Bzowski:2022rlz}.  In contrast, the new shift operators we find can  be applied for any scaling dimensions, and moreover map non-derivative  to non-derivative vertices.  This  enlarges the available arsenal of shift operators for Witten diagrams (and by extension, cosmological correlators), and as such is a useful and nontrivial result.  
We believe these examples provide a first proof of principle that the creation operator method, and the GKZ formalism more generally, holds 
promise for a variety of physical applications.

An outline of this paper is as follows.  Section \ref{sec:GKZformalism} introduces $\mathcal{A}$-hypergeometric functions and the GKZ formalism.  We summarise the PDEs these functions obey, their construction, and their invariance under affine reparametrisations.  In section \ref{sec:sings}, we relate the spectral singularities of GKZ integrals to the Newton polytope of the denominator.
In section \ref{sec:shiftops}, we introduce creation operators and detail their construction based on the spectral singularities of the integral.  In section \ref{sec:Witten}, we construct creation operators for 3- and 4-point contact Witten diagrams in momentum space, as well as a further set of shift operators that preserve the spacetime dimension.  Using these results, we then derive novel shift operators for exchange diagrams.
Section \ref{sec:Feynman}  constructs creation operators for a variety of simple Feynman integrals introducing the use of Gr{\"o}bner bases and convex hulling algorithms to automate the computation.  We conclude in section \ref{sec:Discussion} with a summary of results and open directions.  In the appendices we discuss the conversion of Feynman  to GKZ integrals, creation operators for position-space contact Witten diagrams, and an extension of the minimal construction algorithm outlined above.

\section{\texorpdfstring{$\mathcal{A}$-hypergeometric functions}{A-hypergeometric functions}}
\label{sec:GKZformalism}

The application of the GKZ formalism to Feynman integrals has been explored in a number of recent works \cite{Vanhove:2018mto, de_la_Cruz_2019, Klausen:2019hrg,Klausen:2021yrt,  Klausen:2023gui,Klemm_2020,Feng_2020,Chestnov:2022alh,Ananthanarayan:2022ntm,Zhang:2023fil}.  In addition, many excellent expositions are available in the mathematics literature  \cite{Stienstra:2005nr, cattani2006three, BeukersNotes,saito2013grobner, takayama2020hypergeometric, reichelt2021algebraic}.
Here, we focus on providing a simple and self-contained summary of the key material needed to understand the construction of creation operators.

\subsection{GKZ integrals}

An $\mathcal{A}$-hypergeometric function (or equivalently,  GKZ integral), is a multi-variable hypergeometric function depending on a set of real parameters $\bs{\gamma}=(\gamma_0,\gamma_1,\ldots,\gamma_N)$ and independent variables $\x = (x_1,\ldots, x_n)$, where $n\ge N+1$.  The integral takes the form
\[\label{GKZint}
\mathcal{I}_{\bs{\gamma}} =\Big(\prod_{i=1}^N \int_0^\infty\D z_i\, z_i^{\gamma_i-1} \Big)\mathcal{D}^{-\gamma_0},
\]
where the `denominator' $\mathcal{D}$ can be expressed as a polynomial in the integration variables  $z_i$.  Every term in this polynomial is moreover multiplied by a {\it nonzero} coefficient $x_j$:
\[\label{GKZden}
\mathcal{D} = \sum_{j=1}^n x_j \prod_{i=1}^N z_i^{a_{ij}}
\]
The parameters  $a_{ij}\in \mathbb{Z}^+$ specifying the powers can be assembled into an $N\times n$ matrix $A$,
\[
(A)_{ij} = a_{ij}.
\]
Thus, the $j$th term in the denominator $\mathcal{D}$ corresponds to the column $j$ of the matrix $A$, whose entries are then the powers of the variables $z_i$ appearing in that particular term.
(We will return  to the relation between this matrix $A$ and the larger $\mathcal{A}$-matrix shortly.)

For Feynman integrals, it is useful to consider the Lee-Pomeransky representation \cite{Lee:2013hzt} in which the denominator $\mathcal{G}=\mathcal{U}+\mathcal{F}$ is formed from the sum of the first and second Symanzik polynomials $\mathcal{U}$ and $\mathcal{F}$.  To uplift this to the GKZ integral \eqref{GKZint},
we simply promote the coefficient of every term in $\mathcal{G}$ to a generalised independent variable $x_j$  \cite{de_la_Cruz_2019, Klausen:2019hrg}, as summarised in appendix \ref{LPrep}.   The original Lee-Pomeransky integral can then be restored by returning the $x_j$ to their physical values, namely, unity for any of the terms in $\mathcal{U}$, and the appropriate function of the masses and external momenta for every term in $\mathcal{F}$.

\paragraph{Example:}  
As discussed in appendix \ref{LPrep}, 
the massless triangle Feynman integral
\[\label{masslesstri}
I= 
\int\frac{\mathrm{d}^d\bs{q}}{(2\pi)^d}\frac{1}{q^{2\gamma_3}|\bs{q}-\bs{p}_1|^{2\gamma_2}|\bs{q}+\bs{p}_2|^{2\gamma_1}}
\]
has the Lee-Pomeransky representation
\[
I = c\,\Big( \prod_{i=1}^3\int_0^\infty \D z_i\, z_i^{\gamma_i-1}\Big)(p_1^2 z_2 z_3+p_2^2 z_1 z_3+p_3^2 z_1 z_2+z_1+z_2+z_3)^{-d/2}
\]
 where the coefficient
\[\label{Cdef}
c_{\bs{\g}}=(4\pi)^{-d/2}\frac{\Gamma(d/2)}{\Gamma(d-\gamma_t)\prod_{i=1}^3\Gamma(\gamma_i)},\qquad \gamma_t=\sum_{i=1}^3\gamma_i.
\] 
The corresponding  GKZ integral is 
\[\label{triGKZ}
\mathcal{I}_{\bs{\gamma}} = \Big(\prod_{i=1}^3\int_0^\infty\D z_i z_i^{\gamma_i-1}\Big) \mathcal{D}^{-\gamma_0}
\]
where the denominator
\[\label{triD}
\mathcal{D} =  x_1 z_2 z_3 + x_2 z_1 z_3+x_3 z_1 z_2+x_4 z_1 + x_5 z_2 + x_6 z_3
\]
corresponds to the matrix
\[
A=  \left(\begin{matrix}
0&1&1&1&0&0\\
1&0&1&0&1&0\\
1&1&0&0&0&1
\end{matrix}
\right).
\]
To recover the original Lee-Pomeransky integral, we project to the physical subspace
\[\label{triphys}
\x = (p_1^2,p_2^2,p_3^2,1,1,1),\qquad \bs{\gamma} = (d/2,\gamma_1,\gamma_2,\gamma_3),
\]
after which $I = c_{\bs{\g}}\, \mathcal{I}_{\bs{\g}}$. 

\subsection{The Euler and toric equations}

The primary advantage of uplifting from the original masses and momenta to the generalised GKZ space parametrised by the  variables $\x$ is that the integral now obeys a systematic set of linear partial differential equations.  These can be grouped into two categories, known as the Euler equations and the toric equations.

\subsubsection{Euler equations}

The Euler equations arise  from integrating by parts with respect to the  variables $z_i$, under the assumption that all boundary terms vanish.
For $z_1$, for example, we have
\begin{align}
0 &=\int_0^\infty\D z_1\, \frac{\partial}{\partial z_1}\Big(z_1^{\gamma_1} \Big(\prod_{i=2}^N\int_0^\infty\D z_i\,z_i^{\gamma_i-1}\Big)\mathcal{D}^{-\gamma_0}\Big)\nn\\
&= \gamma_1 \mathcal{I}_{\bs{\gamma}}+\Big(\prod_{i=1}^N\int_0^\infty \D z_i\,z_i^{\gamma_i-1}\Big) z_1\frac{\partial}{\partial z_1} \mathcal{D}^{-\gamma_0}.
\end{align}
In the second term here, we can trade derivatives with respect to the integration variable $z_1$ for derivatives with respect to the external variables $x_j$:
\begin{align}
 z_1\frac{\partial}{\partial z_1} \mathcal{D}^{-\gamma_0} =
-\gamma_0 \mathcal{D}^{-\gamma_0-1}\Big(\sum_{j=1}^n a_{1j}x_j \prod_{i=1}^N z_i^{a_{ij}}\Big)=  
 \Big( \sum_{j=1}^n a_{1j}\theta_j\Big) \mathcal{D}^{-\gamma_0} 
\end{align}
where, here and throughout the paper, we define the Euler operators
\[
\theta_j = x_j\frac{\partial}{\partial x_j}, \qquad j=1,\ldots, n.
\]
Pulling these Euler operators outside the integrals, we obtain the equation
\[
0 = \Big(\gamma_1 + \sum_{j=1}^n a_{1j}\theta_j\Big)\mathcal{I}_{\bs{\gamma}}.
\]
Repeating this exercise for  the remaining $z_i$ then leads to the set of {\it Euler equations}
\[\label{EulereqnsN}
0 = \Big(\gamma_i + \sum_{j=1}^n a_{ij}\theta_j\Big)\mathcal{I}_{\bs{\gamma}}, \qquad i=1,\ldots, N.
\]
We are not quite done, however, since  in addition we have the general identity 
\[
\Big(\sum_{j=1}^n \theta_j \Big) \mathcal{D}^{-\gamma_0} = -\gamma_0 \mathcal{D}^{-\gamma_0}
\]
which, when applied to the GKZ integral, yields 
\[\label{DWI}
0 = \Big(\gamma_0+\sum_{j=1}^n\theta_j\Big)\mathcal{I}_{\bs{\gamma}}.
\]
This equation is effectively a dilatation Ward identity (or DWI, as we will use for short) 
encoding the scaling behaviour of the GKZ integral under a dilatation $\bs{x}\rightarrow \lambda \bs{x}$ of the external variables. 

Evidently this dilatation Ward identity can be placed on the same footing as the Euler equations \eqref{EulereqnsN} by enlarging the matrix $A$ to include a top row consisting of all $1$s.  This construction defines the $\mathcal{A}$-matrix mentioned in the introduction,
\[\label{AtoA}
\mathcal{A} = \left(\begin{matrix} \bs{1}\\ A\end{matrix}\right)\!,
\]
where $\bs{1}$ is the $n$-dimensional row vector with all-$1$ entries, or  equivalently, 
\[
(\mathcal{A})_{0j}=1,\qquad (\mathcal{A})_{ij}=a_{ij}, \qquad i=1,\ldots, N,\qquad  j=1,\ldots, n,
\]
where we henceforth adopt the convention that the top row of $\mathcal{A}$ always carries index $0$.
The $\mathcal{A}$-matrix is thus $(N+1)\times n$ dimensional, and 
 the Euler equations and DWI together correspond to the $(N+1)$ equations
\[\label{allEulers}
0 = \Big(\gamma_i + \sum_{j=1}^n \mathcal{A}_{ij}\theta_j\Big)\mathcal{I}_{\bs{\gamma}}, \qquad i=0,\ldots, N.
\]
This is in effect a single matrix equation,
\[
0 = \Big(\bs{\gamma}+\mathcal{A}\cdot\bs{\theta}\Big)\mathcal{I}_{\bs{\gamma}},
\]
regarding $\bs{\theta}=(\theta_1,\ldots,\theta_n)^T$ and $\bs{\gamma}=(\gamma_0,\gamma_1,\ldots,\gamma_N)^T$ as  $n$- and $(N+1)$-component column vectors respectively.

\paragraph{Example:}  Returning to the  massless triangle integral above, the $\mathcal{A}$-matrix is
\[\label{Atri}
\mathcal{A}=  \left(\begin{matrix}
1&1&1&1&1&1\\
0&1&1&1&0&0\\
1&0&1&0&1&0\\
1&1&0&0&0&1
\end{matrix}
\right)
\]
and the GKZ integral satisfies  the  Euler equations
\begin{align}\label{triEulers}
0=(\gamma_1+\theta_2+\theta_3+\theta_4)\mathcal{I}_{\bs{\gamma}},\quad 0=(\gamma_2+\theta_1+\theta_3+\theta_5)\mathcal{I}_{\bs{\gamma}},\quad 0=(\gamma_3+\theta_1+\theta_2+\theta_6)\mathcal{I}_{\bs{\gamma}}
\end{align}
and DWI
\[\label{triDWI}
0 = (\gamma_0+\sum_{j=1}^6 \theta_j)\mathcal{I}_{\bs{\gamma}}.
\]
Notice the form of these equations can be directly read off from the  rows of the $\mathcal{A}$-matrix.

\subsubsection{Toric equations}  The toric equations arise from  vectors in the {\it kernel} of the $\mathcal{A}$-matrix, and are closely related to the corresponding toric ideal \cite{saito2013grobner}.\footnote{The kernel is the space of vectors $\bs{u}$ such that  $\mathcal{A}\cdot\bs{u}=\bs{0}$, obtained {\it e.g.,} via $\tt{NullSpace}[\mathcal{A}]$  in Mathematica.  The full toric ideal, though not needed here, can be constructed using {\it Singular}  \cite{singular}: see  section \ref{toricidealdisc}.}  
Their origin can 
 easily be grasped using the example of the massless triangle integral above.  
Defining
\[
\partial_j = \frac{\p}{\p x_j},\qquad j=1,\ldots, n
\]
in all that follows, the  denominator \eqref{triD} obeys the  relations
\[
\p_1\p_4\mathcal{D}^{-\gamma_0}=
\p_2\p_5\mathcal{D}^{-\gamma_0}=
\p_3\p_6\mathcal{D}^{-\gamma_0}=-\gamma_0 (-\gamma_0-1)z_1 z_2 z_3\mathcal{D}^{-\gamma_0-2},
\]
giving rise to the two independent (toric) equations
\[\label{tritorics}
0=(\p_1\p_4-\p_3\p_6)\mathcal{I}_{\bs{\gamma}},\qquad
0 = (\p_2\p_5-\p_3\p_6)\mathcal{I}_{\bs{\gamma}}.
\]
For comparison, the kernel of the $\mathcal{A}$-matrix \eqref{Atri} is spanned by two independent vectors, $\bs{u}_{(1)}$ and $\bs{u}_{(2)}$, which we can choose to be
\[\label{trikern}
\bs{u}_{(1)} = (1,0,-1,1,0,-1)^T,\qquad \bs{u}_{(2)} = (1,-1,0,1,-1,0)^T.
\]
Notice that since the top row of the $\mathcal{A}$-matrix is all  $1$s, the sum of the components of any kernel vector is always zero.  
There is now a one-to-one match between kernel vectors and toric equations \eqref{tritorics} as follows. First, for each kernel vector $\bs{u}$, we form a vector $\bs{u}^+$ composed only of the {\it positive} components of $\bs{u}$, and a vector $\bs{u}^-$ composed of only the {\it negative} components.  The components of $\bs{u}^\pm$, for each $j=1,\ldots, n$, are thus 
\[
u^\pm_j=\mathrm{max}(\pm u_j,0).
\]
By inspection, the toric equation corresponding to the kernel vector  $\bs{u} = \bs{u}^+ -\bs{u}^-$ is now
\[\label{torics}
0 = \Big(\prod_{j=1}^n \p_j^{u_j^+} -\prod_{j=1}^n \p_j^{u_j^-}\Big)\,\mathcal{I}_{\bs{\gamma}}.
\]
For example, for $\bs{u}_{(1)}$ in \eqref{trikern}, $\bs{u}_{(1)}^+=(1,0,0,1,0,0)^T$ while $\bs{u}_{(1)}^-=(0,0,1,0,0,1)^T$ hence \eqref{torics} reduces to the first equation in \eqref{tritorics}.

Some investigation shows this construction is a general one. 
First, the action of each differential operator is
\[
\prod_{j=1}^n \p_j^{u_j^\pm}\mathcal{D}^{-\gamma_0} = (-\gamma_0)(-\gamma_0-1)\ldots(-\gamma_0-\mathfrak{u}^\pm+1) \mathcal{D}^{-\gamma_0-\mathfrak{u}^\pm}\Big(\prod_{i=1}^N z_i^{\sum_{j=1}^n a_{ij}u_j^\pm}
\Big)
\]
where $\mathfrak{u}^\pm=\sum_{j=1}^n u_j^\pm$.    Moreover, since the sum of components in any kernel vector vanishes (as the top row of the $\mathcal{A}$-matrix is all $1$s), we have that $\mathfrak{u}^+=\mathfrak{u}^-=\mathfrak{u}$.  Thus, 
\begin{align}\label{toriceval}
& \Big(\prod_{j=1}^n \p_j^{u_j^+} -\prod_{j=1}^n \p_j^{u_j^-}\Big)\,\mathcal{I}_{\bs{\gamma}}\\[-1ex]
 &\qquad =(-\gamma_0)(-\gamma_0-1)\ldots(-\gamma_0-\mathfrak{u}+1) \mathcal{D}^{-\gamma_0-\mathfrak{u}}\Big(\prod_{i=1}^N z_i^{\sum_{j=1}^n a_{ij}u_j^+}-\prod_{i=1}^N z_i^{\sum_{j=1}^n a_{ij}u_j^-}
\Big).\nn
\end{align}
However, for any kernel vector we have $\mathcal{A}\cdot\bs{u} = \mathcal{A}\cdot(\bs{u}^+-\bs{u}^-)=\bs{0}$ and hence
\[
\sum_{j=1}^n a_{ij}u_j^+=\sum_{j=1}^n a_{ij}u_j^-, \qquad i=1,\ldots, N.
\]
The two terms appearing within the final factor of \eqref{toriceval} are thus exactly equal producing a cancellation.
In general, as the $\mathcal{A}$-matrix is $(N+1)\times n$, there are $(n-N-1)$ independent vectors in the kernel, and hence this same number of independent toric equations.

To summarise,  given a GKZ integral defined by an $\mathcal{A}$-matrix and parameters $\bs{\gamma}$,  we have two sets of linear partial differential equations: 
the Euler equations (and DWI) \eqref{allEulers}, and the toric equations \eqref{torics}.  
We can also go in reverse: the Euler equations and DWI fix $\bs{\gamma}$ and the $\mathcal{A}$-matrix, and hence the toric equations and the GKZ integral.  
Note the Euler equations all commute among themselves, as do the toric equations, but an Euler and a toric equation do not in general commute.

\subsection{Projection to physical variables} 

The systematic structure of the Euler and toric equations above is a consequence of uplifting from the Lee-Pomeransky  to the GKZ denominator \eqref{GKZden}.
To recover a set of PDEs satisfied by the original Lee-Pomeransky integral we need to reverse this process.
This requires projecting the Euler and toric equations back to the {\it physical hypersurface} where the $\x$ variables take their true physical values.
Derivatives in directions not tangential to this hypersurface (which therefore cannot be expressed purely in terms of physical variables) can be exchanged for purely tangential derivatives through use of the Euler equations and DWI.  
Together these provide $N+1$ equations, and so for {\it all} unphysical ({\it i.e.,} non-tangential) derivatives to be removable requires the original Lee-Pomeransky polynomial to contain at least $n-N-1$ independent physical variables ({\it i.e.,} masses and external momenta).  This will generally be the case for the examples we consider, but does not hold universally -- particularly for higher-loop Feynman integrals -- as we discuss  in section \ref{sec:Discussion}.

\paragraph{Example:}  For the massless triangle integral, the physical hypersurface is the $3$-dimensional subspace spanned by the momenta in \eqref{triphys}, namely $x_1=p_1^2$, $x_2=p_2^2$ and $x_3=p_3^2$, with $x_4=x_5=x_6=1$.
On this hypersurface, the Euler equations \eqref{triEulers} reduce to
\[\label{linderivs}
0=(\gamma_1+\theta_2+\theta_3+\partial_4)\mathcal{I}_{\bs{\gamma}},\quad 0=(\gamma_2+\theta_1+\theta_3+\partial_5)\mathcal{I}_{\bs{\gamma}},\quad 0=(\gamma_3+\theta_1+\theta_2+\partial_6)\mathcal{I}_{\bs{\gamma}},
\]
where, as always, $\partial_j=\partial/\partial x_j$.
These equations allow us to eliminate the unphysical derivatives $\partial_4$, $\partial_5$ and $\partial_6$ from all remaining equations in which they appear linearly.\footnote{More generally, we can rewrite $\p_4^m = x_4^{-m}\t_4(\t_4-1)\ldots(\t_4-m+1)$, {\it etc.}, then use the full Euler equations to eliminate $\t_4$, $\t_5$ and $\t_6$ before setting $x_4=x_5=x_6=1$.
 Alternatively, we can supplement \eqref{linderivs} with derivatives of the Euler equations (and DWI) evaluated on the physical hypersurface.}
For example, evaluating the first toric equation in \eqref{tritorics} on the physical hypersurface,
\begin{align}\label{physeq1}
0 &= (\partial_1\partial_4-\partial_3\partial_6)\mathcal{I}_{\bs{\gamma}} \nn\\
&= \Big(\partial_1(-\gamma_1-\theta_2-\theta_3) - \partial_3(-\gamma_3-\theta_1-\theta_2)\Big)\mathcal{I}_{\bs{\gamma}} \nn\\&
=\frac{1}{4}\Big[-\Big(2\gamma_1+p_2\frac{\partial}{\partial p_2}+p_3\frac{\partial}{\partial p_3}\Big)\frac{1}{p_1}\frac{\partial}{\partial p_1}+\Big(2\gamma_3+p_1\frac{\partial}{\partial p_1}+p_2\frac{\partial}{\partial p_2}\Big)\frac{1}{p_3}\frac{\partial}{\partial p_3}\Big]\mathcal{I}_{\bs{\gamma}} ,
\end{align}
while for the second toric equation,
\begin{align}
0 &= (\partial_2\partial_5-\partial_3\partial_6)\mathcal{I}_{\bs{\gamma}} \nn\\
&= \Big(\partial_2(-\gamma_2-\theta_1-\theta_3) - \partial_3(-\gamma_3-\theta_1-\theta_2)\Big)\mathcal{I}_{\bs{\gamma}} \nn\\&
=\frac{1}{4}\Big[-\Big(2\gamma_2+p_1\frac{\partial}{\partial p_1}+p_3\frac{\partial}{\partial p_3}\Big)\frac{1}{p_2}\frac{\partial}{\partial p_2}+\Big(2\gamma_3+p_1\frac{\partial}{\partial p_1}+p_2\frac{\partial}{\partial p_2}\Big)\frac{1}{p_3}\frac{\partial}{\partial p_3}\Big]\mathcal{I}_{\bs{\gamma}} .
\end{align}
Finally,  on the physical hypersurface, the DWI \eqref{triDWI} reduces to
\begin{align}\label{physeq3}
0 &= \Big(\frac{d}{2}+\theta_1+\theta_2+\theta_3+\partial_4+\partial_5+\partial_6\Big)\mathcal{I}_{\bs{\gamma}}\nn\\
&=
\Big(\frac{d}{2}-\gamma_1-\gamma_2-\gamma_3-\theta_1-\theta_2-\theta_3\Big)\mathcal{I}_{\bs{\gamma}}\nn\\
&=\frac{1}{2}\Big(d-2\gamma_1-2\gamma_2-2\gamma_3-p_1\frac{\partial}{\partial p_1}-p_2\frac{\partial}{\partial p_2}-p_3\frac{\partial}{\partial p_3}\Big)\mathcal{I}_{\bs{\gamma}}
\end{align}
Equations \eqref{physeq1}-\eqref{physeq3} involve only physical variables, namely, the momentum magnitudes.

\subsection{Affine reparametrisations}

As we have seen, the set of Euler equations associated with a given GKZ integral can be read off from the rows of the  $\mathcal{A}$-matrix: in the $i$th Euler equation \eqref{EulereqnsN}, the coefficient of the operator $\theta_j$ is $a_{ij}=(\mathcal{A})_{ij}$ where $1\le i\le N$ and $1\le j\le n$.  (Recall we are labelling the top all-$1$s row of the $\mathcal{A}$-matrix as $i=0$.)
Viewed in reverse, the set of Euler equations determines both the $\mathcal{A}$-matrix and the set of parameters $\bs{\gamma}$, and hence the GKZ integral.

What happens if we now form a {\it new} set of Euler equations by taking  linear combinations of the old ones?  In the process, we could simultaneously add to each Euler equation some multiple of the DWI.
Together, these operations correspond to left-multiplying  the $\mathcal{A}$-matrix by an $(N+1)\times(N+1)$ matrix 
\[\label{affineM}
\mathcal{M}=
\left(\begin{matrix} 1 & \bs{0}\\
\bs{b} & M\end{matrix}\right)\!,
\]
where $\bs{0}$ is an $N$-dimensional row vector of zeros, $\bs{b}$ is an $N$-dimensional column vector and $M$ an $N\times N$ matrix.  This yields
\[\label{affineA}
\mathcal{A}'=\mathcal{M}\mathcal{A} =
\left(\begin{matrix} 1 & \bs{0}\\
\bs{b} & M\end{matrix}\right)\left(\begin{matrix}\bs{1}\\A\end{matrix}\right)=\left(\begin{matrix}\bs{1}\\A'\end{matrix}\right)\!,
\]
where the components of $A$ undergo the  affine transformation
\[\label{aprime}
(A')_{ij}=a'_{ij}=b_i+ \sum_{k=1}^N m_{ik}a_{kj}.
\]
The new set of Euler equations now corresponds to the rows of $\mathcal{A}'$: the $i$th new Euler equation  is  the sum of $m_{ik}$ times the $k$th old Euler equation 
 plus $b_i$ times the DWI (for which the coefficient of every $\theta_j$ is one).    
In order to have $a'_{ij}\in\mathbb{Z}^+$, so as to form a new denominator polynomial $\mathcal{D}'$ via \eqref{GKZden}, we will restrict the entries of $\mathcal{M}$ to 
 $m_{ij}\in\mathbb{Z}^+$ and $b_i\in\mathbb{Z}^+$.
Note the transformation  \eqref{affineA} leaves the DWI unchanged.

The new set of Euler equations now takes the form
\[
0 = \Big(\bs{\gamma}'+\mathcal{A}'\cdot\bs{\theta}\Big)\mathcal{I}_{\bs{\gamma}'},
\]
where
\begin{align}\label{gammaprime}
\bs{\gamma}' = \left(\begin{matrix}\gamma_0\\\gamma'_1\\\vdots\\\gamma'_N\end{matrix}\right) =\left(\begin{matrix} 1 & \bs{0}\\
\bs{b} & M\end{matrix}\right) \left(\begin{matrix}\gamma_0\\\gamma_1\\\vdots\\\gamma_N\end{matrix}\right)= \mathcal{M}\bs{\gamma}
\end{align}
so that  $\gamma'_i = \gamma_0 b_i+\sum_{k=1}^N m_{ik}\gamma_k$ for $1\le i\le N$ while the DWI \eqref{DWI} remains unchanged.
Provided that $\mathrm{det}(\mathcal{M})$ is nonzero,  the toric equations are also unchanged since the kernel of $\mathcal{A}$ is preserved under multiplication by an invertible matrix.

What is now the relation of this new GKZ integral, defined by $\mathcal{A}'$, to the original?
The new integral is
\[
\mathcal{I}_{\bs{\gamma}'} =\Big(\prod_{i=1}^N \int_0^\infty\D z'_i\, (z'_i)^{\gamma'_i-1} \Big)(\mathcal{D}')^{-\gamma_0},
\]
where 
\[\label{GKZdenprime}
\mathcal{D}' = \sum_{j=1}^n x_j \prod_{i=1}^N (z'_i)^{a'_{ij}}.
\]
Using \eqref{aprime}, and making the identification
\[
z_k = \prod_{i=1}^N (z'_i)^{m_{ik}},
\]
we find
\begin{align}
\mathcal{D}' &= \sum_{j=1}^n x_j \prod_{i=1}^N (z'_i)^{b_i+\sum_{k=1}^N m_{ik}a_{kj}} 
= \Big(\prod_{l=1}^N (z'_l)^{b_l}\Big)\Big(\sum_{j=1}^n x_j
\prod_{i=1}^N\prod_{k=1}^N (z'_i)^{m_{ik}a_{kj}}\Big) 
\nn\\&
=\Big(\prod_{l=1}^N (z'_l)^{b_l}\Big)\Big(\sum_{j=1}^n x_j
\prod_{k=1}^Nz_k^{a_{kj}}\Big)=\Big(\prod_{l=1}^N (z'_l)^{b_l}\Big)\mathcal{D}.
\end{align}
Moving the factor of $\prod_{l=1}^N (z'_l)^{b_l}$ from the denominator to the numerator and using \eqref{gammaprime} then gives
\begin{align}
\mathcal{I}_{\bs{\gamma}'} &=\Big(\prod_{i=1}^N \int_0^\infty\D z'_i\, (z'_i)^{\gamma'_i-\gamma_0 b_i-1} \Big) \mathcal{D}^{-\gamma_0} = \Big(\prod_{i=1}^N \int_0^\infty\D z'_i\, (z'_i)^{\sum_{k=1}^Nm_{ik}\gamma_k-1} \Big) \mathcal{D}^{-\gamma_0} \nn\\
&= \Big(\prod_{i=1}^N \int_0^\infty\frac{\D z'_i}{z'_i}\,\prod_{k=1}^N (z'_i)^{m_{ik}\gamma_k} \Big) \mathcal{D}^{-\gamma_0}=
\Big(\prod_{i=1}^N \int_0^\infty\frac{\D z'_i}{z'_i}\, z_i^{\gamma_i} \Big)\mathcal{D}^{-\gamma_0}.
\end{align}
Finally, since
\[
\frac{\D z_i}{z_i}=\sum_{j=1}^N m_{ji}\frac{\D z_j'}{z'_j}, \qquad \qquad \prod_i \int_0^\infty\frac{\D z_i}{z_i} =  |\mathrm{det}\,M|\prod_i \int_0^\infty\frac{\D z'_i}{z'_i}, 
\]
we find
\[\label{IgIgp}
\mathcal{I}_{\bs{\gamma}'} = |\mathrm{det}\,M|^{-1}\mathcal{I}_{\bs{\gamma}}. 
\]
Thus, choosing a new basis for the Euler equations by taking linear combinations of the old Euler equations and the DWI only rescales the GKZ integral by a constant factor.  As the GKZ system of equations is linear, this overall scaling is in any case not fixed and the solution is effectively unchanged.

\paragraph{Example:} \label{tripleKex} The affine reparametrisation above can be used to show the equivalence  of the massless triangle integral \eqref{masslesstri} with the {\it triple-K integral} (see also \cite{Bzowski:2013sza,Bzowski:2020kfw})
\[\label{tripleKdef}
I_{\alpha,\{\beta_1,\beta_2,\beta_3\}} = \int_0^\infty \D z\, z^{\alpha}\prod_{i=1}^3 p_i^{\beta_i} K_{\beta_i}(p_i z).
\]
For  $\alpha=d/2-1$ and $\beta_i=\Delta_i-d/2$, this integral represents the momentum-space 3-point function of scalars $\O_{\Delta_i}$ in any $d$-dimensional CFT.
The triple-$K$ integral can be put into GKZ form by first Schwinger parametrising the modified Bessel functions as 
\[
p_i^{\beta_i} K_{\beta_i}(p_i z)=\frac{1}{2} \int_0^\infty \D z_i' \, (z_i')^{\beta_i-1}\exp\Big[-\frac{z}{2}\Big(z_i'+\frac{p_i^2}{z_i'}\Big)\Big]
\]
then performing the $z$ integral.  This gives
\[
I_{\alpha,\{\beta_1,\beta_2,\beta_3\}} = 2^{\alpha-2}\Gamma(\alpha+1)\Big(\prod_{i=1}^3\int_0^\infty \D z_i'\,(z'_i)^{\beta_i-1}\Big)\Big[\sum_{j=1}^3\Big( z_j'+\frac{p_j^2}{z_j'}\Big)\Big]
^{-\alpha-1}
\]
which uplifts to the GKZ integral
\[\label{GKZ3rep00}
I_{\alpha,\{\beta_1,\beta_2,\beta_3\}} = 
2^{\alpha-2}\Gamma(\alpha+1)\Big(\prod_{i=1}^3\int_0^\infty \D z_i'\,(z'_i)^{\gamma_i'-1}\Big)(\mathcal{D}')^{-\gamma_0'}
\]
where
\[\label{3Kden}
\mathcal{D}' =  \frac{x_1}{z_1'}+ \frac{x_2}{z_2'}+ \frac{x_3}{z_3'}+x_4 z_1'+x_5 z_2'+x_6 z_3'. 
\]
The physical hypersurface ({\it i.e.,} the original triple-$K$ integral) corresponds to
\[\label{3Kphys}
\gamma_i'=\beta_i,\qquad \gamma_0'=\alpha+1, \qquad
\x = (p_1^2,p_2^2,p_3^2,1,1,1).
\]
Here, we are using primes to distinguish the parameters of the triple-$K$ integral from those of the massless triangle integral earlier.
Also, while the denominator \eqref{3Kden} is not a polynomial, this simple generalisation will nevertheless   turn out to be the most convenient representation for us later.\footnote{
Should a  purely polynomial denominator be required, one can simply pull out an overall factor of $(z_1' z_2' z_3')^{-1}$ from the right-hand side of \eqref{3Kden} then transfer this to the numerator by shifting the $\gamma_i'$.}
The $\mathcal{A}$-matrix corresponding to the triple-$K$ integral is then 
\[\label{3KA}
\mathcal{A}_{\mathrm{3K}} 
=\left(\begin{matrix}
1&1&1&1&1&1\\
-1&0&0&1&0&0\\
0&-1&0&0&1&0\\
0&0&-1&0&0&1
\end{matrix}
\right).
\]
Comparing with the massless triangle $\mathcal{A}$-matrix \eqref{Atri}, we find that 
\[
\mathcal{M}\mathcal{A}_{\mathrm{triangle}} =\mathcal{A}_{\mathrm{3K}},
\]
where
\[\label{Mtrito3K}
\mathcal{M} = \left(\begin{matrix} 1 & 0 & 0 & 0\\1 & 0&-1&-1\\
1&-1&0&-1\\1&-1&-1&0 \end{matrix}\right).
\]
The parameters of the triangle integral are  connected to those of the triple-$K$ integral by
\[\label{affinetransfofg}
\mathcal{M}\bs{\g}_{\mathrm{triangle}} =  \mathcal{M}\left(\begin{matrix}d/2\\ \gamma_1\\ \gamma_2\\ \gamma_3\end{matrix}\right) = \left(\begin{matrix} d/2\\d/2-\gamma_2-\gamma_3\\d/2- \gamma_1-\gamma_3\\
d/2-\gamma_1-\gamma_2\end{matrix}\right)=\left(\begin{matrix} \alpha+1\\ \beta_1\\ \beta_2\\ \beta_3\end{matrix}\right) = \bs{\gamma}_{3K}.
\]
Putting everything together, 
 from \eqref{IgIgp} with $\mathrm{det}\,\mathcal{M}=2$ and \eqref{Cdef}, we have
\[
I_{d/2-1,\{d/2-\gamma_2-\gamma_3,\,d/2-\gamma_1-\gamma_3,\,d/2-\gamma_1-\gamma_2\}} =C'\int\frac{\mathrm{d}^d\bs{q}}{(2\pi)^d}\frac{1}{q^{2\gamma_3}|\bs{q}-\bs{p}_1|^{2\gamma_2}|\bs{q}+\bs{p}_2|^{2\gamma_1}}
\]
where
\[
C' = \pi^{d/2}2^{3d/2-4}\Gamma(d-\gamma_t)\prod_{i=1}^3\Gamma(\gamma_i).
\]
As we saw above, the matrix multiplication here is  just a slick way of executing the change of variables 
\[
z_1 = \frac{1}{z_2'z_3'}, \qquad z_2 =\frac{1}{ z_1'z_3'},\qquad z_3 = \frac{1}{z_1'z_2'},
\]
on the triangle GKZ representation, followed by moving a factor
of $(z_1'z_2'z_3')^{-\gamma_0}$ from the denominator to the numerator.

\section{Spectral singularities and the Newton polytope}
\label{sec:sings}

We now turn to examine the singularities of GKZ integrals arising for special values  of the parameters $\bs{\g}$.  As we will see, these can be viewed geometrically in terms of the {\it Newton polytope} of the GKZ denominator $\mathcal{D}$.

\subsection{The Newton polytope}

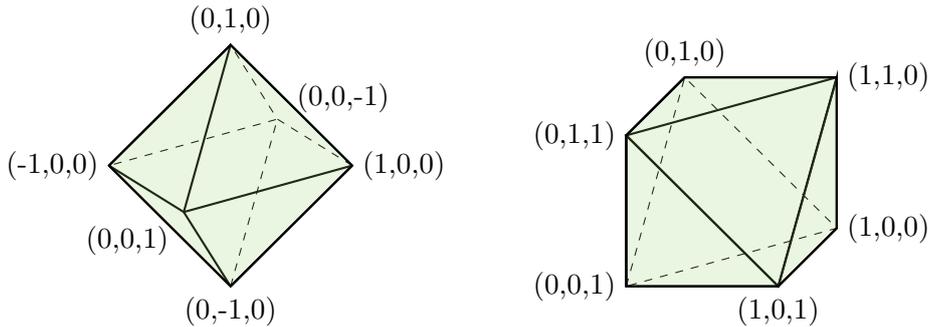
\begin{figure}[t]
\centering
\begin{tikzpicture}[scale=1.6]
   \draw[thick,black] (1,0,0) -- (0,1,0) -- (0,0,1); 
  \draw[thick,black]  (-1,0,0) -- (0,-1,0) ;
   \draw[thick,black] (-1,0,0) -- (0,1,0) ;
    \draw[thick,black] (-1,0,0) -- (0,0,1)--(1,0,0) ;
    \draw[thick,black] (0,0,1) -- (0,-1,0)--(1,0,0) ;
   \draw[style=dashed, color=black] (0,-1,0)-- (0,0,-1)--(-1,0,0);
   \draw[style=dashed, color=black] (1,0,0)-- (0,0,-1)--(0,1,0); 
     \draw node[right] at (1,0,0) {(1,0,0)};   
   \draw node[above] at (0,1,0) {(0,1,0)};
   \draw node[left] at (0,-0.2,1.1) {(0,0,1)};
   \draw node[left] at (-1,0,0) {(-1,0,0)};
    \draw node[below] at (0,-1,0) {(0,-1,0)};
   \draw node[right] at (0,0.1,-1.2) {(0,0,-1)};

\draw[thick,black,fill=YellowGreen, fill opacity=0.2] (-1,0,0)--(0,-1,0) -- (1,0,0) -- (0,1,0) --cycle;

   \end{tikzpicture}
\qquad
\begin{tikzpicture}[scale=2]

   \draw[thick,black] (1,0,0) -- (1,0,1) -- (0,1,1) -- (0,1,0); 
   \draw[thick,black] (0,0,1) -- (0,1,1) -- (0,1,0) ;
   \draw[thick,black] (1,1,0) -- (0,1,1) -- (1,0,1)--(1,1,0) ;
   \draw[thick,black] (1,0,1) -- (1,1,0) -- (1,0,0) ;
   \draw[thick,black] (1,0,1) -- (0,0,1) -- (0,1,1) ;
   \draw[thick,black] (1,1,0) -- (1,0,0) ;
   \draw[thick,black] (0,1,0)--(1,1,0) ;
 \draw[style=dashed, color=black] (0,0,1) -- (0,1,0) -- (1,0,0)--(0,0,1);
    \draw node[right] at (1,0,0) {(1,0,0)};   
   \draw node[above] at (0,1,0) {(0,1,0)};
   \draw node[left] at (0,0,1) {(0,0,1)};
   \draw node[right] at (1,1,0) {(1,1,0)};
    \draw node[below] at (1,0,1) {(1,0,1)};
   \draw node[left] at (0,1,1) {(0,1,1)};

\draw[thick,black,fill=YellowGreen, fill opacity=0.2] (0,0,1)--(1,0,1) -- (1,0,0) -- (1,1,0) --(0,1,0)--(0,1,1)--cycle;

   \end{tikzpicture}
   \caption{The Newton polytopes corresponding to the denominators of the triple-$K$ integral  \eqref{3Kden} (left) and the massless triangle integral \eqref{triD} (right).}
\label{fig:tripolytope}
\end{figure}

A defining feature of the GKZ representation is that only a single denominator \eqref{GKZden} is present:
\[
\mathcal{D} = \sum_{j=1}^n x_{j} \prod_{i=1}^N z_i^{a_{ij}}.
\] 
The exponents of the $j$th term in this denominator 
define a vector $\bs{a}_j$ living in an $N$-dimensional space, whose components  are
\[
(\bs{a}_j)_i = a_{ij}, \qquad i=1,\ldots N.
\]
Thus, $\bs{a}_j$ is  the $j$th column of the $\mathcal{A}$-matrix after stripping off the top row of all $1$s.
Constructing the convex hull of these exponent vectors then defines the $N$-dimensional  Newton polytope of $\mathcal{D}$:
\[
\mathrm{Newt}(\mathcal{D}) =  \sum_{j=1}^n\alpha_j \bs{a}_j, \quad\mathrm{with}\quad \sum_{j=1}^n \alpha_j = 1, \,\quad \alpha_j\ge 0\,\,\, \forall\,\, \, j.
\]
For the denominator \eqref{3Kden} of the triple-$K$ integral, for example,  we obtain the regular octahedron shown on the left of  figure \ref{fig:tripolytope}.  For the denominator of the massless triangle integral \eqref{triD}, we also obtain an octahedron, but now with vertices as shown on the right of the figure.  The  vertices of each polytope are related by the affine transformation  \eqref{aprime},
\[
\bs{a}_j^{(3K)} = \bs{b}+M \bs{a}_j^{\mathrm{(triangle)}}, \qquad j=1,\ldots, 6
\]
where, from \eqref{affineM} and  \eqref{Mtrito3K}, 
\[
\bs{b} = \left(\begin{matrix} 1\\1\\1\end{matrix}\right),
\qquad M = \left(\begin{matrix} 0 & -1 &-1\\- 1 & 0 &- 1\\ -1 & -1&0\end{matrix}\right)\!.
\]
As we saw above,  for any two $\mathcal{A}$-matrices (and hence any two Newton polytopes) related by an affine transformation, the corresponding GKZ integrals are proportional to each other and satisfy the same system of  equations ({\it i.e.,} DWI, Euler and toric equations). 
Thus, Newton polytopes such as these related by affine transformations  are effectively equivalent.

\subsection{Spectral singularities}

The physical significance of the Newton polytope becomes apparent when we consider the {\it spectral singularities} of the GKZ integral.
These are the divergences that arise for special values of the parameters $\bs{\gamma}$, with general kinematics, and are distinct from the {\it kinematic} (or Landau) singularities (discussed, {\it e.g.,} in \cite{Klausen:2021yrt}) which arise for general $\bs{\gamma}$ but special kinematics. 
Remarkably, it can be shown \cite{nilsson2010mellin,berkesch2013eulermellin} that the spectral singularities 
are closely related to the 
facets ({\it i.e.,} co-dimension one faces) 
  of the Newton polytope. 
As this polytope lives in an $N$-dimensional space, let us first define the $N$-dimensional parameter vector 
\[
\hat{\bs{\gamma}}=(\gamma_1,\ldots,\gamma_N)^T,
\]
where the hat serves to distinguish from the $(N+1)$-dimensional parameter vector $\bs{\g} = (\g_0,\hat{\bs{\g}})^T$.
In addition, we define the {\it rescaled} Newton polytope to be the convex hull of the  vertex vectors $\g_0\bs{a}_j$.  This corresponds to a linear rescaling\footnote{The significance of this rescaling can be anticipated by noting that the Newton polytope of the GKZ denominator $\mathcal{D}^{\g_0}$, in the special cases where $\g_0\in\mathbb{N}$ so that $\mathcal{D}^{\g_0}$ is itself a polynomial when expanded out, is simply the Newton polytope of $\mathcal{D}$ linearly rescaled by $\g_0$. } of the original Newton polytope by a factor of $\g_0$.
 The GKZ integral is then finite for all parameter values $\hat{\bs{\g}}$ lying {\it within} this rescaled Newton polytope.  On the hyperplanes corresponding to the facets of the rescaled Newton polytope, as well as on an infinite set of further hyperplanes both parallel and exterior to these facets, the integral is singular.

An exact formula for all  singular hyperplanes will be derived below in \eqref{hypsings}.  The location of these singularities will then be the main ingredient in our subsequent construction of creation operators.
Two key steps are needed to establish the result \eqref{hypsings}. 
 The first is to show that the GKZ integral converges for all $\hat{\bs{\gamma}}$ values lying in the interior of the rescaled Newton polytope.  Rather than 
recounting  the 
formal proof of \cite{nilsson2010mellin,berkesch2013eulermellin},  we will instead pursue a more informal approach based on a tropical analysis of the GKZ integral \cite{Arkani-Hamed:2022cqe,Matsubara-Heo:2023ylc}.   Many closely related constructions appear in sector decomposition, see {\it e.g.,} \cite{Kaneko:2009qx, Schultka:2018nrs}.
 The second step in the analysis is to construct a series of  meromorphic continuations across each of the singular hyperplanes.  This can be achieved by a scaling argument due to  \cite{nilsson2010mellin,berkesch2013eulermellin}.
 Here, we present a further variation of this argument involving a special linear combination of the Euler equations and DWI.

\paragraph{Example:}  \label{Ex1} As an initial check of the picture  above, we recall that the spectral singularities of the triple-$K$ integral \eqref{tripleKdef} are already known from conformal field theory \cite{Bzowski:2015pba}.\footnote{The argument in \cite{Bzowski:2015pba} involves  expanding the integrand of the triple-$K$ integral about its lower limit and looking for the appearance of $z^{-1}$ poles.}
The condition for  the triple-$K$ integral $I_{\alpha,\{\beta_1,\beta_2,\beta_3\}}$ to be singular is
\[
\alpha+1\pm\beta_1\pm\beta_2\pm\beta_3 = -2m,\qquad m\in \mathbb{Z}^+
\]
where any independent choice of the three $\pm$ signs can be made, and any value $m=0,1,2,\ldots$ is permitted.
(Throughout this paper, we will take $\mathbb{Z}^+$ to be the set of all  non-negative integers {\it including} zero.)
Re-expressing this condition in terms of the $\bs{\gamma}$ parameters  \eqref{3Kphys} appearing in the GKZ integral, and dropping the primes, this is
\[\label{3Kgammasings}
\gamma_0 \pm \gamma_1\pm\gamma_2\pm\gamma_3=-2m.
\]
We see immediately that the $m=0$ singularities indeed correspond to 
the equations of the hyperplanes containing the eight facets of the regular octahedron on the left of figure \ref{fig:tripolytope}, where the vertices in the figure correspond to $(\gamma_1,\gamma_2,\gamma_3) = \gamma_0(\pm 1,0,0),$ $\gamma_0(0,\pm 1,0)$ and $\gamma_0(0,0,\pm 1)$.
 The remaining singularities for  $m>0$ then correspond to an infinite series of regularly spaced hyperplanes, both parallel, and exterior, to the facets of the octahedron.

\subsubsection{Tropical analysis: an example}

To appreciate the role of the Newton polytope, let us start with a simple example introduced in \cite{nilsson2010mellin}.  This is the  GKZ integral 
\[\label{houseint}
\mathcal{I}_{\bs{\gamma}} = \int_0^\infty\D z_1\int_0^\infty\D z_2 \,z_1^{\g_1-1}z_2^{\g_2-1} (x_1+x_2 z_2+x_3 z_1^2+x_4 z_1 z_2^2)^{-\gamma_0},
\]
whose $\mathcal{A}$-matrix is
\[\label{houseA}
\mathcal{A} = \left(\begin{matrix} 1 & 1 & 1& 1\\ 0 & 0 & 2 & 1\\ 0 & 1 & 0 & 2\end{matrix}\right)\!.
\]
The singularities of the integral derive from regions where the $z_i$ (for $i=1,2$) either vanish or tend to infinity.  Setting $z_i=e^{\tau_i}$, these regions are mapped to $|\tau_i|\rightarrow \infty$ and 
\[
\mathcal{I}_{\bs{\gamma}} = \int_{-\infty}^\infty\D \tau_1\int_{-\infty}^\infty\D \tau_2 \,e^{\gamma_1\tau_1+\gamma_2\tau_2}(x_1+x_2e^{\tau_2}+x_3 e^{2\tau_1}+x_4 e^{\tau_1+2\tau_2})^{-\gamma_0}.
\]
For large $|\tau_i|$, we can approximate this integral by  its {\it tropicalisation} as discussed in  \cite{Arkani-Hamed:2022cqe},
\[\label{trophouse}
\mathcal{I}_{\bs{\gamma}}^{\mathrm{trop.}} =x_j^{-\gamma_0} \int_{-\infty}^\infty\D \tau_1\int_{-\infty}^\infty\D \tau_2 \exp\Big[\gamma_1\tau_1+\gamma_2\tau_2- \gamma_0\, \mathrm{max}(0,\,\tau_2,\,2\tau_1,\,\tau_1+2\tau_2)\Big],
\]
which corresponds to retaining only the leading exponential in the GKZ denominator.  Which term this is will depend on which sector of the  $(\tau_1,\tau_2)$ plane we are in.  If the dominant term is, say, the $j$th one, then the overall prefactor is $x_j^{-\gamma_0}$ as shown. 
If all $x_k>0$ for $k=1,\ldots, 4$,  the tropicalisation of the denominator in fact provides a lower bound and so, for real $\gamma_0>0$ and real $\gamma_1$ and $\gamma_2$, we have $\mathcal{I}_{\bs{\gamma}}< \mathcal{I}_{\bs{\gamma}}^{\mathrm{trop.}}$.   The convergence of $\mathcal{I}_{\bs{\gamma}}^{\mathrm{trop.}}$ then establishes that of $ \mathcal{I}_{\bs{\gamma}}$.   (For rigorous bounds allowing complex $\gamma_i$, see \cite{nilsson2010mellin,berkesch2013eulermellin}.)

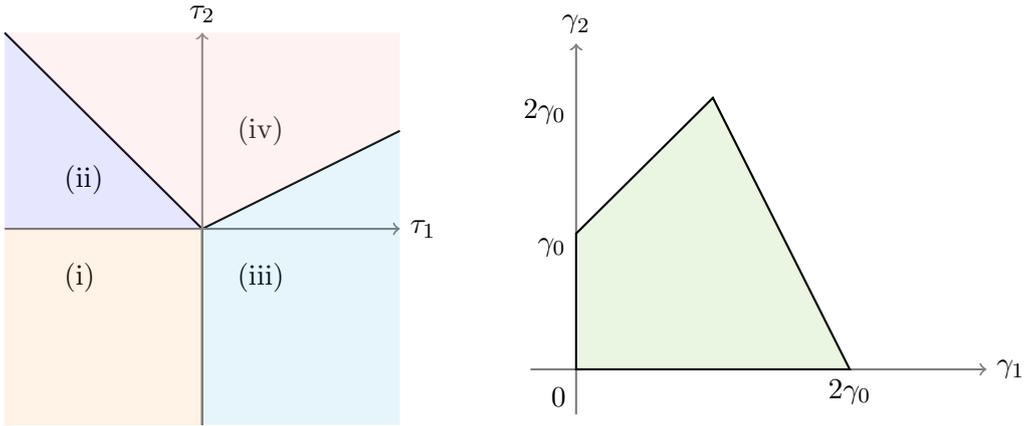
\begin{figure}[t]
\centering
\begin{tikzpicture}[scale=0.65]

\draw[thick](-4,4)--(0,0);
\draw[thick](0,-4)--(0,0);
\draw[thick](0,0)--(4,2)
;\draw[->,thick,gray] (-4,0)--(4,0);
\draw[->,thick,gray] (0,-4)--(0,4);
    \draw node[above] at (0,4) {$\tau_2$};
     \draw node[right] at (4,0) {$\tau_1$};
      \draw node[right] at (-3,-1) {(i)};
            \draw node[right] at (-3,1) {(ii)};
                  \draw node[right] at (0.5,2) {(iv)};
                        \draw node[right] at (0.5,-1) {(iii)};
                        
\fill[blue,opacity=0.1](-4,0)--(0,0)--(-4,4);
\fill[orange,opacity=0.1](-4,0)--(0,0)--(0,-4)--(-4,-4);
\fill[Cerulean,opacity=0.1](0,0)--(0,-4)--(4,-4)--(4,2);
\fill[pink,opacity=0.2](-4,4)--(0,0)--(4,2)--(4,4);

\end{tikzpicture}
\qquad
\begin{tikzpicture}[scale=1.8]
\draw[->,thick,gray] (2,0) -- (3,0); 
\draw[thick,gray] (-1/3,0) -- (0,0); 
\draw[->,thick,gray] (0,1) -- (0,2.4); 
\draw[thick,gray] (0,0) -- (0,-1/3); 
\draw[white] (0,-1/3) -- (0,-1/3-0.08);   

\draw[thick,black,fill=YellowGreen, fill opacity=0.2] (0,0)--(2,0) -- (1,2) -- (0,1) --cycle;

     \draw node[left] at (0,-0.2) {0};   
   \draw node[left] at (0,1-0.1) {$\gamma_0$};
   \draw node[left] at (0,2-0.1) {$2\gamma_0$};
   \draw node[below] at (2,0) {$2\gamma_0$};
    \draw node[right] at (3,0) {$\gamma_1$};
     \draw node[above] at (0,2.4) {$\gamma_2$};
   
  \end{tikzpicture}
\caption{{\it Left:}  The integration sectors for the tropicalised GKZ integral \eqref{trophouse}, where each sector corresponds to the dominance of a different term in the denominator. 
The sector boundaries are simultaneously the normals to the facets of the Newton polytope shown on the right.  {\it Right:} Combining the conditions on $\gamma_1$ and $\gamma_2$ for the convergence of each sector, we obtain the interior of the Newton polytope (rescaled by $\gamma_0$) as shaded.
 \label{fig:sectors}}
\end{figure}

The various integration sectors, as illustrated in figure \ref{fig:sectors}, are then as follows:
\begin{itemize}
\item[(i)] $\tau_1<0$ and $\tau_2<0$ so $j=1$ and $\mathrm{max}(0,\,\tau_2,\,2\tau_1,\,\tau_1+2\tau_2)=0$.
\item[(ii)] $\tau_1+\tau_2<0$ and $\tau_2>0$ so $j=2$ and $\mathrm{max}(0,\,\tau_2,\,2\tau_1,\,\tau_1+2\tau_2)=\tau_2$.
\item[(iii)] $\tau_1>0$ and $\tau_1-2\tau_2>0$ so $j=3$ and $\mathrm{max}(0,\,\tau_2,\,2\tau_1,\,\tau_1+2\tau_2)=2\tau_1$.
\item[(iv)] $\tau_1+\tau_2>0$ and $\tau_1-2\tau_2<0$ so
$j=4$ and $\mathrm{max}(0,\,\tau_2,\,2\tau_1,\,\tau_1+2\tau_2)=\tau_1+2\tau_2$.
\end{itemize}
Each sector forms a cone within which we can reparametrise  $\bs{\tau}=(\tau_1,\tau_2)$ as 
\[
\bs{\tau} = \lambda_1\bs{n}_1+ \lambda_2\bs{n}_2,\qquad \lambda_1,\lambda_2\ge 0
\]
where $\bs{n}_1$ and $\bs{n}_2$ are the outward-pointing vectors forming the boundary of that particular sector.   By inspection, these are simultaneously the normal vectors to the facets of the Newton polytope  shown in the right-hand panel of figure \ref{fig:sectors}, where the two normals chosen are those for the two facets  containing the leading vertex $j$. 
For the sector $j=3$, for example, we have $\bs{n}_1 = (0,-1)$ and $\bs{n}_2 = (2,1)$ and so $\tau_1=2\lambda_2$ and $\tau_2=-\lambda_1+\lambda_2$.  This third sector of the tropicalised integral is then 
\[
\mathcal{I}_{\bs{\gamma}}^{\mathrm{trop.}}\Big|_{j=3} = 2x_3^{-\gamma_0}\int_0^\infty\D\lambda_1\int_0^\infty\D\lambda_2 \exp\Big[ - \gamma_2\lambda_1+(2\gamma_1+\gamma_2-4\gamma_0)\lambda_2\Big]. 
\]
The linearity of the tropicalised exponent means that the integrals over $\lambda_1$ and $\lambda_2$ factorise, and for convergence as $\lambda_i\rightarrow \infty$, both exponents must separately be negative:
\[\label{houserules}
\gamma_2>0,\qquad -2\gamma_1-\gamma_2+4\gamma_0>0.
\]
This corresponds to the interior region bounded by the two lines intersecting the vertex $(\gamma_1,\gamma_2)=(2\gamma_0,0)$ in the right-hand panel of figure \ref{fig:sectors}.  This vertex is precisely that corresponding to the dominant $j=3$ term  (namely, $x_3z_1^2$) in the GKZ denominator, after rescaling by $\gamma_0$.  On the boundary of the convergence region, 
the integral has either a single or a double pole according to how many of the inequalities in \eqref{houserules} are saturated.

Repeating this exercise for the remaining  sectors,
we obtain the additional constraints
\[\label{houserules2}
\gamma_1>0, \qquad \gamma_1-\gamma_2+\gamma_0>0.
\]
Combining all these conditions, the full integral $\mathcal{I}_{\bs{\gamma}}^{\mathrm{trop.}}$ then converges for $(\gamma_1,\gamma_2)$ within the polytope shown in the figure.  This is indeed the Newton polytope for the GKZ denominator after rescaling all vertex vectors by $\gamma_0$.

\subsubsection{Tropical analysis: general case}

The  analysis above clearly  generalises. 
Setting again $z_i = e^{\tau_i}$, the general GKZ integral \eqref{GKZint} has the tropical approximation
\[\label{GKZtropical}
\mathcal{I}_{\bs{\gamma}}^{\mathrm{trop.}} =
\int_{\mathbb{R}^N}\D\bs{\tau}\,
\exp\Big[\sum_{i=1}^N \gamma_i\tau_i - \gamma_0\,\mathrm{max}_k\,\Big(\ln x_k+\sum_{i=1}^N a_{ik}\tau_i\Big)\Big].
\]
In particular, this is a good approximation precisely for the large $|\tau_i|$ regions where any singularities of the GKZ integral must arise, and so convergence of the tropical approximation implies convergence of the full GKZ integral.\footnote{
For real $\gamma_i$, $\gamma_0>0$ and $x_j>0$, the tropical approximation provides an upper bound on the GKZ integral as noted in the previous example.  Cases where the $\gamma_i$ can be complex and the $x_j$ are not constrained to be positive can be handled by establishing a rigorous bound on the GKZ denominator, see \cite{nilsson2010mellin,berkesch2013eulermellin}.}

The different integration sectors of the tropical integral \eqref{GKZtropical} correspond to when  different terms dominate and are selected as the maximum  in the exponent.  For sufficiently large $|\tau_i|$, this depends only on the {\it direction} in the $\bs{\tau}=(\tau_1,\ldots,\tau_N)$ plane and we can neglect any contribution from the $\ln x_k$ terms.
Let us consider then the sector  where, say, the $j$th term  forms the maximum.  This sector can be parametrised as
\[\label{gentau}
\bs{\tau} = \sum_{J\in \Phi_j} \lambda^{(J)}\bs{n}^{(J)},\qquad \lambda^{(J)}\ge 0,
\]
where $\Phi_j$ denotes the set of facets containing the vertex $j$, the $\lambda^{(J)}$ are the new integration variables, and
\[\label{nJdef}
\bs{n}^{(J)}=(n_1^{(J)},\ldots, n_N^{(J)})^T
\] 
is the outward-pointing normal to the facet $J$.  We will assume that $\Phi_j$ contains precisely $N$ facets so that \eqref{gentau} holds.\footnote{If there are fewer than this, we can factor out a finite integral over a transverse subspace following appendix A of \cite{Arkani-Hamed:2022cqe} then apply the argument above for the remaining integral over a lower-dimensional cone.} 
The contribution of this sector is then
\[
\mathcal{I}_{\bs{\gamma}}^{\mathrm{trop.}}\Big|_j = x_j^{-\gamma_0}
\prod_{J\in\Phi_j}\int_0^\infty\D\lambda^{(J)}
\exp\Big[\lambda^{(J)}\sum_{i=1}^N n_i^{(J)}(\gamma_i- \gamma_0 a_{ij})\Big].
\]
As in the previous example, convergence then requires each of these exponents to be negative giving
\[
\sum_{i=1}^N n_i^{(J)}(\gamma_i- \gamma_0 a_{ij})<0 \qquad \forall \, J\in \Phi_j.
\]
Viewed geometrically, these conditions state that the  parameter vector $\hat{\bs{\gamma}}$ lies to the {\it inside} of the $(N-1)$-dimensional hyperplane containing  facet $J$ of
the rescaled Newton polytope,
\[\label{inside}
\bs{n}^{(J)}\cdot (\hat{\bs{\g}}-\g_0\bs{a}_j) <0,
\]
and that this holds for all facets $J$ containing the $j$th vertex vector $\g_0\bs{a}_j$.
Convergence of the {\it full} tropicalised GKZ integral requires convergence in every integration sector, and hence for every vertex $j$ of the rescaled Newton polytope. The condition \eqref{inside} must thus hold for {\it all} facets $J$, meaning  $\hat{\bs{\gamma}}$ must lie completely inside the rescaled Newton polytope.

\subsection{Meromorphic continuation}

Having shown  the convergence of GKZ integrals for $\hat{\bs{\gamma}}$ lying within the rescaled Newton polytope, the existence of further infinite sets of singular hyperplanes parallel to each facet  can be established by meromorphic continuation \cite{nilsson2010mellin,berkesch2013eulermellin}.
Once again, the idea is most easily seen in the context of an example, after which we resume our  general analysis.

\subsubsection{Example}

Returning the GKZ integral \eqref{houseint}, let us construct a continuation across, say,  the upper-right facet of the Newton polytope shown on the right of figure \ref{fig:sectors}.  The relevant outward normal is $\bs{n}=(2,1)$.  Following \cite{nilsson2010mellin}, we perform a  rescaling $z_i\rightarrow \lambda^{-n_i}z_i$, namely $z_1\rightarrow \lambda^{-2}z_1$ and $z_2\rightarrow\lambda^{-1}z_2$, where $\lambda$ is some fixed parameter. The integral \eqref{houseint} becomes
\[
\mathcal{I}_{\bs{\gamma}} = \lambda^{-2\gamma_1-\gamma_2+4\gamma_0}\int_0^\infty\D z_1\int_0^\infty\D z_2 \,z_1^{\g_1-1}z_2^{\g_2-1} (x_1\lambda^4+x_2 z_2\lambda^3+x_3 z_1^2+x_4 z_1 z_2^2)^{-\gamma_0}
\]
but its {\it value} remains unchanged.  We can therefore differentiate to find
\[\label{lambdadiff}
0 = \frac{\D}{\D\lambda}\mathcal{I}_{\bs{\gamma}} \Big|_{\lambda=1}=(-2\gamma_1-\gamma_2+4\gamma_0)\mathcal{I}_{\bs{\gamma}} - 4\gamma_0 x_1\mathcal{I}_{\bs{\gamma}'}-3\gamma_0 x_2\mathcal{I}_{\bs{\gamma}''} 
\]
where
\begin{align}
\gamma_0'&=\gamma_0+1,\qquad \gamma_1'=\gamma_1,\qquad \gamma_2'=\gamma_2\nn\\
\gamma_0''&=\gamma_0+1,\qquad \gamma_1''=\gamma_1,\qquad \gamma_2''=\gamma_2+1.
\end{align}
Alternatively, \eqref{lambdadiff} can be obtained by taking a linear combination of the  Euler equations and DWI for \eqref{houseint}, namely
\begin{align}
0 &= \Big(-2(\gamma_1+2\theta_3+\theta_4)-(\gamma_2+\theta_2+2\theta_4)+4\big(\gamma_0+\sum_{j=1}^4\theta_j\big)\Big)\mathcal{I}_{\bs{\gamma}}\nn\\
&=\Big(4\gamma_0-2\gamma_1-\gamma_2+4\theta_1+3\theta_3\Big)\mathcal{I}_{\bs{\gamma}},
\end{align}
where evaluating the action of the $\theta_i=x_i\partial_{x_i}$ yields \eqref{lambdadiff}.

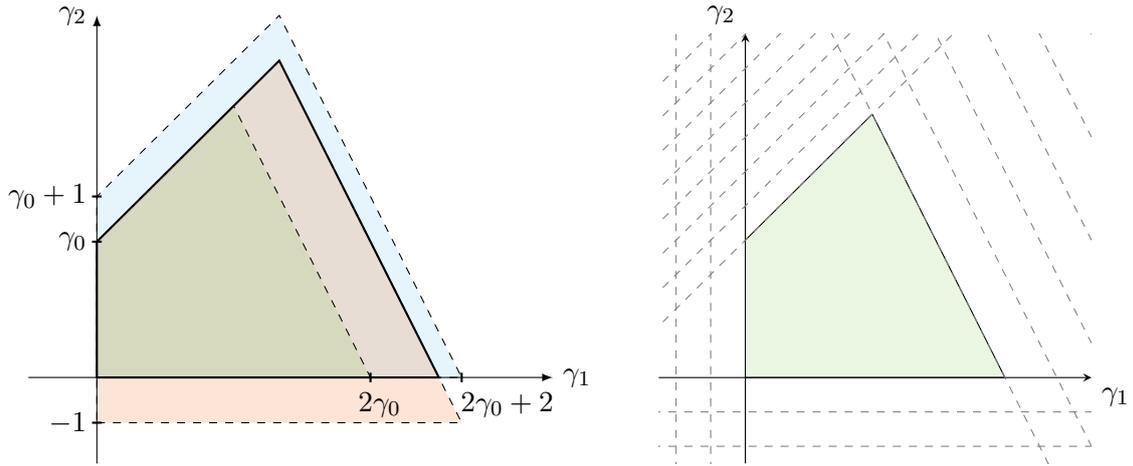
\begin{figure}[t]
\centering
\hspace{-0.5cm}\begin{tikzpicture}[scale=0.6]
\draw[-latex] (-1.5,0) -- (10,0) node[right]{$\gamma_1$};
\draw[-latex] (0,-1.9) -- (0,8) node[left]{$\gamma_2$};

\draw[dashed,black,fill=Cerulean, fill opacity=0.1] (0,0)--(8,0) -- (4,8) -- (0,4) --cycle;
\draw[dashed,black,fill=Orange, fill opacity=0.2] (0,-1)--(8,-1) -- (4,7) -- (0,3) --cycle;
\draw[dashed,black,fill=YellowGreen, fill opacity=0.2] (0,0)--(6,0) -- (3,6) -- (0,3) --cycle;

\draw[thick,black] (0,0)--(7.5,0) -- (4,7) -- (0,3)--cycle ;
\draw node[below] at (9,-0.1) {$2\gamma_0+2$};
\draw node[below] at (6.2,-0.1) {$2\gamma_0$};
\draw node[left] at (0,3) {$\gamma_0$};
\draw node[left] at (0,4) {$\gamma_0+1$};
\draw node[left] at (0,-1) {$-1$};

\foreach \x in {6,8} {%
    \draw [thick]($(\x,0) + (0,-\TickSize)$) -- ($(\x,0) + (0,\TickSize)$);
 }

 \foreach \y in {-1,3,4} {%
    \draw [thick]($(0,\y) + (-\TickSize,0)$) -- ($(0,\y) + (\TickSize,0)$);
}

\end{tikzpicture}
\hspace{5mm}
\begin{tikzpicture}[scale=1][>=latex]
\begin{axis}[
  axis x line=middle,
  axis y line=middle,
  xtick,
  ytick,
  xlabel={$\gamma_1$},
  ylabel={$\gamma_2$},
  xlabel style={below right},
  ylabel style={above left},
  xmin=-2.5,
  xmax=10,
  ymin=-2.5,
  ymax=10,
  axis equal image]

\addplot [mark=none,domain=0:11/3,name path=A] {x+3+1};
\addplot [mark=none,domain=11/3:7.5] {-2*x+12+3};
 \addplot [mark=none,domain=0:7.5,name path=B] {0};

\addplot [YellowGreen, fill opacity=0.2] fill between[of=A and B,soft clip={domain=0.01:8}];

\foreach \k in {1,...,8} {
 
\addplot [mark=none,dashed,gray,domain=-5:10] {x+3+\k};
\addplot [mark=none,dashed,gray,domain=-5:10] {-2*x+12+3*\k};
\addplot [mark=none,dashed,gray,domain=-5:10] {-\k};
\addplot [mark=none,dashed,gray,domain=-5:10] (-\k,x);

}
\end{axis}

\end{tikzpicture}

\caption{ {\it Left:} Convergence regions of $\mathcal{I}_{\bs{\gamma}}$ (green), $\mathcal{I}_{\bs{\gamma}'}$ (blue) and $\mathcal{I}_{\bs{\gamma}''}$ (orange). The meromorphic continuation \eqref{rearrlambdadiff} holds for the intersection of the convergence regions for $\mathcal{I}_{\bs{\gamma}'}$ and $\mathcal{I}_{\bs{\gamma}''}$.
This extends the domain of convergence of  $\mathcal{I}_{\bs{\gamma}}$ across the upper-right facet with normal $(2,1)$ to form the region bounded by the solid line. {\it Right:} Dashed lines indicate the complete set of singular hyperplanes \eqref{housesing} of the GKZ integral \eqref{houseint}.
\label{shiftedhouses}}
\end{figure}

As  both $\mathcal{I}_{\bs{\gamma}'}$ and $\mathcal{I}_{\bs{\gamma}''}$ in \eqref{lambdadiff} take the same form as the original integral $\mathcal{I}_{\bs{\gamma}}$, except with shifted parameters, the convergence regions are given by \eqref{houserules} and \eqref{houserules2} replacing $\bs{\gamma}$ with $\bs{\gamma}'$ or $\bs{\gamma}''$.  In terms of the unshifted parameters, $\mathcal{I}_{\bs{\gamma}'}$ thus converges for 
\begin{align}\label{gamshift1}
\gamma_1>0, \qquad \gamma_2>0,\qquad \gamma_0+\gamma_1-\gamma_2+1>0,\qquad 4\gamma_0-2\gamma_1-\gamma_2+4>0,
\end{align}
while
$\mathcal{I}_{\bs{\gamma}''}$  converges for 
\begin{align}\label{gamshift2}
\gamma_1>0, \qquad \gamma_2+1>0,\qquad \gamma_0+\gamma_1-\gamma_2>0,\qquad 4\gamma_0-2\gamma_1-\gamma_2+3>0.
\end{align}
In each case,  the size of the Newton polytope is rescaled from $\gamma_0\rightarrow\gamma_0+1$, while for $\mathcal{I}_{\bs{\gamma}''}$ we also translate by the vector $(0,-1)$ as shown in figure \ref{shiftedhouses}.  Note that neither of these operations change the normals to the facets.
Re-arranging \eqref{lambdadiff}, we now have
\[\label{rearrlambdadiff}
\mathcal{I}_{\bs{\gamma}} = (4\gamma_0-2\gamma_1-\gamma_2)^{-1}\big(4\gamma_0 x_1\mathcal{I}_{\bs{\gamma}'}+3\gamma_0 x_2\mathcal{I}_{\bs{\gamma}''} \big),
\]
where the sum of shifted integrals on the right-hand side converges only for the {\it intersection} of the two shifted polytopes \eqref{gamshift1} and \eqref{gamshift2}, namely
\begin{align}\label{gamshift3}
\gamma_1>0, \qquad \gamma_2>0,\qquad \gamma_0+\gamma_1-\gamma_2>0,\qquad 4\gamma_0-2\gamma_1-\gamma_2+3>0.
\end{align}
Comparing with the original polytope formed by \eqref{houserules} and \eqref{houserules2}, only the final inequality has changed.  Now, the region of convergence  extends across the facet with normal $(2,1)$ as shown in figure \ref{shiftedhouses}.  Equation \eqref{rearrlambdadiff} thus gives a meromorphic continuation of $\mathcal{I}_{\bs{\gamma}}$ around the pole at $4\gamma_0-2\gamma_1-\gamma_2=0$ (corresponding to the facet of the original Newton polytope normal to $(2,1)$) to the larger region \eqref{gamshift3}.  

This process can then be repeated for the boundary of the new region \eqref{gamshift3} by applying the same procedure (namely, rescaling $z_i\rightarrow \lambda^{-n_i}z_i$, differentiating with respect to $\lambda$ then setting $\lambda=1$) to the integrals on the right-hand side of \eqref{rearrlambdadiff}.  Alternatively, we can extend \eqref{rearrlambdadiff}  iteratively by using shifted analogues of \eqref{rearrlambdadiff} 
to replace $\mathcal{I}_{\bs{\gamma}'}$ and $\mathcal{I}_{\bs{\gamma}''}$ on the right-hand side of  \eqref{rearrlambdadiff} itself.
Repeating such calculations for all the facet normals of the original Newton polytope, we obtain an infinite set of singular hypersurfaces parallel to the facets of the Newton polytope.  The integral \eqref{houseint} is thus singular on the hyperplanes 
\begin{align}\label{housesing}
\gamma_1=-m_1,\qquad \gamma_2= -m_2,\qquad \gamma_0+\gamma_1-\gamma_2 = -m_3,\qquad
4\gamma_0-2\gamma_1-\gamma_2=-3m_4
\end{align}
for any (independent) choice of non-negative integers $m_i\in \mathbb{Z}^+$, as illustrated in the right-hand panel of figure \ref{shiftedhouses}.

\subsubsection{General analysis}\label{sec:genmero}

The analysis in this last example readily extends to general GKZ integrals.
We begin by defining a few useful quantities.  First, we have
 the $(N+1)$-dimensional vectors 
\[
\bs{\g} = \left(\begin{matrix}\g_0\\[0.5ex]\hat{\bs{\g}}\end{matrix}\right),  \qquad \bs{\mathcal{A}}_j = \left(\begin{matrix}1,\\[0.3ex]\bs{a}_j\end{matrix}\right), \qquad \bs{N}^{(J)} = \left(\begin{matrix}n_0^{(J)}\\[0.5ex]\bs{n}^{(J)}\end{matrix}\right),
\]
where $\bs{\g}$ is the usual GKZ parameter vector, $\bs{\mathcal{A}}_j$ is the $j$th column of the full $\mathcal{A}$-matrix (including the top row of $1$s), and, as above, $\bs{n}^{(J)}$ is the $N$-dimensional outwards-pointing normal to facet $J$ of the Newton polytope.  The additional component $n_0^{(J)}$ is fixed by requiring that
\[\label{NdotA}
0=\bs{N}^{(J)}\cdot\bs{\mathcal{A}}_j, \qquad  j\in\vphi_J
\]
where $\vphi_J$ denotes the set of vertices lying within the facet $J$, giving
\[\label{muJdef}
n_0^{(J)} =- \bs{n}^{(J)}\cdot\bs{a}_j,  \qquad  j\in\vphi_J.
\] 
The condition that $\hat{\bs{\g}}$ lies in the hyperplane containing facet $J$ of the rescaled Newton polytope,
\[
0=\bs{n}^{(J)}\cdot(\hat{\bs{\g}}-\g_0\bs{a}_j)
\] 
can now be compactly re-expressed as 
\[\label{gdotN}
0 = \bs{\g}\cdot \bs{N}^{(J)}
\]
and the domain of convergence \eqref{inside} corresponds to $\bs{\g}\cdot \bs{N}^{(J)}<0$ for all facets $J$.  (From an $(N+1)$-dimensional perspective, the Newton polytope therefore corresponds to a cone.)   
In addition, we define the distance function
\[\label{ddef}
d_i^{(J)} = -\bs{\mathcal{A}}_i\cdot \bs{N}^{(J)}
=\bs{n}^{(J)}\cdot(\bs{a}_j-\bs{a}_i),\qquad j\in\vphi_J.
\]
If  $\bs{n}^{(J)}$ is a unit vector, $d_i^{(J)}$ is the normal distance from vertex $i$  to facet $J$ of the Newton polytope.
Rather than choosing $\bs{n}^{(J)}$ to be a unit vector, however, it will be more convenient in practice to choose  $\bs{n}^{(J)}$ (and hence $\bs{N}^{(J)}$) to have integer components.

We now proceed to construct a meromorphic continuation of the GKZ integral across a chosen facet $J$ of the rescaled Newton polytope.  
To this end, we form a linear combination of 
 $n_0^{(J)}$ times the DWI plus the sum of $n_k^{(J)}$ times the $k$th Euler equation, namely
\begin{align}\label{euldwisum}
0&= \Big[n_0^{(J)}\Big(\gamma_0+\sum_{l=1}^n\theta_l\Big)+\sum_{k=1}^N n_k^{(J)}\Big(\gamma_k+\sum_{l=1}^n a_{kl}\theta_l\Big)\Big]\mathcal{I}_{\bs{\gamma}}\nn\\
&=\Big(\bs{\g}\cdot\bs{N}^{(J)}-
\sum_{l=1}^nd_l^{(J)}\, \theta_l \Big)\mathcal{I}_{\bs{\gamma}}.
\end{align}
The sum over $l$ on the second line here can be restricted to values $l\notin\vphi_J$, corresponding to vertices $l$ not in the facet $J$, since $d_l^{(J)}$ vanishes for all $l\in\vphi_J$.
Moreover, by direct differentiation of the GKZ integral as we will discuss further in section \ref{sec:annihilate}, one can show that 
\[
\theta_l \mathcal{I}_{\bs{\gamma}} 
= -\gamma_0 x_l \mathcal{I}_{\bs{\gamma}+\bs{\mathcal{A}}_l}.
\]
Here, the parameter vector of the right-hand integral has been shifted from $\bs{\g}\rightarrow\bs{\g}+\bs{\mathcal{A}}_l$.
Rearranging, this immediately gives the desired meromorphic continuation:\footnote{Alternatively, this equation can be derived by rescaling all $z_i\rightarrow \lambda^{-n_i^{(J)}}z_i$ in $\mathcal{I}_{\bs{\gamma}}$ and extracting a prefactor of $\lambda^{-\bs{\g}\cdot\bs{N}^{(J)}}$.  We then differentiate with respect to $\lambda$ and set $\lambda=1$ analogously to in \eqref{lambdadiff}.}
\begin{align}\label{genmero}
\mathcal{I}_{\bs{\gamma}} = -\frac{\g_0}{\bs{\g}\cdot\bs{N}^{(J)}}
\Big(\sum_{l\notin\vphi_J} \,x_l \, d_l^{(J)}\,\mathcal{I}_{\bs{\gamma}+\bs{\mathcal{A}}_l}\Big).
\end{align}
The denominator $\bs{\g}\cdot\bs{N}^{(J)}$ generates a pole at the hyperplane containing the facet $J$, while the sum of shifted integrals has a larger domain of convergence extending across the facet $J$ of the original rescaled Newton polytope for $\mathcal{I}_{\bs{\gamma}}$.

To see this,
for each shifted integral labelled by an $l\notin\vphi_J$ in the sum \eqref{genmero}, the domain of convergence \eqref{inside} is
\[
(\bs{\g}+\bs{\mathcal{A}}_l)\cdot\bs{N}^{(K)} = \bs{\g}\cdot\bs{N}^{(K)}-d_l^{(K)}<0 \qquad \forall\,\, K.
\]
This is equivalent to 
\[
\bs{n}^{(K)}\cdot((\hat{\bs{\gamma}}+\bs{a}_{l})-(\gamma_0+1)\bs{a}_{k})<0, \qquad k\in\vphi_K,
\]
{\it i.e.,} for every facet $K$, the shifted parameter vector $\hat{\bs{\gamma}}'=\hat{\bs{\gamma}}+\bs{a}_{l}$ must lie inside the Newton polytope rescaled by $\gamma_0'=\gamma_0+1$.
The common overlap of these domains for every $l\notin\vphi_J$ then corresponds to
\[\label{fulldom}
\bs{\g}\cdot\bs{N}^{(K)}<\delta^{(K)} \qquad \forall \,\, K ,
\]
where 
\[
\delta^{(K)} =\mathrm{min}_{l\notin\vphi_J}\,\big[ d_{l}^{(K)}\big] \ge 0.
\]
For any facet $K\neq J$, the set of vertices $l\notin\vphi_J$ includes vertices $l\in\vphi_K$ lying {\it in} the facet $K$.
For such vertices, $d_l^{(K)}$ and hence $\delta^{(K)}$ is then zero.  
Just as in our earlier example, the domain of convergence for the sum of shifted integrals in \eqref{genmero} is therefore unchanged for all facets $K\neq J$, 
\[\label{deltaKneqJ}
\delta^{(K)}=0 \quad \forall\,\,K\neq J.
\]  
The only facet across which the domain of convergence is extended is the facet $K=J$, for which we obtain an extension 
\begin{align}\label{Jshift}
\delta^{(J)} 
&= \mathrm{min}_{l\notin\vphi_J}\,\big[ d_{l}^{(J)}\big].
\end{align}
Geometrically, $\delta^{(J)}>0$ is the normal distance to the facet $J$ of the  (non-rescaled) Newton polytope starting from the {\it nearest} vertex $l$ not belonging to $J$, multiplied by $|\bs{n}^{(J)}|$.  If we choose $\bs{n}^{(J)}$ to have integer components, then as the components of the $\mathcal{A}$-matrix are also integer, $\delta^{(J)}$ will be a positive integer.

Equation \eqref{fulldom},  together with \eqref{deltaKneqJ} and \eqref{Jshift}, thus give us the domain of convergence of the meromorphic continuation \eqref{genmero}.  
Repeating the argument to construct further meromorphic continuations, one  finds that the GKZ integral $\mathcal{I}_{\bs{\gamma}}$ has an infinite series of singular hyperplanes lying parallel to each facet $J$ of the original Newton polytope.  These hyperplanes are given by 
\[\label{hypsings}
\bs{\g}\cdot\bs{N}^{(J)}=m_J\,\delta^{(J)}, \qquad m_J\in\mathbb{Z}^+,
\]
where $m_J$ is any non-negative integer $m=0,1,2,\ldots$.

\paragraph{Example:} Let us check \eqref{Jshift}
against our previous example.  Taking $J$ to be the facet with outward normal $\bs{n}^{(J)}=(2,1)$, we have $\vphi_J = \{3,4\}$ and so using the $\mathcal{A}$-matrix \eqref{houseA}, 
\[
\delta^{(J)} = \mathrm{min}_{l\in \{1,2\}} \sum_{i=1}^2 n_i^{(J)}(a_{i3}-a_{il}) =  \mathrm{min}(4,3) = 3.
\]
The sole shifted boundary
\[
\bs{\g}\cdot\bs{N}^{(J)} =
\gamma_0 n_0^{(J)}+\hat{\bs{\gamma}}\cdot\bs{n}^{(J)}<\delta^{(J)},
\]
where $n_0^{(J)} = -\sum_{i=1}^2 n_i^{(J)}a_{i3}=-4$, 
then evaluates to
\[
-4\gamma_0+2\gamma_1+\gamma_2-3<0
\]
in agreement with \eqref{gamshift3}, and the singular hyperplanes in \eqref{hypsings} match those in \eqref{housesing}.

\subsubsection{Implementation}

In higher-dimensional examples, a convenient way to determine the singular hyperplanes \eqref{hypsings} is to apply a convex hulling algorithm (see, {\it e.g.,} \cite{Nhull}) to identify which sets of vertex vectors $\bs{a}_{j}$ form the facets of the Newton polytope.  We will discuss this explicitly in section \ref{sec:hulling}.
The condition \eqref{gdotN} that $\hat{\bs{\g}}$ lies in the hyperplane containing facet $J$ of the rescaled Newton polytope is then equivalent to 
\[\label{fullgammaAdet}
0=\bs{\g}\cdot\bs{N}^{(J)}=\mathrm{\det}\,(\bs{\g}\,|\,\bs{\mathcal{A}}_{j_1}\,|\ldots |\,\bs{\mathcal{A}}_{j_N}),
\]
where $j_1,\ldots, j_N\in\vphi_J$ are the $N$ vertices belonging to  facet $J$, and the $\bs{\mathcal{A}}_{j}$ are the corresponding $\mathcal{A}$-matrix columns.  To see this, note that from \eqref{NdotA}  we have $\bs{\mathcal{A}}_{j}\cdot \bs{N}^{(J)}=0$ for all the $N$ vectors $j\in\vphi_J$.  As the total dimension of the vector space is $N+1$, the condition $\bs{\g}\cdot\bs{N}^{(J)}=0$ implies that $\bs{\g}$ lies in the span of the $\bs{\mathcal{A}}_{j}$ with $j\in\vphi_J$, and hence the determinant above vanishes.
The components  $n_i^{(J)}$ 
of $\bs{N}^{(J)}$, for $i=0,\ldots,N$, can thus be identified by expanding out the determinant and extracting the coefficient of $\g_i$. 
This tells us that $n_i^{(J)}$ is given by the $(i,1)$th cofactor of the matrix, for example
\[
n_0^{(J)} = \mathrm{det}\,(\bs{a}_{j_1}\,|\ldots|\,\bs{a}_{j_N}).
\]
 One must however also check that $\bs{n}^{(J)}$ corresponds to the outwards-pointing normal
by verifying that $d_k^{(J)}=-\bs{\mathcal{A}}_k\cdot\bs{N}^{(J)}>0$ for some $k\notin\vphi_J$, and swapping two  columns of \eqref{fullgammaAdet} if not. 
The spacing $\delta^{(J)}$ of the singular hyperplanes can then be computed using  \eqref{Jshift} and \eqref{ddef}.

\section{Shift operators}
\label{sec:shiftops}

Let us now examine the shift operators associated with $\mathcal{A}$-hypergeometric functions. 
Two natural classes present themselves: the `annihilation' operators which correspond to the simple derivative $\partial_j = \partial/\partial x_j$, and the `creation' operators which are purely polynomial differential operators ({\it i.e.,} operators in the Weyl algebra) that invert this operation.  

\subsection{Annihilation operators}
\label{sec:annihilate}

From the GKZ integral \eqref{GKZint} and denominator \eqref{GKZden}, we see by direct differentiation that 
\[\label{annihilateeq}
\partial_j \mathcal{I}_{\bs{\gamma}} = -\gamma_0 \mathcal{I}_{\bs{\gamma}'},\qquad j=1,\ldots, n
\]
where
\[\label{gshift1}
\gamma_0' = \gamma_0+1,\qquad \gamma_i' =\gamma_i+a_{ij},\qquad i=1,\ldots, N.
\]
In other words, differentiating with respect to $x_j$ increases the power of the denominator by one, and adds to the numerator all powers of $z_i$ multiplying $x_j$ in the denominator.
From the $\mathcal{A}$-matrix perspective, the shift of the parameter vector $\bs{\gamma}$ is given by the $j$th column of the full $\mathcal{A}$-matrix, 
\[
\bs{\gamma}' = \bs{\gamma} + \bs{\mathcal{A}}_{j},
\]
combining the two formulae in \eqref{gshift1}.

One can naturally think of the toric equations \eqref{torics} as representing the difference of two products of annihilation operators, such that the total shift generated by each product is the same leading to a cancellation.  Namely, each factor 
\[
\prod_{j=1}^n \partial_j^{u_j^\pm}
\]
produces an overall parameter shift
\[
\bs{\gamma}\rightarrow \bs{\gamma} + \sum_{j=1}^n \bs{\mathcal{A}}_{j}u_{j}^\pm,
\]
but since
\[
\sum_{j=1}^n \bs{\mathcal{A}}_{j}u_j^+= \sum_{j=1}^n\bs{\mathcal{A}}_{j}u_j^-
\]
the final shifted integral is the same in both cases and the difference vanishes.

Notice also that knowledge of the full set of $n$ annihilation operators, plus the parameter shifts they produce, is equivalent to knowledge of all columns of the $\mathcal{A}$-matrix and hence the full GKZ integral itself.\footnote{
Prior to the work of GKZ, this approach was pioneered by Miller {\it et al}  \cite{MillerBook, Miller} 
for various  Lauricella and Horn-type  
hypergeometric functions for which the annihilators can be identified from the series definition.}

\paragraph{Example:} The annihilation operators for the GKZ uplift \eqref{GKZ3rep00} of the triple-$K$ integral \eqref{tripleKdef} are $\partial_j$ for $j=1,\ldots, 6$.
The triple-$K$ integral  itself corresponds to evaluating the GKZ integral on the physical hypersurface $\x = (p_1^2,p_2^2,p_3^2,1,1,1)$ according to \eqref{3Kphys}.  The first three annihilators thus become 
\[
\partial_j= \frac{\p}{\p x_j}=\frac{\partial}{\partial p_j^2} =\frac{1}{p_j}\frac{\partial}{\partial p_j},\qquad j = 1,2,3.
\]
while for the remaining three we need to use the Euler equations
following from the $\mathcal{A}$-matrix \eqref{3KA}.  These are
\[
0=\beta_1 -\theta_1+\theta_4,\qquad0= \beta_2-\theta_2+\theta_5,\qquad0=\beta_3-\theta_3+\theta_6,
\]
and projecting to the physical hypersurface by setting $x_4=x_5=x_6=1$ gives
\[
\partial_4 = \theta_1-\beta_1 = \frac{p_1}{2}\frac{\partial}{\partial p_1} - \beta_1, \quad
\partial_5 = \theta_2-\beta_2 = \frac{p_2}{2}\frac{\partial}{\partial p_2} - \beta_2, \quad
\partial_6 = \theta_3-\beta_3 = \frac{p_3}{2}\frac{\partial}{\partial p_3} - \beta_3.
\]
Up to trivial numerical factors, these are the shift operators 
\[\label{LRdef0}
\mathcal{L}_j = -\frac{1}{p_j}\frac{\partial}{\partial p_j},\qquad
\mathcal{R}_j = 2\beta_j-p_j\frac{\partial}{\partial p_j}, \qquad j=1,2,3,
\]
introduced in \cite{Bzowski:2013sza, Bzowski:2015yxv}.
The action of these operators on the triple-$K$ integral \eqref{tripleKdef} can be obtained from their action  on the individual Bessel functions in the integrand giving
\[
\mathcal{L}_1 I_{\alpha,\{\beta_1,\beta_2,\beta_3\}} = -(\alpha+1)I_{\alpha+1,\{\beta_1-1,\beta_2,\beta_3\}},\qquad 
\mathcal{R}_1 I_{\alpha,\{\beta_1,\beta_2,\beta_3\}} =-(\alpha+1) I_{\alpha+1,\{\beta_1+1,\beta_2,\beta_3\}},
\]
with the others following by permutation.
This is consistent with the expected action for the annihilation operators: from the columns of the $\mathcal{A}$-matrix \eqref{3KA}, this is 
\[
\mathcal{L}_j:\quad\gamma_0'\rightarrow\gamma_0'+1,\qquad \gamma_j' \rightarrow \gamma_j' -1,\qquad
\mathcal{R}_j:\quad \gamma_0'\rightarrow\gamma_0'+1,\qquad\gamma_j' \rightarrow\gamma_j'+1,
\]
which from \eqref{3Kphys} is 
\[
\mathcal{L}_j:\quad\alpha\rightarrow\alpha+1,\qquad
\beta_j\rightarrow \beta_j-1,\qquad
\mathcal{R}_j:\quad\alpha\rightarrow\alpha+1,\qquad
\beta_j\rightarrow \beta_j+1.
\]

\subsection{Creation operators}
\label{sec:balgorithm}

Over the next three subsections, we present a construction of creation operators motivated by consideration of the spectral singularities.  These ideas are then illustrated using  the Gauss hypergeometric function.   
Originally, creation operators were first proposed by Saito in 
\cite{Saito_param_shift, Saito_restrictions}; for further discussion, see  \cite{Saito_Lauricella, saito_sturmfels_takayama_1999, smeets2000}.

By definition,  when acting on a GKZ integral, the creation operator $\mathcal{C}_j$ produces the {\it inverse} parameter shift to the annihilation operator $\partial_j=\partial/\partial x_j$.
If we act with one operator followed by the other, therefore, we must  arrive back at the original integral up to some function  of the parameters:
\[\label{bfndef}
\mathcal{C}_j\partial_j \mathcal{I}_{\bs{\gamma}}= b_j(\bs{\gamma}) \mathcal{I}_{\bs{\gamma}}.
\]
As we will see shortly, this `$b$-function' $b_j(\bs{\gamma})$ is a polynomial whose 
zeros correspond to a specific subset of the singular hyperplanes of  $\mathcal{I}_{\bs{\gamma}}$ given in \eqref{hypsings}.
First, however,  let us sketch how knowing $b_j(\bs{\gamma})$ enables a direct construction of the creation operator $\mathcal{C}_j$.

The first step is to replace all the parameters $\bs{\gamma}$ appearing in the $b$-function with linear combinations of Euler operators using the DWI and Euler equations \eqref{allEulers}.  
This defines a new polynomial $B_j(\theta)$ in the Euler operators, 
\[
 B_j(\theta)=b_j(\bs{\g})\Big|_{\bs{\gamma}\rightarrow -\sum_{k=1}^n \bs{\mathcal{A}}_{k}\theta_k}
\]
such that 
\[\label{CdB}
\mathcal{C}_j\partial_j \mathcal{I}_{\bs{\gamma}}= B_j(\theta) \mathcal{I}_{\bs{\gamma}}.
\]
As all Euler operators commute with one another, there are no ordering ambiguities here.

Next, we expand out $B_j(\theta)$  and re-arrange so that, in every term, all factors of  $x_k$ are to  the left of all derivatives $\partial_k$.  Up to a constant coefficient, each term of $B_j(\theta)$ is thus of the form
\[\label{Bterm}
\prod_{k=1}^n x_k^{\mathfrak{b}_k}\partial_k^{\mathfrak{b}_k}
\]
for some  set of powers $\mathfrak{b}_k$.  In certain cases, the  product  $\prod_{k}\partial_k^{\mathfrak{b}_k}$ will already contain an explicit factor of $\partial_j$.  
Otherwise, we can use the toric equations \eqref{torics} to replace the product  $\prod_{k}\partial_k^{\mathfrak{b}_k}$  (which acts on the GKZ integral $\mathcal{I}_{\bs{\gamma}}$ as per \eqref{CdB}) with an equivalent product that {\it does} contain an explicit factor of $\partial_j$.   Such a replacement will always be possible provided the $b$-function is correctly chosen.  After completing this operation for every term,  the right-hand side of \eqref{CdB}  now matches the form of the left-hand side allowing the operator $\mathcal{C}_j$ to be read off.
Thus, with the aid of the toric equations,   $B_j(\theta)$ acting on $\mathcal{I}_{\bs{\gamma}}$ can be explicitly factorised into the form $\mathcal{C}_j\partial_j$.

As a final step, the creation operator  $\mathcal{C}_j$, which is a differential operator 
with polynomial coefficients 
defined in the $n$-dimensional GKZ space, must be projected back to the physical hypersurface.  For this,  we restore all $x_k$ to their physical values (noting the $x_k$ are  positioned to the left of all derivatives), and use the Euler equations evaluated on the physical hypersurface to replace derivatives in directions lying off the physical hypersurface with derivatives tangential to this hypersurface.  This replacement also restores a dependence on the parameters $\bs{\gamma}$.
Many examples of this projection procedure will appear in subsequent sections.

\subsection{Action of the creation operator}

Returning to \eqref{bfndef} and using the action of the annihilator $\partial_j$ as given in \eqref{annihilateeq}, the action of the creation operator is
\[\label{creationeq}
\mathcal{C}_j\mathcal{I}_{\bs{\gamma}'} = -\gamma_0^{-1} b_j(\bs{\gamma}) \mathcal{I}_{\bs{\gamma}}.
\]
As the shift here is acting in the direction $\bs{\gamma}'\rightarrow\bs{\gamma}=\bs{\g}'-\bs{\mathcal{A}}_j$, rearranging \eqref{gshift1} we have
\[\label{gshift2}
\gamma_0 = \gamma_0'-1,\qquad \gamma_i = \gamma_i' - a_{ij},\qquad i=1,\ldots N.
\]
We will retain this allocation of prime and unprimed variables in the following  for compatibility with the algorithm  in the previous section based on \eqref{bfndef}.

Before discussing the $b$-function itself, 
a crucial point to notice is that the parameter shift \eqref{gshift2} can potentially take us from a {\it finite} to a {\it divergent} GKZ integral.  In contrast, the reverse shift \eqref{gshift1} associated with the annihilation operator $\partial_j$, when acting on a finite integral, will always produce another finite integral.

To see this, let us start with an integral 
$\mathcal{I}_{\bs{\gamma}'}$ for which 
the vector $\hat{\bs{\gamma}}'=(\gamma_1',\ldots,\gamma_N')$ lies  inside the rescaled Newton polytope with vertices $ \gamma_0' \bs{a}_{j}$.  In the notation of section \ref{sec:genmero}, this means that for every facet $K$ we  have 
\[\label{fincond}
\bs{\gamma}'\cdot\bs{N}^{(K)} < 0
\]
and the GKZ representation for $\mathcal{I}_{\bs{\gamma}'}$ converges without meromorphic continuation.  
For the {\it shifted} integral $\mathcal{I}_{\bs{\gamma}}$ in \eqref{creationeq},  we then have
\[\label{shiftedcond}
\bs{\gamma}\cdot\bs{N}^{(K)} =\bs{\gamma}'\cdot\bs{N}^{(K)} +d_j^{(K)}
\]
where
\[
d_j^{(K)}=\bs{n}^{(K)}\cdot (\bs{a}_{k} -\bs{a}_{j}),\qquad k\in\vphi_K
\] 
 is proportional to the 
normal distance from vertex $j$ to facet $K$ of the (non-rescaled) Newton polytope.
Now, for any facet $K$ {\it containing} the vertex $j$, 
$d_j^{(K)}$  vanishes and hence  $\bs{\g}\cdot\bs{N}^{(K)}<0$.
For the remaining facets {\it not} containing the vertex $j$, 
however, $d_j^{(K)}>0$ since $\bs{n}^{(K)}$ is the outward normal and  $j$ lies to the inside of the facet.  
Consequently, we  cannot be sure that $\bs{\g}\cdot\bs{N}^{(K)}<0$ 
for all facets $K$, and hence that $\mathcal{I}_{\bs{\gamma}}$ is finite.
Rather, if there are any facets for which $\bs{\g}\cdot\bs{N}^{(K)}\ge 0$, the shifted integral $\mathcal{I}_{\bs{\gamma}}$ will diverge whenever  the singularity condition \eqref{hypsings}, 
\[\label{Ksings}
\bs{\g}\cdot\bs{N}^{(K)}=m_K\delta^{(K)}, \qquad m_K\in\mathbb{Z}^+ 
\]
is satisfied for some non-negative integer $m_K$.
Combined with \eqref{shiftedcond}, this condition allows us to identify the initial parameter values $\bs{\g}'$ for which the shifted integral $\mathcal{I}_{\bs{\g}}$ diverges.

For the annihilation operator $\partial_j$, the direction of the parameter shifts  are reversed and so if the starting integral is finite,  the shifted integral is also necessarily finite.

\subsection{Finding the b-function}
\label{sec:findingb}

An apparent puzzle now arises for cases where the shifted integral $\mathcal{I}_{\bs{\gamma}}$ in \eqref{creationeq} is divergent, since the action of a differential operator $\mathcal{C}_j$ with polynomial coefficients  on any finite integral $\mathcal{I}_{\bs{\gamma}'}$ must clearly be finite.  The resolution is that, for such cases, the $b$-function in \eqref{creationeq} must have a {\it zero} cancelling the divergence in $\mathcal{I}_{\bs{\gamma}}$  such that the right-hand side is finite.\footnote{In a `dimensional' regularisation scheme where all parameters  are shifted infinitesimally $\bs{\gamma}\rightarrow \bs{\gamma}+\ep\,\bar{\bs{\gamma}}$, this requires $b_j(\bs{\gamma})\sim\ep^k$ while $\mathcal{I}_{\bs{\gamma}}\sim \ep^{-k}$ for some $k\in \mathbb{Z}^+$ such that $b_j(\bs{\g})\mathcal{I}_{\bs{\g}}$ is finite as $\ep\rightarrow 0$.}

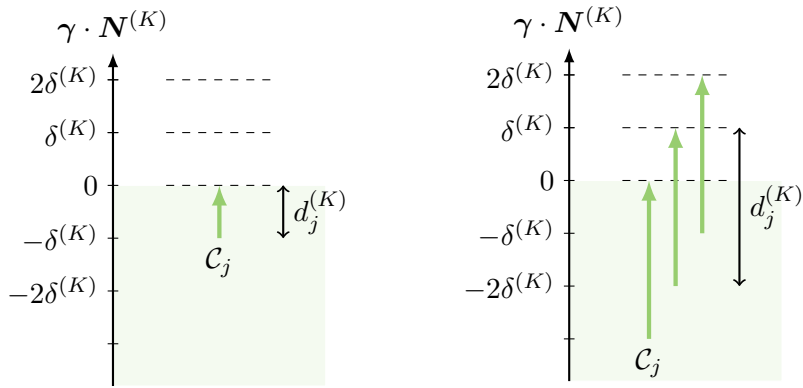
\begin{figure}[t]
\centering
\hspace{-1cm}
\begin{tikzpicture}[scale=0.7]
\draw[white,fill=YellowGreen, fill opacity=0.1]   (-2,0) -- (-2,-3.8)--(2,-3.8)--(2,0)--cycle ;
\draw[-latex,thick] (-2,-3.8) -- (-2,2.5) node[above]{$\bs{\g}\cdot \bs{N}^{(K)}$};
\draw (-2.1,2) -- (-1.9,2) node[left]{$2\delta^{(K)}\,\,$};
\draw (-2.1,1) -- (-1.9,1) node[left]{$\delta^{(K)}\,\,$};
\draw (-2.1,0) -- (-1.9,0) node[left]{$0\,\,$};
\draw (-2.1,-1) -- (-1.9,-1) node[left]{$-\delta^{(K)}\,\,$};
\draw (-2.1,-2) -- (-1.9,-2) node[left]{$-2\delta^{(K)}\,\,$};
\draw (-2.1,-3) -- (-1.9,-3) ;
\draw[dashed] (-1,0) -- (1,0);
\draw[dashed] (-1,1) -- (1,1);
\draw[dashed] (-1,2) -- (1,2);
\draw (-2.1,1) -- (-1.9,1); 
\draw (-2.1,2) -- (-1.9,2);  
\draw[-latex,ultra thick,YellowGreen] (0,-1) -- (0,0);
\draw[<->,thick] (1.2,-1)--(1.2,0);
\draw  node[right] at (1.2,-0.5){$d_j^{(K)}$};
\draw[black] node[below] at (0,-1) {$\mathcal{C}_j$} ;
\end{tikzpicture}
\hspace{1cm}
\begin{tikzpicture}[scale=0.7]
\draw[white,fill=YellowGreen, fill opacity=0.1]   (-2,0) -- (-2,-3.8)--(2,-3.8)--(2,0)--cycle ;
\draw[-latex,thick] (-2,-3.8) -- (-2,2.5) node[above]{$\bs{\g}\cdot \bs{N}^{(K)}$};
\draw (-2.1,2) -- (-1.9,2) node[left]{$2\delta^{(K)}\,\,$};
\draw (-2.1,1) -- (-1.9,1) node[left]{$\delta^{(K)}\,\,$};
\draw (-2.1,0) -- (-1.9,0) node[left]{$0\,\,$};
\draw (-2.1,-1) -- (-1.9,-1) node[left]{$-\delta^{(K)}\,\,$};
\draw (-2.1,-2) -- (-1.9,-2) node[left]{$-2\delta^{(K)}\,\,$};
\draw (-2.1,-3) -- (-1.9,-3) ;
\draw[dashed] (-1,0) -- (1,0);
\draw[dashed] (-1,1) -- (1,1);
\draw[dashed] (-1,2) -- (1,2);
\draw (-2.1,1) -- (-1.9,1); 
\draw (-2.1,2) -- (-1.9,2); 
\draw[-latex,ultra thick,YellowGreen] (-0.5,-3) -- (-0.5,0);
\draw[black] node[below] at (-0.5,-3) {$\mathcal{C}_j$} ;
\draw[-latex,ultra thick,YellowGreen] (0,-2) -- (0,1);
\draw[<->,thick] (1.2,-2)--(1.2,1);
\draw  node[right] at (1.2,-0.5){$d_j^{(K)}$};
\draw[-latex,ultra thick,YellowGreen] (0.5,-1) -- (0.5,2);
\end{tikzpicture}
\caption{Mapping of  finite  to divergent integrals under the action of the creation operator $\mathcal{C}_j$ as per \eqref{shiftedcond}, and construction of the corresponding $b$-functions.  
{\it Left:} If $d_j^{(K)}=\delta^{(K)}$, facet $K$ contributes only the factor $\bs{\g}\cdot \bs{N}^{(K)}$ to the $b$-function.  The zero of this factor cancels the pole of the only singular integral (dashed line) that can be reached starting from a finite integral.  
{\it Right:} If $d_j^{(K)}=3\delta^{(K)}$, the facet contributes three factors, $\prod_{m_K=0}^2(\bs{\g}\cdot \bs{N}^{(K)}-m_K\delta^{(K)})$ whose zeros cancel the poles of the three singular integrals reachable from a finite starting integral.  The shaded region indicates the rescaled Newton polytope. 
\label{Cshiftfig}}
\end{figure} 

The $b$-function for the creation operator $\mathcal{C}_j$ must thus have zeros corresponding to  every singular hyperplane that can be reached by a single application of $\mathcal{C}_j$ to any finite starting integral, as illustrated in figure \ref{Cshiftfig}. 
The minimal $b$-function, containing just these factors alone, is 
\[\label{stdbfn}
b_j(\bs{\gamma}) = \prod_{K\notin \Phi_j}\prod_{m_K=0}^{F_j^{(K)}-1}(\bs{\g}\cdot\bs{N}^{(K)}-m_K\delta^{(K)})
\]
where the first product runs over all facets $K$ not containing the vertex $j$ and the upper limit in the second product is set by 
\[\label{Fdef}
F_j^{(K)} =\frac{\vphantom{\big|} d_j^{(K)}}{\vphantom{\big|} \delta^{(K)}}.
\]
This counts by how many  steps (in units of $\delta^{(K)}$, the spacing between singular hyperplanes) the creation operator $\mathcal{C}_j$ 
raises $\bs{\g}'\cdot\bs{N}^{(K)}$ according to \eqref{shiftedcond}.  
 Effectively, if we {\it define} an initial $m_K'$  by the relation $\bs{\g}'\cdot\bs{N}^{(K)}=m_K'\delta^{(K)}$, the creation operator $\mathcal{C}_j$ acts to raise this to $m_K = m_K'+F_j^{(K)}$.
Thus, if $F_j^{(K)}=1$ for some particular facet $K$, only the singularity in \eqref{Ksings} with $m_K=0$ can be reached by the action of $\mathcal{C}_j$ on a finite starting integral (namely, that  with $m_K'=-1$).  The product over $m_K$ in \eqref{stdbfn} is thus capped at $F_j^{(K)}-1=0$.  Alternatively, if $F_j^{(K)}=2$ for some facet, both the $m_K=0$ and $m_K=1$ singularities can be reached by acting with $\mathcal{C}_j$ on the finite starting integrals  with $m_K'=-2$ and $m_K'=-1$ respectively.  The product over $m_K$ in \eqref{stdbfn} then runs up to $F_j^{(K)}-1=1$, and so on.

For all the Feynman and Witten diagrams we analyse in the remainder of the paper,  $F_j^{(K)}$ is an integer for all $K$ and 
 the minimal $b$-function \eqref{stdbfn} (containing only the zeros necessary to cancel out the singularities of $\mathcal{I}_{\bs{\gamma}}$) is sufficient to find all creation operators.  These operators are moreover of the lowest possible order in derivatives, since the $b$-function has the fewest factors. 
Nevertheless, in certain exceptional cases, the factorisation step of the algorithm in section \ref{sec:balgorithm} can fail when using
the minimal $b$-function.  Such cases, which arise when the associated toric ideal is non-normal \cite{Saito_param_shift, Saito_restrictions, saito_sturmfels_takayama_1999, smeets2000}, can be handled by supplementing  \eqref{stdbfn} with additional factors.  An example, which also features a non-integer $F_j^{(K)}$, is discussed in appendix \ref{sec:nonnormal}.

Despite its formal appearance, the formula \eqref{stdbfn} is straightforward to evaluate in practice as will become clear in the examples to follow. All that is required is to identify the singular hyperplanes \eqref{hypsings} for a given GKZ integral, along with the shift produced by the creation operator $\mathcal{C}_j$, 
and then to form the $b$-function from the product of all singular hyperplanes that can be reached by one application of $\mathcal{C}_j$ on any finite starting integral.  
Also, while consideration of singular cases has been used to construct the $b$-function, the creation operators we obtain can be used to map finite integrals to finite integrals.

\subsection{Example}
As a simple first example before turning to Witten diagrams and Feynman integrals in the following sections, we
compute creation operators for  the GKZ integral \cite{nilsson2010mellin, de_la_Cruz_2019}
\begin{align}\label{2f1example}
\mathcal{I}_{\bs{\gamma}} &= \int_{\mathbb{R}_+^2}\dd z_1 \dd z_2 \frac{z_1^{\gamma_1-1} z_2^{\gamma_2-1}}{(x_1+x_2 z_1+x_3 z_2+x_4 z_1 z_2)^{\gamma_0}}.
\end{align}
On the hypersurface $(x_1,x_2,x_3,x_4) = (1,1,1,y)$, this can be  evaluated in terms of the Gauss hypergeometric function 
\begin{align}\label{2f1eval}
\mathcal{I}_{\bs{\gamma}}(y)  &= \frac{\Gamma(\gamma_1) \Gamma(\gamma_2)
\Gamma(\gamma_0-\gamma_1) \Gamma(\gamma_0-\gamma_2)}{\Gamma(\gamma_0)^2}\; {}_2F_1(\gamma_1,\gamma_2,\gamma_0;1-y).
\end{align}
The shift operators for this function are well known allowing an easy check of our results.

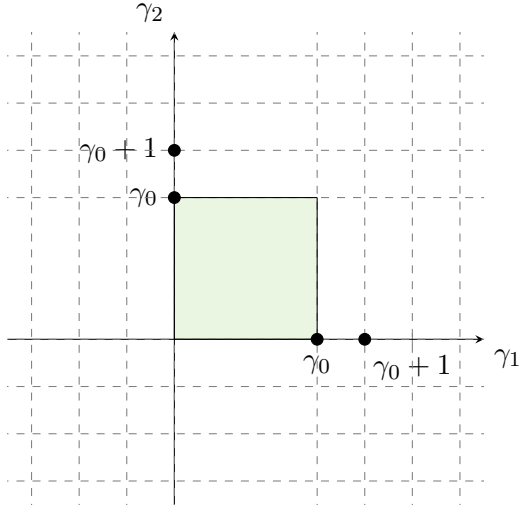
\begin{figure}[t]
\centering
 \begin{tikzpicture}[scale=1.1][>=latex]
\begin{axis}[
  axis x line=middle,
  axis y line=middle,
  xtick={3,5},
  ytick={3,4},
  xticklabels={$\gamma_0$, $\gamma_0+1$},
  yticklabels={$\gamma_0$, $\gamma_0+1$},
   xlabel={$\gamma_1$},
  ylabel={$\gamma_2$},
  xlabel style={below right},
  ylabel style={above left},
  xmin=-3.5,
  xmax=6.5,
  ymin=-3.5,
  ymax=6.5,
  axis equal image]

\foreach \k in {0,...,8} {
 
\addplot [mark=none,dashed,gray,domain=-5:10] {3+\k};
\addplot [mark=none,dashed,gray,domain=-5:10] (\k+3,x);
\addplot [mark=none,dashed,gray,domain=-5:10] {-\k};
\addplot [mark=none,dashed,gray,domain=-5:10] (-\k,x);

}

\addplot [mark=none,domain=0:3,name path=A] {3};
\addplot [mark=none,domain=0:3] (0,x);
\addplot [mark=none,domain=0:3] (3,x);
\addplot [mark=none,domain=0:3,name path=B] {0};

\addplot [YellowGreen, fill opacity=0.2] fill between[of=A and B,soft clip={domain=0.01:8}];

\addplot[holdot] coordinates{(3,0)(4,0)};
\addplot[holdot] coordinates{(0,3)(0,4)};

\end{axis}

\end{tikzpicture}
   \caption{The singular hyperplanes of \eqref{2f1example}.  \label{fig:2f1sings}}
\end{figure}

From the $\mathcal{A}-$matrix 
\begin{align}
\mathcal{A}= \left(\begin{matrix}
    1&1&1&1\\
    0&1&0&1\\
    0&0&1&1
   \end{matrix}\right)\!,
 \end{align}
we can read off the DWI and Euler equations
\[\label{2f1Eulers}
0 = \left(\gamma_0 +\t_1+\t_2+\t_3+\t_4\right)\mathcal{I}_{\bs{\gamma}},\quad 0 = \left(\gamma_1+\t_2+\t_4\right)\mathcal{I}_{\bs{\gamma}},\quad 0=\left(\gamma_2+\t_3+\t_4\right)\mathcal{I}_{\bs{\gamma}},
\]
where $\theta_i=x_i\partial_i$ and, from the kernel of the $\mathcal{A}$-matrix, we find a single toric equation,
\[\label{2f1toric}
0=\left(\p_2\p_3-\p_1\p_4\right)\mathcal{I}_{\bs{\gamma}}.
\]
From \eqref{hypsings}, the singular hyperplanes are
\[\label{2f1sings}
\gamma_1=-m_1,\qquad \gamma_2=-m_2,\qquad \gamma_0-\gamma_1 = -m_3,\qquad \gamma_0-\gamma_2 = -m_4, \qquad m_i\in \mathbb{Z}^+
\]
as displayed in figure \ref{fig:2f1sings}.  These singularities are consistent with the poles of the gamma functions in the numerator of the projected integral 
\eqref{2f1eval}.

The annihilation operators $\partial_j$ send $\bs{\gamma}\rightarrow\bs{\gamma}'$ while the creation operators $\mathcal{C}_j$ send $\bs{\gamma}'\rightarrow\bs{\gamma}$,
where for each $j$ these parameters are related by
\begin{align}\label{2f1Cshifts}
j&=1: & \gamma_0'&=\gamma_0+1,&\gamma_1'&=\gamma_1, & \gamma_2'&=\gamma_2,\nn\\
j&=2: & \gamma_0'&=\gamma_0+1, &\gamma_1'&= \gamma_1+1, & \gamma_2'&=\gamma_2,\nn\\
j&=3: & \gamma_0'&=\gamma_0+1, & \gamma_1'&=\gamma_1,&\gamma_2'&=\gamma_2+1,\nn\\
j&=4: & \gamma_0'&=\gamma_0+1, & \gamma_1'&=\gamma_1+1, & \gamma_2'&=\gamma_2+1.
\end{align}
The corresponding $b$-functions are
\begin{align}\label{2f1bs}
b_1&=(\gamma_0-\gamma_1)(\gamma_0-\gamma_2),\nn\\
b_2&=\gamma_1(\gamma_0-\gamma_2),\nn\\
b_3&=\gamma_2(\gamma_0-\gamma_1),\nn\\
b_4&=\gamma_1\gamma_2.
\end{align}
In each case, the factors appearing provide the zeros needed to cancel the poles that arise when the creation operator shifts us from a finite to a singular integral.  For $b_1$, for example, the shift produced by $\mathcal{C}_1$ can take a finite integral with $\gamma_0'-\gamma_i'=1$ to a singular integral with $\gamma_0-\gamma_i=0$ for both $i=1$ and $i=2$, as we see from \eqref{2f1sings}.   The zeros of $b_1$ then cancel these singularities so that the action \eqref{creationeq}  of $\mathcal{C}_1$ on a finite integral is always finite.
For $b_2$, the shifts  produced by $\mathcal{C}_2$ 
can take a finite integral with $\gamma_1'=1$ to a singular integral with $\gamma_1=0$, and a finite integral with $\gamma_0'-\gamma_2'=1$ to a singular integral with $\gamma_0-\gamma_2=0$, with these singularities again being cancelled by the zeros of $b_2$.  Note that the action of $\mathcal{C}_2$ leaves $\gamma_2'$ and $\gamma_0'-\gamma_1'$ unchanged hence no further singularities arise and there are no further factors in $b_2$.
One can also check that the $b$-functions \eqref{2f1bs} are consistent with the general formula \eqref{stdbfn}.

From \eqref{CdB} plus the DWI and Euler equations \eqref{2f1Eulers}, we now have, for example,
\begin{align}
\mathcal{C}_1 \partial_1 \mathcal{I}_{\bs{\gamma}} &= (\gamma_0-\gamma_1)(\gamma_0-\gamma_2)\mathcal{I}_{\bs{\gamma}} =(\theta_1+\theta_3)(\theta_1+\theta_2)\mathcal{I}_{\bs{\gamma}} \nn\\
&=(x_1\p_1+x_1^2\p_1^2+x_1x_2\p_1\p_2+x_1x_3\p_1\p_3+x_2x_3\p_2\p_3)
\mathcal{I}_{\bs{\gamma}}.
\end{align}
By inspection, every term in the final line contains an explicit factor of $\p_1$ except for the last, but this can be replaced by $x_2x_3\p_1\p_4$ using the toric equation \eqref{2f1toric}.
This gives us the desired factorisation
\begin{align}
\mathcal{C}_1 &=x_1+ x_1^2\p_1+x_1x_2\p_2+x_1x_3\p_3
+x_2x_3\p_4 \nn\\ &= x_1(1+\theta_1+\theta_2+\theta_3)+x_2x_3\p_4.
\end{align}
In the same fashion, we obtain
\begin{align}
\mathcal{C}_2 &=x_2(1+\theta_1+\theta_2+\theta_4)+x_1x_4\p_3,\nn\\
\mathcal{C}_3 &=x_3(1+\theta_1+\theta_3+\theta_4)+x_1x_4\p_2,\nn\\
\mathcal{C}_4 &=x_4(1+\theta_2+\theta_3+\theta_4)+x_2x_3\p_1.
\end{align}

Finally, in order to understand their action on \eqref{2f1eval}, these creation operators can be projected to the `physical' hypersurface $(x_1,x_2,x_3,x_4) = (1,1,1,y)$.  For this we use the DWI and Euler equations \eqref{2f1Eulers} evaluated on this hypersurface,
which can be re-arranged so as to eliminate all derivatives apart from $\p_y$:
\begin{align}
\p_1\mathcal{I}_{\bs{\gamma}'}(y) &= (-\gamma_0'+\gamma_1'+\gamma_2'+\theta_y)\mathcal{I}_{\bs{\gamma}'}(y),\nn\\
\p_2\mathcal{I}_{\bs{\gamma}'}(y) &= -(\gamma_1'+\theta_y)\mathcal{I}_{\bs{\gamma}'}(y),\nn\\
\p_3\mathcal{I}_{\bs{\gamma}'}(y) &= -(\gamma_2'+\theta_y)\mathcal{I}_{\bs{\gamma}'}(y).
\end{align}
Notice here that as the creation operators act on the integral with parameters $\bs{\gamma}'$ by our definition \eqref{creationeq}, we need to use these parameters here.
With the aid of these equations, the creation operators project to
\begin{align}
\mathcal{C}_1^{\mathrm{ph}} &= 1-\gamma_0' + (1-y)\p_y,\nn\\
\mathcal{C}_2^{\mathrm{ph}} &= 1-\gamma_0' + (1-y)(\gamma_2'+\theta_y),\nn\\
\mathcal{C}_3^{\mathrm{ph}} &=1-\gamma_0' + (1-y)(\gamma_3'+\theta_y),\nn\\
\mathcal{C}_4^{\mathrm{ph}} &=1-\gamma_0' +(1-y)(\gamma_1'+\gamma_2'-1+\theta_y),
\end{align}
where the `ph' superscript indicates the operators expressed in physical variables.  
From \eqref{creationeq}, we then have, for example,
\begin{align}
\mathcal{C}_1^{\mathrm{ph}}\mathcal{I}_{\gamma_0',\gamma_1',\gamma_2'}(y)&=
-\gamma_0^{-1}(\gamma_0-\gamma_1)(\gamma_0-\gamma_2)
\mathcal{I}_{\gamma_0,\gamma_1,\gamma_2}(y)\nn\\&=
-(\gamma_0'-1)^{-1}(\gamma_0'-\gamma_1'-1)(\gamma_0'-\gamma_2'-1)\mathcal{I}_{\gamma_0'-1,\gamma_1',\gamma_2'},
\end{align}
since here the creation operator shifts $\gamma_0'\rightarrow\gamma_0=\gamma_0'-1$ while  $\gamma_i'=\gamma_i$ for $i=1,2$.  Accounting for the gamma functions in \eqref{2f1eval}, this corresponds to
\[
\big(1-\gamma_0'+(1-y)\p_y)\,{}_2F_1(\gamma_1',\gamma_2',\gamma_0';1-y) = (1-\gamma_0')\,{}_2F_1(\gamma_1',\gamma_2',\gamma_0'-1;1-y)
\]
which indeed follows from standard relations for ${}_2F_1$ (see {\it e.g.,} equation 15.5.4 of  \cite{NIST:DLMF}).

Taking into account the shifts \eqref{2f1Cshifts}, for the remaining operators we find
\begin{align}
\mathcal{C}_2^{\mathrm{ph}}\mathcal{I}_{\gamma_0',\gamma_1',\gamma_2'}(y)&=-\gamma_0^{-1}\gamma_1(\gamma_0-\gamma_2)\mathcal{I}_{\gamma_0,\gamma_1,\gamma_2}\nn\\&
=-(\gamma_0'-1)^{-1}(\gamma_1'-1)(\gamma_0'-\gamma_2'-1)\mathcal{I}_{\gamma_0'-1,\gamma_1'-1,\gamma_2'}(y)
,\nn\\
\mathcal{C}_3^{\mathrm{ph}}\mathcal{I}_{\gamma_0',\gamma_1',\gamma_2'}(y)&=-\gamma_0^{-1}\gamma_2(\gamma_0-\gamma_1)\mathcal{I}_{\gamma_0,\gamma_1,\gamma_2}\nn\\&
=-(\gamma_0'-1)^{-1}(\gamma_2'-1)(\gamma_0'-\gamma_1'-1)\mathcal{I}_{\gamma_0'-1,\gamma_1',\gamma_2'-1}(y)
,\nn\\
\mathcal{C}_4^{\mathrm{ph}}\mathcal{I}_{\gamma_0',\gamma_1',\gamma_2'}(y)&=-\gamma_0^{-1}\gamma_1\gamma_2\mathcal{I}_{\gamma_0,\gamma_1,\gamma_2}\nn\\&=
-(\gamma_0'-1)^{-1}(\gamma_1'-1)(\gamma_2'-1)\mathcal{I}_{\gamma_0'-1,\gamma_1'-1,\gamma_2'-1}(y).
\end{align}
These can again be verified using standard  shift identities and contiguity relations for the Gauss hypergeometric function.

\section{Creation operators for Witten diagrams}
\label{sec:Witten}

At strong coupling, the correlators of holographic conformal field theories 
can be  computed via Witten diagrams in anti-de Sitter spacetime.  
As the evaluation of these diagrams is frequently challenging,  it is  important to identify classes of shift operators connecting known `seed' solutions to a wider family of correlators.

In this section, we construct creation operators for Witten diagrams in momentum space.\footnote{Results for the position-space contact diagram, or holographic $D$-function, are given in appendix \ref{sec:Dfn}.}  Starting with the contact diagram, we derive explicit results at  $3$- and $4$-points where the expressions remain relatively compact, though the construction itself is valid at any number of points.  
In principle, 
as the momentum-space contact diagram can be expressed as a linear combination of Lauricella $F_C$ hypergeometric functions \cite{prudnikov, Coriano:2019sth}, the  creation operators we find here should be related to those for $F_C$ in \cite{Saito_Lauricella}.  However, our present construction is more direct. 
We then show,  again at $3$- and $4$-points, how to construct operators that shift the scaling dimensions while preserving the spacetime dimension.

A case of particular interest, given the close connection to cosmological correlators,  is the 4-point exchange  diagram.
Here, a class of weight-shifting operators is known connecting exchange diagrams with different external scaling dimensions  \cite{Karateev:2017jgd, Baumann:2019oyu}, but subject to two  restrictions \cite{Bzowski:2022rlz}: first, these operators map an exchange diagram with {\it non-derivative} vertices to one with {\it derivative} vertices; and second, they work only for a special set of initial scaling dimensions.  While these results are sufficient for cosmologies where the inflaton is a derivatively-coupled massless scalar, finding further generalisations is highly desirable.

A key problem, therefore, is 
to find a  shift operator connecting exchange diagrams with {\it non-derivative} vertices to new exchange diagrams, with shifted scaling dimensions, but still with {\it non-derivative} vertices. 
Such an operator should moreover be applicable to diagrams with arbitrary initial scaling dimensions.
In section \ref{sec:exch}, we construct an operator with precisely these properties.

\subsection{Definitions}

In momentum space, the $n$-point contact Witten diagram is
\begin{align}\label{icont}
& i_{[d;\,\Delta_1, \,\ldots,\, \Delta_n]}
 = \int_0^\infty \D z \, z^{-d-1} \prod_{i=1}^n \mathcal{K}_{[\Delta_i]}(z, p_i) 
\end{align}
where $d$ is the boundary spacetime dimension of the CFT, $\Delta_i$ is the scaling dimension of the scalar operator $\O_i$, 
and the bulk-to-boundary propagator
\begin{align} \label{Kprop}
	\mathcal{K}_{[\Delta_i]}(z, p_i) = \frac{ z^{\frac{d}{2}} p_i^{\beta_i}K_{\beta_i}(p_i z)}{2^{\beta_i - 1} \Gamma (\beta_i)}, \qquad \beta_i=\Delta_i-\frac{d}{2}.
\end{align}
Since the Bessel-$K$ function is invariant under reversing the sign of its index, we have the shadow relation
\begin{align}\label{contshadow}
i_{[d;\,\Delta_1,\,\ldots, \,\Delta_n]}\Big|_{\Delta_i\rightarrow d-\Delta_i} =\frac{4^{\beta_i}\Gamma(\beta_i)}{\Gamma(-\beta_i)}p_i^{-2\beta_i} i_{[d;\,\Delta_1,\,\ldots,\,\Delta_n]}.
\end{align}

\begin{figure}[t]
\begin{tikzpicture}[scale=0.8]
\draw [darkgray] (0,0) circle [radius=3];
\draw [fill=black] (-2.121,-2.121) circle [radius=0.1];
\draw [fill=black] (-2.121, 2.121) circle [radius=0.1];
\draw [fill=black] ( 2.121,-2.121) circle [radius=0.1];
\draw [fill=black] ( 2.121, 2.121) circle [radius=0.1];
\draw [fill=black] ( 0, 0) circle [radius=0.1];
\draw (-2.121,-2.121) -- ( 2.121, 2.121);
\draw ( 2.121,-2.121) -- (-2.121, 2.121);	
\node [left] at (-2.121, 2.2) {$\O_1(\bs{p}_1)$}; 
\node [left] at (-2.121,-2.2) {$\O_2(\bs{p}_2)$}; 	
\node [right] at ( 2.121,-2.2) {$\O_3(\bs{p}_3)$}; 
\node [right] at ( 2.121, 2.2) {$\O_4(\bs{p}_4)$}; 	
\node [above] at (-0.9, 1.06) {$\mathcal{K}_{[\Delta_1]}$};
\node [above] at (-1.3,-1.06) {$\mathcal{K}_{[\Delta_2]}$};
\node [above] at ( 1.2,-1.06) {$\mathcal{K}_{[\Delta_3]}$};
\node [above] at ( 0.8, 1.06) {$\mathcal{K}_{[\Delta_4]}$};
\end{tikzpicture}
\qquad
\begin{tikzpicture}[scale=0.8]
\draw [darkgray] (0,0) circle [radius=3];
\draw [fill=black] (-2.121,-2.121) circle [radius=0.1];
\draw [fill=black] (-2.121, 2.121) circle [radius=0.1];
\draw [fill=black] ( 2.121,-2.121) circle [radius=0.1];
\draw [fill=black] ( 2.121, 2.121) circle [radius=0.1];
\draw [fill=black] (-1, 0) circle [radius=0.1];
\draw [fill=black] ( 1, 0) circle [radius=0.1];
\draw (-2.121,-2.121) -- (-1,0) -- (-2.121, 2.121);
\draw ( 2.121, 2.121) -- ( 1,0) -- ( 2.121,-2.121);
\draw (-1,0) -- (1,0);
\node [left] at (-2.121, 2.2) {$\O_1(\bs{p}_1)$}; 
\node [left] at (-2.121,-2.2) {$\O_2(\bs{p}_2)$}; 	
\node [right] at ( 2.121,-2.2) {$\O_3(\bs{p}_3)$}; 
\node [right] at ( 2.121, 2.2) {$\O_4(\bs{p}_4)$}; 	
\node [right] at (-1.5, 1.2) {$\mathcal{K}_{[\Delta_1]}$};
\node [right] at (-1.5, -1.2) {$\mathcal{K}_{[\Delta_2]}$};
\node [left] at ( 1.5, -1.2) {$\mathcal{K}_{[\Delta_3]}$};
\node [left] at ( 1.5, 1.2) {$\mathcal{K}_{[\Delta_4]}$};
\node [above] at (0,0) {$\mathcal{G}_{[\Delta_x]}$};
\end{tikzpicture}
\centering
\caption{Witten diagrams representing the contact and exchange 4-point diagram $\ino_{[\Delta_1 \Delta_2 \Delta_3 \Delta_4]}$ and $\ino_{[\Delta_1 \Delta_2, \Delta_3 \Delta_4 x \Delta_x]}$  given by the integrals \eqref{icont} and \eqref{iexch}.\label{Wittendias}}
\end{figure}
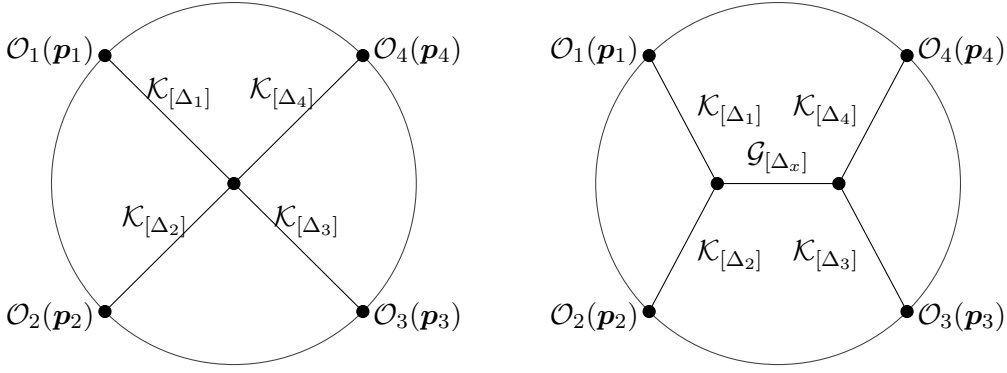

In addition to the contact diagram, we will discuss the $4$-point $s$-channel exchange diagram shown in figure \ref{Wittendias},
\begin{align}\label{iexch}
 i_{[d;\,\Delta_1, \Delta_2; \,\Delta_3, \Delta_4; \,\Delta_x]}
&= \int_0^\infty \D z \, z^{-d-1}\mathcal{K}_{[\Delta_1]}(z, p_1) \mathcal{K}_{[\Delta_2]}(z, p_2) 
\\ &\qquad \times
 \int_0^\infty \D \z \, \z^{-d-1} \mathcal{G}_{[\Delta_x]}(z, s; \z) \mathcal{K}_{[\Delta_3]}(\z, p_3) \mathcal{K}_{[\Delta_4]}(\z, p_4),\nn
\end{align}
where $\Delta_x$ is the dimension of the exchanged operator and  $s^2 = (\bs{p}_1+\bs{p}_2)^2$.  The bulk-to-bulk propagator in this expression is
\begin{align} \label{Gprop}
	\mathcal{G}_{[\Delta_x]}(z, s; \z) = \left\{ \begin{array}{ll}
		(z \z)^{\frac{d}{2}} I_{\beta_x}(s z) K_{\beta_x}(s \z) & \text{ for } z < \z, \\
		(z \z)^{\frac{d}{2}} K_{\beta_x}(s z) I_{\beta_x}(s \z) & \text{ for } z > \z,
	\end{array} \right.	
\end{align}
with $I_{\beta}$ and $K_{\beta}$ representing modified Bessel functions and $\beta_x=\Delta_x-d/2$.  Where necessary, these integrals can be regulated by infinitesimally shifting the operator dimensions and spacetime dimension $d$ so as to ensure convergence \cite{Bzowski:2022rlz}.

\subsection{GKZ representation of the contact diagram}

The GKZ representation for the $n$-point momentum-space contact diagram can be evaluated analogously to that for the triple-$K$ integral (see page \pageref{tripleKex}).
This yields the GKZ integral
\begin{align}\label{contactGKZ}
 \mathcal{I}_{\bs{\gamma}} =\Big(\prod_{i=1}^n\int_0^\infty \D z_i \,z_i^{\gamma_i-1}\Big)\Big[\sum_{j=1}^n \Big(\frac{x_j}{ z_j} + \bar{x}_j z_j\Big)\Big]^{-\gamma_0}
\end{align}
with the contact diagram  being
\[\label{icontIg}
 i_{[d;\,\Delta_1, \,\ldots, \,\Delta_n]} =  2^{\gamma_0}\Gamma(\gamma_0)\Big(\prod_{i=1}^n \frac{1}{2^{\gamma_i}\Gamma(\gamma_i)}\Big)  \mathcal{I}_{\bs{\gamma}}
 \]
with parameters
\[\label{contparams}
\gamma_0 = \big(\frac{n}{2}-1\big)d,\qquad 
\gamma_i = \Delta_i - \frac{d}{2} = \beta_i,
\]
and physical hypersurface
\[\label{contphys}
x_i = p_i^2, \qquad \bar{x}_i = 1, \qquad i=1,\ldots, n.
\]
Our notation $\x = (x_i,\bar{x}_i)$ for the GKZ variables here and in \eqref{contactGKZ} is designed to simplify the form of the Euler and toric equations as we will see below; $\bar{x}_i$  should be regarded as an independent dynamical variable equivalent to $x_{i+n}$ in the notation of the previous section.

The $(n+1)\times 2n$ dimensional $\mathcal{A}$-matrix for the integral \eqref{contactGKZ}  is now
\begin{align}
\mathcal{A} &= \left(\begin{matrix}
&\bs{1} & \bs{1}& \\
 &-\mathbb{I}_n &\mathbb{I}_n&
\end{matrix}\right)
\end{align}
where $\bs{1}$ is the $n$-dimensional row vector of $1$s and $\mathbb{I}_n$ is the $n\times n$ identity matrix.  (Again, 
we are departing from the notation of the previous section where $n$ referred to the number of columns in the $\mathcal{A}$-matrix, reserving $n$ now for the number of points.)
Writing
\[
\partial_i = \frac{\partial}{\partial x_i}, \qquad \bar{\partial}_i = \frac{\partial}{\partial \bar{x}_i},\qquad \theta_i = x_i\partial_i,\qquad \bar{\theta}_i = \bar{x}_i\bar{\partial}_i,
\]
the Euler equations are
\[\label{eulercont}
0 = (\gamma_i -  \theta_i +\bar{\theta}_i)\mathcal{I}_{\bs{\gamma}},\qquad i=1,\ldots,n,
\]
while the DWI is 
\[\label{DWIcont}
0 = \Big(\gamma_0 + \sum_{i=1}^n (\theta_i+\bar{\theta}_i)\Big)\mathcal{I}_{\bs{\gamma}}.
\]
In addition, we have the toric equations
\[\label{toriccont}
0 = (\p_i\bar{\p}_i - \p_j\bar{\p}_j)\mathcal{I}_{\bs{\gamma}},\qquad i\neq j=1,\ldots,n.
\]
These can easily be verified by noting that $\partial_i\bar{\partial}_i$ sends $\gamma_0\rightarrow\gamma_0+2$ but makes no change to the power of $z_i$ appearing in the numerator of \eqref{contactGKZ}, hence the two terms in \eqref{toriccont} cancel.

It is well known that the contact diagram satisfies the equation, 
\[
0=(K_i-K_j)i_{[d,\,\Delta_1,\ldots,\Delta_n]}\qquad \forall \,\,i\neq j
\]
where $K_i$ is the Bessel operator 
\[
K_i = \partial_{p_i}^2 + \frac{(1-2\g_i)}{p_i}\p_{p_i}= \p_i(\t_i-\g_i).
\]
To see this from a GKZ perspective, we use the Euler and toric equations to show that 
\[
(K_i-K_j)\mathcal{I}_{\bs{\g}} = 
(\p_i\tb_i-\p_j\tb_j)\mathcal{I}_{\bs{\g}}=(\bar{x}_i\p_i\bar{\p}_i-\bar{x}_j\p_j\bar{\p}_j)\mathcal{I}_{\bs{\g}}=(\bar{x}_i-\bar{x}_j)\p_i\bar{\p}_i\mathcal{I}_{\bs{\g}}.
\]
Upon projecting to the physical hypersurface \eqref{contphys}, the right-hand side now vanishes.

We note too that the shadow relation \eqref{contshadow} uplifts to
\[\label{GKZshadow}
\mathcal{I}_{\bs{\g}}\Big|_{\gamma_i\rightarrow - \gamma_i} =
\Big(\frac{\bar{x}_i}{x_i}\Big)^{\g_i}\mathcal{I}_{\bs{\g}}
\]
in GKZ variables, for any $i=1,\ldots, n$. This  can be seen by evaluating the right-hand side of \eqref{contactGKZ} with the substitution $z_i =  x_i/(\bar{x}_i z'_i)$.

\subsection{Creation and annihilation operators}

The action of the annihilation operators is
\[\label{contannih}
\partial_i \mathcal{I}_{\bs{\g}} =-\g_0 \mathcal{I}_{\bs{\g}}\Big|_{\g_0\rightarrow \gamma_0+1,\,\g_i\rightarrow\gamma_i-1},\qquad
\bar{\partial}_i \mathcal{I}_{\bs{\g}} =-\g_0 \mathcal{I}_{\bs{\g}}\Big|_{\g_0\rightarrow \gamma_0+1,\,\g_i\rightarrow\gamma_i+1}
\]
for any $i=1,\ldots, n$.  After projecting to the physical hypersurface \eqref{contphys}, up to numerical factors $\partial_i$ and $\bar{\partial}_i$ become the operators $\mathcal{L}_i$ and $\mathcal{R}_i$ respectively, as defined in \eqref{LRdef0}.
Due to the shadow relation \eqref{GKZshadow}, or by re-arranging the Euler equation \eqref{eulercont}, we have
\[
\bar{\partial}_i \mathcal{I}_{\bs{\g}}= \Big(\frac{\bar{x}_i}{x_i}\Big)^{-\g_i-1} \partial_i\, \Big(\frac{\bar{x}_i}{x_i}\Big)^{\g_i}\mathcal{I}_{\bs{\g}}. 
\]
In physical variables, this projects  to 
\[
\mathcal{R}_i = p_i^{2(\beta_i+1)}\mathcal{L}_i \,p_i^{-2\beta_i}.
\]

The action of the creation operators is the inverse of that in   \eqref{contannih}, namely
\[\label{actionofCi}
\mathcal{C}_i: \quad \gamma_0\rightarrow\gamma_0-1,\quad\gamma_i\rightarrow\gamma_i +1,\qquad
\bar{\mathcal{C}}_i: \quad\gamma_0\rightarrow\gamma_0-1,\quad\gamma_i\rightarrow\gamma_i-1,
\]
where all remaining $\gamma_j$ for $j\neq i$ stay the same. 
As a result of the shadow relation \eqref{contshadow}, however, it suffices to construct only $\mathcal{C}_i$ since
\[\label{Cshadow}
\bar{\mathcal{C}}_i \,\mathcal{I}_{\bs{\g}}= \Big(\frac{\bar{x}_i}{x_i}\Big)^{1-\g_i} \mathcal{C}_i\Big|_{\gamma_i\rightarrow -\gamma_i}\, \Big(\frac{\bar{x}_i}{x_i}\Big)^{\g_i}\mathcal{I}_{\bs{\g}}.
\]

To construct $\mathcal{C}_i$, first we need to identify the singular hyperplanes of $\mathcal{I}_{\bs{\g}}$.   These can be found either by expanding the integrand of \eqref{icont} about the lower limit $z=0$ and looking for the appearance of $z^{-1}$ pole terms (see \cite{Bzowski:2015pba}),  or by using the formula \eqref{hypsings} based on the Newton polytope.  Here, the Newton polytope takes the form of an $n$-dimensional cross-polytope with vertices at 
$\pm \bs{e}_j$ for every basis vector $(\bs{e}_j)_k=\delta_{jk}$ and $2^n$ facets with outward normals
$\bs{n}= (\sigma_1, \ldots, \sigma_n)^T$ for every possible independent choice of  $\sigma_j=\pm 1$.  From \eqref{muJdef} and \eqref{Jshift}, $n_0=-1$ and $\delta=2$ 
for every facet, hence the singular hyperplanes are
\[\label{contsing}
0 = -\gamma_0 +\sum_{j=1}^n \sigma_j\gamma_j  -2m, \qquad m\in \mathbb{Z}^+.
\]

Given the action of $\mathcal{C}_i$ in \eqref{actionofCi},
the only way this operator can shift us from a finite to a singular integral is if $\sigma_i=+1$ so that 
$m$ increases by one.  
The corresponding $b$-function is then
\[\label{littlebdef0}
b_i(\bs{\gamma}) = \prod_{\{\sigma_j=\pm 1\}}\frac{1}{2}(-\gamma_0+\gamma_i +\sum_{j\neq i}\sigma_j\gamma_j),
\]
where the product runs over every possible choice of signs for all $j\neq i$.  
Using the Euler equations,  this gives
\begin{align}\label{bigBdef}
B_i(\theta,\bar{\theta}) &= \prod_{\{\sigma_j=\pm 1\}}\Big(\theta_i + \sum_{j\neq i}( \delta_{\sigma_j,+1}\theta_j+\delta_{\sigma_j,-1}\bar{\theta}_j)\Big).
\end{align}
For convenience, we have chosen to eliminate an overall numerical factor in this expression by inserting factors of one-half  in \eqref{littlebdef0}. 
This is simply a trivial  rescaling of both the creation operator and the $b$-function.
For the 3-point function,  for example, we then have
\[\label{3ptcontb}
b_1(\bs{\gamma}) =\frac{1}{16}(-\gamma_0+\gamma_1+\gamma_2+\gamma_3)(-\gamma_0+\gamma_1-\gamma_2+\gamma_3)(-\gamma_0+\gamma_1+\gamma_2-\gamma_3)(-\gamma_0+\gamma_1-\gamma_2-\gamma_3)
\]
and
\begin{align}\label{3ptbigBex}
B_1(\theta,\bar{\theta}) = (\theta_1+\theta_2+\theta_3) (\theta_1+\bar{\theta}_2+\theta_3) (\theta_1+\theta_2+\bar{\theta}_3) (\theta_1+\bar{\theta}_2+\bar{\theta}_3).
\end{align}
Recalling the creation operators obey
\begin{align}
\mathcal{C}_i\partial_i \mathcal{I}_{\bs{\gamma}} = b_i(\bs{\gamma}) \mathcal{I}_{\bs{\gamma}}  = B_i(\theta,\bar{\theta}) \mathcal{I}_{\bs{\gamma}}
\end{align}
the idea is now to expand out as\footnote{A factorisation of this form always exists as can be seen recursively in the number of points $n$.  
Once all $\t_i$-dependence has been gathered into $Q_i\t_i$, we write the remainder at $n$-points as $\tilde{B}_i^{(n)} = B_i^{(n)}|_{\t_i\rightarrow 0}$.   Multiplying out all the factors containing $\t_n$, and, separately, all the factors containing $\tb_n$, we obtain  $\tilde{B}_i^{(n)} = ((\ldots)\t_n+\tilde{B}_i^{(n-1)})((\ldots)\tb_n + \tilde{B}_i^{(n-1)}) = (\ldots)\t_n\tb_n + (\ldots)\tilde{B}_i^{(n-1)}$, where $\tilde{B}_i^{(n-1)}$ is independent of $\t_n$ and $\tb_n$.  Thus, if the decomposition $\tilde{B}_i^{(n-1)}=\sum_{j\neq i}^{n-1} Q^{(n-1)}_j\t_j\tb_j$ exists at $(n-1)$-points, then it also exists at $n$-points, $\tilde{B}_i^{(n)}=\sum_{j\neq i}^n Q^{(n)}_j\t_j\tb_j$.}  
\begin{align}
B_i(\theta,\bar{\theta}) &= Q_i(\theta,\bar{\theta})\theta_i + \sum_{j\neq i}Q_j(\theta,\bar{\theta}) \theta_j\bar{\theta}_j,
\end{align}
where without loss of generality we can choose all $Q_j(\theta,\bar{\theta})$  
for $j\neq i$ to be  independent of  both $\theta_i$ and $\bar{\theta}_i$.  (Note from \eqref{bigBdef} 
that $B_i(\theta,\bar{\theta})$ is automatically independent of $\bar{\theta}_i$.) 
We then use the toric equations \eqref{toriccont} to re-express
\[\label{useoftorics}
\theta_j\bar{\theta}_j \mathcal{I}_{\bs{\gamma}}= x_j\bar{x}_j\partial_j\bar{\partial}_j\mathcal{I}_{\bs{\gamma}}= x_j\bar{x}_j\partial_i\bar{\partial}_i\mathcal{I}_{\bs{\gamma}}
\]
so that 
\begin{align}
B_i(\theta,\bar{\theta})\mathcal{I}_{\bs{\gamma}} &= \Big[Q_i(\theta,\bar{\theta})x_i + \sum_{j\neq i}Q_j(\theta,\bar{\theta}) x_j\bar{x}_j \bar{\partial}_i\Big]\partial_i\mathcal{I}_{\bs{\gamma}},
\end{align}
This yields the creation operators
\begin{align}
\mathcal{C}_i &= Q_i(\theta,\bar{\theta})x_i + \sum_{j\neq i}Q_j(\theta,\bar{\theta}) x_j\bar{x}_j \bar{\partial}_i.
\end{align}
To project 
from the GKZ space to the physical hypersurface spanned by the momenta, first we  re-write (suppressing  arguments  for clarity)
\begin{align}\label{re1}
Q_ix_i &= x_i Q_i\Big|_{\theta_i\rightarrow\theta_i+1}\\
Q_j x_j\bar{x}_j \bar{\partial}_i &= 
x_j\bar{x}_j\bar{\partial}_iQ_j\Big|_{\theta_j\rightarrow\theta_j+1,\,\bar{\theta}_j\rightarrow\bar{\theta}_j+1}
\label{re2}
\end{align} 
where for  \eqref{re2} we recall the $Q_j$ are independent of both $\theta_i$ and $\bar{\theta}_i$. 
We then project to the physical hypersurface \eqref{contphys} by
using the Euler equations \eqref{eulercont} to replace $\bar{\theta}_k \rightarrow  \theta_k-\gamma_k$ for all $k=1,\ldots, n$ (which is justified since after the re-arrangements \eqref{re1}-\eqref{re2} all $\bar{\theta}_k$ act directly on $\mathcal{I}_{\bs{\gamma}}$) and set all $\bar{x}_k\rightarrow 1$.
Note also that  $\bar{\partial}_i$ and $\tb_i$ are equivalent on the physical hypersurface since $\bar{x}_i=1$, hence we can also replace $\bar{\partial}_i\rightarrow \theta_i-\gamma_i$. 
 The result is
 \begin{align}\label{Cres}
\mathcal{C}_i^{\mathrm{ph}} &= x_i Q_i\Big|_{\theta_i\rightarrow\theta_i+1,\,\bar{\theta}_k\rightarrow \theta_k-\gamma_k} + (\theta_i-\gamma_i)\sum_{j\neq i}x_j Q_j\Big|_{\theta_j\rightarrow\theta_j+1,\,\bar{\theta}_k\rightarrow\theta_k-\gamma_k+\delta_{kj}}
\end{align}
 where the replacement on $\bar{\theta}_k$ applies to all the $\bar{\theta}$ variables present.   As previously, the superscript `ph' denotes the operator expressed in physical variables. 
From the shadow relation \eqref{contshadow}, we also have
\begin{align}\label{Cbarshadow}
\bar{\mathcal{C}}_i{}^{\mathrm{ph}}&= x_i^{\g_i-1} \mathcal{C}_i^{\mathrm{ph}}\Big|_{\g_i\rightarrow-\g_i}\, x_i^{-\g_i}\nn\\&
=Q_i\Big|_{\t_i\rightarrow\t_i-\g_i+1,\,
\bar{\theta}_k\rightarrow \theta_k-\gamma_k} + \sum_{j\neq i}x_j\partial_i Q_j\Big|_{\theta_j\rightarrow\theta_j+1,\,\bar{\theta}_k\rightarrow\theta_k-\gamma_k+\delta_{kj}}.
\end{align}
Together, these expressions give us the creation operators in terms of the physical variables 
\[\label{badnotation}
x_k=p_k^2, \qquad \theta_k = x_k\partial_k = \frac{1}{2}p_k\partial_{p_k},
\]  
From \eqref{creationeq}, their action is 
\begin{align}\label{creationactioncont}
\mathcal{C}_i\mathcal{I}_{\{\gamma_0,\,\gamma_i\}} &= -(\gamma_0-1)^{-1}b_i(\gamma_0-1,\g_i+1)\mathcal{I}_{\{\gamma_0-1,\,\gamma_i+1\}},\nn\\[0.5ex]
\bar{\mathcal{C}}_i\mathcal{I}_{\{\gamma_0,\,\gamma_i\}} &= -(\gamma_0-1)^{-1}\bar{b}_i(\gamma_0-1,\,\g_i-1)\mathcal{I}_{\{\gamma_0-1,\,\gamma_i-1\}}.
\end{align}
The shift in the $b$-function arguments on the right-hand sides here reflects the fact that, in replacing $\tb_k\rightarrow\t_k-\g_k$ in the projection step above, we are taking the creation operator to act on the integral $\mathcal{I}_{\{\gamma_0,\,\gamma_i\}}$.  This is equivalent to eliminating $\bs{\g}$ from \eqref{creationeq} using  \eqref{gshift2} then relabelling $\bs{\g}'\rightarrow\bs{\g}$.

\subsubsection{3-point creation operator}
\label{sec:3KC}

Let us find the creation operator $\mathcal{C}_1$ for the 3-point function via the procedure given above.
Starting from the expression for 
 $B_1(\theta,\bar{\theta})$ in \eqref{3ptbigBex}, we  decompose
\begin{align}
B_1(\theta,\bar{\theta}) = Q_1 \theta_1+Q_2\theta_2\bar{\theta}_2+Q_3\theta_3\bar{\theta}_3
\end{align}
where 
\begin{align}
&Q_1=(\t_1+u_2+u_3)\bigl( (\t_1+u_2)(\t_1+u_3)+2(v_2+v_3)\bigr),\\
&Q_2=(u_2+u_3)u_3+v_2-v_3,\\
&Q_3=(u_2+u_3)u_2-v_2+v_3,
\end{align}
with 
\[
u_i = \theta_i+\bar{\theta}_i,\qquad v_i= \theta_i\bar{\theta}_i,\qquad i=2,3.
\]
Note that $Q_2$ and $Q_3$ are independent of $\theta_1$ and all coefficients are independent of $\bar{\theta}_1$.  We have also chosen $Q_2$ and $Q_3$ to preserve the $2\leftrightarrow 3$ symmetry though this is not essential.

Following the steps outlined above, making use of \eqref{useoftorics}, we have
\begin{align}
B_1(\theta,\bar{\theta})\mathcal{I}_{\bs{\gamma}}
=\big(Q_1 x_1 + Q_2 x_2\bar{x}_2\bar{\partial}_1+Q_3 x_3\bar{x}_3\bar{\partial}_1\big)\p_1\mathcal{I}_{\bs{\gamma}}=\mathcal{C}_1\p_1 \mathcal{I}_{\bs{\gamma}}
\end{align}
yielding the creation operator $\mathcal{C}_1$ in GKZ space.  Moving the $Q_k$ to the right, this can equivalently be written
\begin{align}
\mathcal{C}_1&= x_1 Q_1\Big|_{\theta_1\rightarrow\theta_1+1} +  x_2\bar{x}_2\bar{\partial}_1 Q_2\Big|_{\theta_2\rightarrow\theta_2+1,\,\bar{\theta}_2\rightarrow\bar{\theta}_2+1}+  x_3\bar{x}_3\bar{\partial}_1 Q_3\Big|_{\theta_3\rightarrow\theta_3+1,\,\bar{\theta}_3\rightarrow\bar{\theta}_3+1} 
\end{align}
Since shifting $\theta_i\rightarrow \theta_i+1$ and $\bar{\theta}_i\rightarrow\bar{\theta}_i+1$ is equivalent to shifting $u_i\rightarrow u_i+2$ and $v_i\rightarrow 1+u_i+v_i$, this is 
\begin{align} \label{3KC1gkz}
\mathcal{C}_1&
 = x_1(\t_1+1+u_2+u_3)\big((\t_1+1+u_2)(\t_1+1+u_3)+2(v_2+v_3)\big)\notag\\
&\quad +x_2\bar{x}_2\bar{\p}_1\bigl(
1+u_2+v_2-v_3+(u_2+u_3+2)u_3\bigr)\notag\\
&\quad +x_3\bar{x}_3\bar{\p}_1\bigl(1+u_3+v_3-v_2+(u_2+u_3+2)u_2\bigr).
\end{align}
Finally, to project to the physical hypersurface, we set
\[\label{3Kphysrepl}
\bar{x}_i\rightarrow 1,\qquad
\bar{\theta}_i\rightarrow \theta_i-\gamma_i,\qquad
\bar{\p}_i\rightarrow \theta_i-\gamma_i
\]
which sends $u_i\rightarrow 2\theta_i-\gamma_i$ and $v_i\rightarrow \theta_i(\theta_i-\gamma_i)$, yielding
\begin{align}
\mathcal{C}_1 &=x_1(\t_1+1+2\t_2+2\t_3-\g_2-\g_3)\times \\&\quad \times \bigl[(\t_1+1+2\t_2-\g_2)(\t_1+1+2\t_3-\g_3)
 +2\t_2(\t_2-\g_2)+2\t_3(\t_3-\g_3)
\bigr]\nn\\ &\quad +(\t_1-\g_1)\times\notag\\
&\quad\times\bigl[x_2\bigl(1+2\t_2-\g_2+\t_2(\t_2-\g_2)-\t_3(\t_3-\g_3)+(2\t_2-\g_2+2\t_3-\g_3+2)(2\t_3-\g_3)\bigr)\notag\\
&\quad +x_3\bigl(1+2\t_3-\g_3+\t_3(\t_3-\g_3)-\t_2(\t_2-\g_2)+(2\t_2-\g_2+2\t_3-\g_3+2)(2\t_2-\g_2)\bigr)\bigr]\nn
\end{align}
This result also follows from \eqref{Cres} directly. 
We can simplify somewhat further by using the DWI evaluated on the physical hypersurface,
\[
0 = \Big(\t_1+\t_2+\t_3+\frac{1}{2}(\g_0-\g_1-\g_2-\g_3)\Big)\mathcal{I}_{\bs{\g}}.
\]
This gives the alternative form
\begin{align}\label{3KC1}
&\mathcal{C}_1
=
- \frac{x_1}{2}(\t_1-1+\g_0-\g_1)\Big[2\t_1^2+2(\g_0-\g_1)\t_1+(\g_0-\g_1-1)^2+1-\g_2^2-\g_3^2\Big] \\&\qquad
+(\t_1-\g_1)
\times\nn\\&\qquad \times
\Big[x_2\bigl(1+2\t_2-\g_2+\t_2(\t_2-\g_2)-\t_3(\t_3-\g_3)-(2\t_1-\g_1+\g_0-2)(2\t_3-\g_3)\bigr)\notag\\
&\qquad +x_3\bigl(1+2\t_3-\g_3+\t_3(\t_3-\g_3)-\t_2(\t_2-\g_2)-(2\t_1-\g_1+\g_0-2)(2\t_2-\g_2)\bigr)\Big]. \nn
\end{align}
The action of this creation operator is
\[
\mathcal{C}_1\mathcal{I}_{\{\g_0,\g_1\}} = -(\g_0-1)^{-1} b_1(\bs{\g})\Big|_{\g_0\rightarrow\g_0-1,\,\g_1\rightarrow\g_1+1}\mathcal{I}_{\{\g_0-1,\g_1+1\}}
\]
where, using \eqref{3ptcontb}, 
\[\label{evalbfn3pt}
 b_1(\bs{\g})\Big|_{\g_0\rightarrow\g_0-1,\,\g_1\rightarrow\g_1+1}
=
\frac{1}{16} \Big[ \big((2 -\g_0+\g_1)^2 - \g_2^2-\g_3^2\big)^2 - 4 \g_2^2\g_3^2\Big].
\]

\subsubsection{4-point creation operator}

From \eqref{littlebdef0}, the 4-point $b$-function is 
\begin{align}\label{b1contdef}
b_1(\bs{\gamma}) &= 2^{-8}(-\gamma_0+\gamma_1+\gamma_2+\gamma_3+\g_4)(-\gamma_0+\gamma_1-\gamma_2+\gamma_3+\g_4)
(-\gamma_0+\gamma_1+\gamma_2-\gamma_3+\g_4)
\nn\\&\qquad \times(-\g_0+\g_1+\g_2+\g_3-\g_4)
(-\gamma_0+\gamma_1-\gamma_2-\gamma_3+\g_4)
(-\gamma_0+\gamma_1-\gamma_2+\gamma_3-\g_4)
\nn\\&\qquad \times(-\g_0+\g_1+\g_2-\g_3-\g_4)
(-\g_0+\g_1-\g_2-\g_3-\g_4).
\end{align}
After use of the Euler equations and DWI, this corresponds to
\begin{align}\label{Bfacs}
B_1 
&=\,
 (\theta_1+\theta_2+\theta_3+\t_4) (\theta_1+\bar{\theta}_2+\theta_3+\t_4) (\theta_1+\theta_2+\bar{\theta}_3+\t_4)
 (\theta_1+\theta_2+\t_3+\bar{\t}_4)\nn\\&\quad\times
  (\theta_1+\bar{\theta}_2+\bar{\theta}_3+\t_4)
   (\theta_1+\bar{\theta}_2+\theta_3+\bar{\t}_4)
    (\theta_1+\theta_2+\bar{\theta}_3+\bar{\t}_4)
       (\theta_1+\bar{\theta}_2+\bar{\theta}_3+\bar{\t}_4)
\end{align}
consistent with \eqref{bigBdef}.
We wish to decompose this as
\begin{align}\label{4ptBdec}
B_1(\theta,\bar{\theta}) = Q_1 \theta_1+Q_2\theta_2\bar{\theta}_2+Q_3\theta_3\bar{\theta}_3+Q_4\theta_4\bar{\theta}_4,
\end{align}
where the functions $Q_2$, $Q_3$ and $Q_4$ are independent of $\theta_1$.

Let us deal with $Q_1$ first.
Denoting the eight factors in \eqref{Bfacs} as $R_m$ for $m=1,\ldots, 8$, we have
\[
B_1\Big|_{\t_1=0} = \prod_{m=1}^8 (-\t_1+R_m) = \sum_{m=0}^8 \sigma_{(m)}(R) (-\t_1)^{8-m} = B_1 +  \sum_{m=0}^7 \sigma_{(m)}(R) (-\t_1)^{8-m} 
\]
where $\sigma_{(m)}(R)$ is the $m$th elementary symmetric polynomial in the $R_m$.  Rearranging then gives
\[
Q_1 = \t_1^{-1}\big(B_1-B_1\big|_{\t_1=0}\big) = \sum_{m=0}^7 \sigma_{(m)}(R) (-\t_1)^{7-m},
\]
and since $\t_1$ appears in each of the factors in \eqref{Bfacs},
\[
Q_1 x_1 = x_1 Q_1\Big|_{\theta_1\rightarrow\theta_1+1} = \sum_{m=0}^7 \sigma_{(m)}(1+R) (-1-\t_1)^{7-m}.
\]
When acting on the GKZ integral $\mathcal{I}_{\bs{\g}} $, 
we can now use the Euler equations \eqref{eulercont} and DWI  \eqref{DWIcont} to rewrite this expression in terms of elementary symmetric polynomials  of just the parameters $\bs{\g}$ alone, namely
\[
x_1 Q_1\Big|_{\theta_1\rightarrow\theta_1+1}\mathcal{I}_{\bs{\g}} = x_1\sum_{m=0}^7 \sigma_{(m)}(r) (-1-\t_1)^{7-m}\mathcal{I}_{\bs{\g}},
\]
where the eight variables
\[
r_{\{m\}} = 1-\g_0+\g_1 \pm \g_2 \pm \g_3 \pm \g_4
\]
are formed by making all possible independent choices of $\pm$ signs.

We now turn to the remaining $Q_k$ coefficients in \eqref{4ptBdec} for $k=2,3,4$.
Defining the auxiliary functions
\begin{align}
S(\t)&=(\t+\t_3+\t_4)(\t+\tb_3+\t_4)(\t+\t_3+\tb_4)(\t+\tb_3+\tb_4),\\
T(\t) &= \t^{-1}(S(\t)-S(0)) = (\t+u_3+u_4)\big((\t+u_3)(\t+u_4)+2v_3+2v_4\big),
\end{align}
where
\[
u_k = \t_k+\tb_k,\qquad v_k=\t_k\tb_k,\qquad k=3,4
\]
we can decompose
\begin{align}
Q_2 &= T(\t_2)T(\tb_2),\\
Q_3 &= \big((u_3+u_4)u_4+v_3-v_4)(S(\t_2)+S(\tb_2)-S(0)),\\
Q_4 &= \big((u_3+u_4)u_3+v_4-v_3)(S(\t_2)+S(\tb_2)-S(0)).
\end{align}
Noting that for $k=3,4$,
\[
(S(\t_2)+S(\tb_2)-S(0))\Big|_{\t_k\rightarrow\t_k+1,\,\tb_k\rightarrow\tb_k+1} = (S(\t_2+1)+S(\tb_2+1)-S(1)),
\]
and using \eqref{Cres}, the creation operator is then
\begin{align}\label{4KC1}
&\mathcal{C}_1^{\mathrm{ph}} = 
x_1\sum_{m=0}^7 \sigma_{(m)}(r) (-1-\t_1)^{7-m}
 +(\t_1-\g_1)\Big[ x_2\, \hat{T}(\t_2+1)\hat{T}(\t_2-\g_2+1) \\&\quad\,
+\Big(x_3\big((2+\hat{u}_3+\hat{u}_4)\hat{u}_4+\hat{u}_3+1+\hat{v}_3-\hat{v}_4\big)
+x_4\big((2+\hat{u}_3+\hat{u}_4)\hat{u}_3+\hat{u}_4+1+\hat{v}_4-\hat{v}_3\big)\Big)
\nn\\[0.5ex]&\quad\,
\times \Big(\hat{S}(\t_2+1)+\hat{S}(\t_2-\g_2+1)-\hat{S}(1)\Big)\Big] \nn
\end{align}
where all hatted quantities are defined by replacing $\tb_k\rightarrow \t_k-\g_k$ for  $k=3,4$ in the corresponding unhatted quantities.
Its action is 
\[
\mathcal{C}_1\mathcal{I}_{\{\g_0,\g_1\}} = -(\g_0-1)^{-1} b_1(\bs{\g})\Big|_{\g_0\rightarrow\g_0-1,\,\g_1\rightarrow\g_1+1}
\mathcal{I}_{\{\g_0-1,\g_1+1\}}
\]
where, using $b_1(\bs{\g})$ as given in \eqref{b1contdef}, 
\begin{align}
 b_1(\bs{\g})\Big|_{\g_0\rightarrow\g_0-1,\,\g_1\rightarrow\g_1+1}
&=
\frac{1}{256 }\Big[\Big((2 -\g_0+\g_1)^2 - \mathcal{S}_{(1)}\Big)^4 - 8 \Big((2 -\g_0+\g_1\big)^2 - \mathcal{S}_{(1)}\Big)^2 \mathcal{S}_{(2)} \nn\\&\qquad + 16 \mathcal{S}_{(2)}^2 - 
   64 (2 -\g_0+\g_1)^2 \mathcal{S}_{(3)}\Big]
\end{align}
where the $\mathcal{S}_{(m)}$ are elementary symmetric polynomials in $\g_2^2$, $\g_3^2$ and $\g_4^2$.

\subsection{Examples}

Taking into account the additional gamma function factors in \eqref{icontIg}, the action of these creation operators on contact diagrams is
\begin{align}
\mathcal{C}_1 \,i_{[d;\,\Delta_1,\,\ldots,\,\Delta_n]} &= -4\g_1 b_1(\bs{\g})\Big|_{\g_0\rightarrow\g_0-1,\,\g_1\rightarrow\g_1+1}\,i_{[\tilde{d};\,\tilde{\Delta}_1,\Delta_2\,\ldots,\,\Delta_n]} 
\end{align}
where
\[\label{Cdshift}
\tilde{d} = d - \frac{2}{n-2},\qquad \tilde{\Delta}_1 = \Delta_1+\frac{n-3}{n-2},
\]
Alternatively, in terms of the multiple-Bessel integral 
\[\label{nKdef0}
I_{\g_0\,\{\g_1,\,\ldots,\,\g_n\}} = 
\int_0^\infty \D z\, z^{\g_0}\prod_{i=1}^n p_i^{2\g_i}K_{\g_i}(p_iz),
\]
from \eqref{icont} and \eqref{icontIg} we have
\[\label{nKIrel}
\mathcal{I}_{\bs{\g}}=\frac{2^{n-\g_0}}{\Gamma(\g_0)}\,I_{\g_0\,\{\g_1,\,\ldots,\,\g_n\}} 
\]
and hence
 \begin{align}\label{C13K}
\mathcal{C}_1 I_{\g_0\,\{\g_1,\,\g_2,\,\ldots,\,\g_n\}}  &= -2 b_1(\bs{\g})\Big|_{\g_0\rightarrow\g_0-1,\,\g_1\rightarrow\g_1+1} \,I_{\g_0-1\,\{\g_1+1,\,\g_2,\,\ldots,\,\g_n\}}.
\end{align}
Here we can either use \eqref{badnotation} to rewrite $\mathcal{C}_1$ in terms of the momenta $p_i$, or more easily, re-express \eqref{nKdef0} using $p_i=\sqrt{x_i}$ then convert back to $p_i$ after acting with $\mathcal{C}_1$.

A quick check of these results can be obtained by examining cases where  all the  Bessel indices $\g_i$ take half-integer values allowing direct evaluation of the contact diagrams.  (We restrict to cases where both the initial and the shifted integral are finite; for the analysis of renormalised cases see \cite{Bzowski:2022rlz}.)
For example,
at three points, the triple-$K$ integrals  \eqref{nKdef0}
\begin{align}
&I_{4\{\frac{1}{2}\frac{1}{2}\frac{1}{2}\}}=\frac{15 \pi^2}{16\sqrt{2}}\left(p_1+p_2+p_3\right)^{-7/2},\qquad
I_{3\{\frac{3}{2}\frac{1}{2}\frac{1}{2}\}}=\frac{\pi^2(5p_1+2p_2+2p_3)}{8\sqrt{2}(p_1+p_2+p_3)^{5/2}},
\end{align}
and one can verify that 
\[
\mathcal{C}_1 I_{4\{\frac{1}{2}\frac{1}{2}\frac{1}{2}\}}=
-\frac{45}{128}\,
I_{3\{\frac{3}{2}\frac{1}{2}\frac{1}{2}\}},
\]
consistent with \eqref{C13K} using \eqref{evalbfn3pt} for the 3-point $b$-function.  We have performed many similar checks at both $3$- and $4$-points.

More non-trivially, many  triple-$K$ integrals with integer indices can be evaluated \cite{Bzowski:2015yxv}
by acting with the annihilators $\mathcal{L}_i$ and $\mathcal{R}_i$ 
given in \eqref{LRdef0}
on the known `seed' integral $I_{1\,\{000\}}$
which can be evaluated in terms of 
the Bloch-Wigner dilogarithm.  These relations enable computation of all the necessary triple-$K$ integrals arising in 3-point functions of conserved currents and stress tensors in even spacetime dimensions \cite{Bzowski:2017poo, Bzowski:2018fql}.
Since the creation operators $\mathcal{C}_i$ and $\bar{\mathcal{C}}_i$ are the inverse of $\mathcal{L}_i$ and $\mathcal{R}_i$, this allows us to reverse the direction of all operations linking different triple-$K$ integrals within the reduction scheme.  Thus, for example, we find
\[
\mathcal{R}_1 I_{1\{000\}} = I_{2\{100\}},\qquad
-8 C_1  I_{2\{100\}} =   I_{1\{000\}}
\]
where the integrals 
\begin{align}
 I_{1\{000\}} &= \frac{1}{2p_3^2(z-\bar{z})}\Big[\mathrm{Li}_2\, z
-\mathrm{Li}_{2} \, \bar{z}+\frac{1}{2}\ln (z\bar{z}) \ln\Big(\frac{1-z}{1-\bar{z}}\Big)\Big],\\
I_{2\{100\}}&= 
\frac{1}{2 p_3^2 (z - \bar{z})^2}\Big[  
  4 p_3^2 z\bar{z} (-2 + z + \bar{z})  I_{1\{000\}}  - 2 z \bar{z} \ln(z \bar{z})\nn\\ &\qquad\qquad\qquad\quad - (z + \bar{z} - 
      2 z\bar{z})  \ln[(1 - z) (1 - \bar{z})] \Big]
\end{align}
and the variables 
\[\label{zexp}
z = \frac{1}{2p_3^2}\Big(p_1^2-p_2^2+p_3^2+\sqrt{-J^2}\Big), \qquad
\bar{z} = \frac{1}{2p_3^2}\Big(p_1^2-p_2^2+p_3^2-\sqrt{-J^2}\Big)
\]
or equivalently
\[
\frac{p_1^2}{p_3^2} = z\bar{z}, \qquad
\frac{p_2^2}{p_3^2}= (1-z)(1-\bar{z})
\]
with 
\begin{align}
J^2 & = (p_1 + p_2 + p_3) (- p_1 + p_2 + p_3) (p_1 - p_2 + p_3) (p_1 + p_2 - p_3)\nn\\
&= -p_1^4-p_2^4-p_3^4+2p_1^2p_2^2+2p_2^2p_3^2+2p_3^2p_1^2. 
\end{align}

\subsection{Shift operators preserving the spacetime dimension} 

The creation operators  constructed above decrease the spacetime dimension according to \eqref{Cdshift}.
 For many applications, we would prefer an operator capable of changing the scaling dimensions of a contact diagram while preserving the spacetime dimension.  Thus, we seek an operator $W_{12}^{\sigma_1,\sigma_2}$ such that 
\[\label{desiredaction}
W_{12}^{\sigma_1,\sigma_2}i_{[d;\,\Delta_1, \Delta_2, \,\Delta_3,\,\ldots,\, \Delta_n]} \propto\, i_{[d;\,\Delta_1+\sigma_1,\, \Delta_2+\sigma_2, \,\Delta_3,\,\ldots,\, \Delta_n]} 
\] 
for any independent choice of  signs $\sigma_1=\pm 1$ and $\sigma_2=\pm 1$.
Operators of this type are known at three points \cite{Karateev:2017jgd,Baumann:2019oyu}, but  their analogue at four points acts on contact diagrams to generate shifted contact diagrams with derivative vertices \cite{Bzowski:2022rlz}. Instead,  our discussion of creation operators above can be modified to enable operators with the action \eqref{desiredaction} to be identified.\footnote{The shift operators that we identify will moreover be of minimal order, unlike the $d$-preserving combination of an annihilator $\p_i$ or $\bar{\p}_i$ followed by a creation operator  $\mathcal{C}_j$ or $\bar{\mathcal{C}}_j$. For example, the combination $\bar{\mathcal{C}}_1\p_2-\bar{\mathcal{C}}_2\p_1$ produces the same shift as $W_{12}^{--}$ but is of seventh order in derivatives for the 4-point function, since each product is eighth order and taking the difference lowers the order by one.  In contrast, the 4-point operator  $W_{12}^{--}$ we find will be of only fourth order.}
  At three points, these will coincide with the operators of \cite{Karateev:2017jgd,Baumann:2019oyu}, but at four points and higher they are novel.  Using these operators we will then  construct new shift operators  for exchange diagrams.

Our starting point is the observation that, for the GKZ integral \eqref{contactGKZ} corresponding to the contact diagram, 
\[\label{Weqn}
W_{12}^{--}\bar{\p}_1 \mathcal{I}_{\bs{\g}}= b_W(\bs{\g}) \p_2  \mathcal{I}_{\bs{\g}}.
\]
Recalling the parameter identifications \eqref{contparams}, the action of the  operators here is
\begin{align}\label{Wshift0}
&W_{12}^{--}:& \gamma_0&\rightarrow\gamma_0, & \gamma_1&\rightarrow\g_1-1, &\g_2&\rightarrow\g_2-1,\nn\\
&\bar{\p}_1: & \gamma_0&\rightarrow\gamma_0+1, & \gamma_1&\rightarrow\g_1+1, & \g_2&\rightarrow\g_2,\nn\\
&\p_2: & \gamma_0&\rightarrow\gamma_0+1, & \gamma_1&\rightarrow\g_1, & \g_2&\rightarrow\g_2-1,
\end{align}
with all remaining $\g_k$ for $k=3,\ldots, n$ staying  the same.
As  the shifts produced by the operators on each side of \eqref{Weqn} are the same, both sides involve the same integral $\mathcal{I}_{\bs{\g}}$.   As previously, the $b$-function $b_W(\bs{\g})$ should be a product of linear factors that vanishes whenever $W_{12}^{--}$ maps us from a finite to a singular integral.  
Taking into account the action \eqref{contannih} of the annihilators in \eqref{Weqn}, we have
\begin{align}\label{Weqn2}
W_{12}^{--}\bar{\p}_1 \mathcal{I}_{\bs{\g}}
&=-\g_0 W_{12}^{--}\mathcal{I}_{\bs{\g}}\Big|_{\g_0\rightarrow\g_0+1,\,\g_1\rightarrow\g_1+1}\nn\\
&=-\g_0 b_W(\bs{\g})   \mathcal{I}_{\bs{\g}}\Big|_{\g_0\rightarrow\g_0+1,\g_2\rightarrow \g_2-1}
= b_W(\bs{\g}) \p_2  \mathcal{I}_{\bs{\g}}
\end{align}
and so the zeros of $b_W(\bs{\g})$ must cancel the singularities of 
$ \mathcal{I}_{\bs{\g}}|_{\g_0\rightarrow\g_0+1,\g_2\rightarrow \g_2-1}$.
From \eqref{contsing},  this means
\begin{align}\label{bWres}
b_W(\bs{\g}) &=\prod_{\sigma_k\in\pm1}\frac{1}{2} \Big(-(\g_0+1)-\g_1-(\g_2-1)+ \sigma_3\g_3+\ldots \sigma_n\g_n\Big)\nn\\
&=\prod_{\sigma_k\in\pm1}\frac{1}{2} \Big(-\g_0-\g_1-\g_2+ \sigma_3\g_3+\ldots \sigma_n\g_n\Big).
\end{align}
Only the singularities with $\sigma_1=\sigma_2=-1$ in \eqref{contsing} appear here since these are the only cases for which $ \mathcal{I}_{\bs{\g}}|_{\g_0\rightarrow\g_0+1,\g_2\rightarrow \g_2-1}$ is singular but the integral $ \mathcal{I}_{\bs{\g}}|_{\g_0\rightarrow\g_0+1,\g_1\rightarrow \g_1+1}$ on which $W_{12}^{--}$ acts is finite.  Every possible independent choice of $\sigma_k\in\pm 1$ for all $k=3,\ldots, n$ is permitted, however, and gives rise to a corresponding factor in \eqref{bWres}.
Once again, we have also chosen to include trivial factors of one-half in $b_W(\bs{\g})$ to simplify the subsequent form of $W_{12}^{--}$.
Replacing the parameters $\bs{\g}$ in $b_W(\bs{\g})$ using the Euler equations \eqref{eulercont} and DWI \eqref{DWIcont}, we find
\begin{align}\label{Weqn3}
W_{12}^{--}\bar{\p}_1 \mathcal{I}_{\bs{\g}}
=\p_2 \big(b_W(\bs{\g})\mathcal{I}_{\bs{\g}}\big) = \p_2 B_W(\t,\tb)\mathcal{I}_{\bs{\g}} 
\end{align}
where
\begin{align}
 B_W(\t,\tb)&=\prod_{\sigma_k\in\pm1}\frac{1}{2} \Big(\sum_{j=1}^n (\t_j+\tb_j)-(\t_1-\tb_1)-(\t_2-\tb_2)+ \sigma_3(\t_3-\tb_3)+\ldots \sigma_n(\t_n-\tb_n)\Big)\nn\\
&= \prod_{\sigma_k\in\pm1}\Big(\tb_1+\tb_2+(\delta_{\sigma_3,+1}\t_3+\delta_{\sigma_3,-1}\tb_3)+\ldots +(\delta_{\sigma_n,+1}\t_n+\delta_{\sigma_n,-1}\tb_n)\Big).
\end{align}
Since $B_W(\t,\tb)$ is in fact independent of $\t_2$ the ordering of $\p_2$ and $B_W(\t,\tb)$ on the right-hand side of \eqref{Weqn3} is in fact immaterial, but had this not been the case the ordering shown would be the correct one when using the unshifted Euler equations and DWI to replace the $\bs{\g}$ parameters.

To identify $W_{12}^{--}$, all that is needed now is to start with $\p_2B_W(\t,\tb)$ and, using the toric equations \eqref{toriccont}, pull out a right factor of $\bar{\p}_1$ according to \eqref{Weqn3}.
As usual, the resulting operator can then be projected down to the physical hypersurface using the Euler equations and DWI.  These procedures are illustrated for the 3- and 4-point function below. 
Finally, given $W_{12}^{--}$ in physical variables, all the remaining operators in \eqref{desiredaction} can be found by shadow conjugation using \eqref{GKZshadow}, namely
\begin{align}\label{Wshadows}
(W_{12}^{+-})_{\mathrm{ph}} &= x_1^{1+\g_1} (W_{12}^{--})_{\mathrm{ph}}\,  x_1^{-\g_1},\\
(W_{12}^{-+})_{\mathrm{ph}}\,  &= x_2^{1+\g_2} (W_{12}^{--})_{\mathrm{ph}}\,  x_2^{-\g_2},\\
(W_{12}^{++})_{\mathrm{ph}}\,  &= x_1^{1+\g_1} x_2^{1+\g_2} (W_{12}^{--})_{\mathrm{ph}}\,  x_1^{-\g_1}x_2^{-\g_2}.
\end{align}

\subsubsection{3-point function}
\label{sec:3KW}

To illustrate the above discussion, for the 3-point function we have
\[
b_W(\bs{\g}) = \frac{1}{4}(-\g_0-\g_1-\g_2+\g_3)(-\g_0-\g_1-\g_2-\g_3)
\]
and 
\[
B_W(\t,\tb) = (\tb_1+\tb_2+\t_3)(\tb_1+\tb_2+\tb_3).
\]
The operator $W_{12}^{--}$ can now be extracted from
\[
W_{12}^{--}\bar{\p}_1\mathcal{I}_{\bs{\g}}  = \p_2 B_W(\t,\tb)\mathcal{I}_{\bs{\g}}.
\]
For this, we write
\begin{align}
&\p_2(\tb_1+\tb_2+\t_3)(\tb_1+\tb_2+\tb_3)\mathcal{I}_{\bs{\g}}\nn\\
&\qquad=\p_2\big[(\tb_1+\tb_2+\tb_3+\t_3)(\tb_1+\tb_2)+\t_3\tb_3]\mathcal{I}_{\bs{\g}}\nn\\
&\qquad=\big[(\tb_1+\tb_2+\tb_3+\t_3)(\bar{x}_1\p_2\bar{\p}_1+\bar{x}_2\p_2\bar{\p}_2)+x_3\bar{x}_3 \p_2\p_3\bar{\p}_3]\mathcal{I}_{\bs{\g}}\nn\\
&\qquad= \big[(\tb_1+\tb_2+\tb_3+\t_3)(\bar{x}_1\p_2+\bar{x}_2\p_1)+x_3\bar{x}_3 \p_2\p_1]\bar{\p}_1 \mathcal{I}_{\bs{\g}}
\end{align}
where in the penultimate line we used the toric equations \eqref{toriccont}.  Thus
\begin{align}\label{W3ptgkz}
W_{12}^{--}&=(\tb_1+\tb_2+\tb_3+\t_3)(\bar{x}_1\p_2+\bar{x}_2\p_1)+x_3\bar{x}_3 \p_2\p_1 \nn\\
&=(\bar{x}_1\p_2+\bar{x}_2\p_1)(1+\tb_1+\tb_2+\tb_3+\t_3)+x_3\bar{x}_3 \p_2\p_1, 
\end{align}
and using the DWI \eqref{DWIcont} to project  to the physical hypersurface \eqref{contphys},  we obtain
\begin{align}\label{W3ptphys}
(W_{12}^{--})_{\mathrm{ph}}&=
(\p_2+\p_1)(1-\g_0-\t_1-\t_2)+x_3\p_2\p_1\nn\\&=
-(\g_0+\t_1+\t_2)(\p_1+\p_2)+x_3\p_1\p_2
\end{align}
where for the 3-point function $\g_0=d/2$ from \eqref{contparams}.
A short calculation shows that
\[\label{curlyWdef}
(W_{12}^{--})_{\mathrm{ph}} = -\frac{1}{4}\Big( \p_{p_1}^2+\partial_{p_2}^2+\frac{(d-1)}{p_1}\p_{p_1}+\frac{(d-1)}{p_2}\p_{p_2}+(p_1^2+p_2^2-p_3^2)\frac{1}{p_1p_2}\p_{p_1}\p_{p_2}\Big)
\]
which, up to a factor of $-2$, is the 3-point shift operator studied in \cite{Baumann:2019oyu, Bzowski:2022rlz}.

The action of $W_{12}^{--}$ is
\begin{align}\label{Weqn4}
W_{12}^{--}\mathcal{I}_{\bs{\g}} = b_W(\bs{\g})\Big|_{\g_0\rightarrow\g_0-1,\,\g_1\rightarrow\g_1-1}\mathcal{I}_{\bs{\g}}\Big|_{\g_1\rightarrow\g_1-1,\,\g_2\rightarrow\g_2-1}
\end{align}
where the shift on the $b$-function derives from the fact that, in the projection step going from \eqref{W3ptgkz} to \eqref{W3ptphys}, we have chosen that $W_{12}^{--}$ acts on the integral $\mathcal{I}_{\bs{\g}}$ requiring us to shift the $\bs{\g}$ parameters present in \eqref{Weqn2}.  Evaluating, this gives
\begin{align}
b_W(\bs{\g})\Big|_{\g_0\rightarrow\g_0-1,\,\g_1\rightarrow\g_1-1} &= \frac{1}{4}(2-\g_0-\g_1-\g_2+\g_3)(2-\g_0-\g_1-\g_2-\g_3)\nn\\&
=\frac{1}{4}\Big((\g_0+\g_1+\g_2-2)^2-\g_3^2\Big)
\end{align}
such that \eqref{Weqn4} is consistent with the action of $\mathcal{W}_{12}^{--}$ obtained  in  \cite{Bzowski:2022rlz}.
Acting on the holographic contact diagram, from \eqref{icontIg}
we have
\[
W_{12}^{--}i_{[d,\,\Delta_1,\,\Delta_2,\,\Delta_3]} = 
\frac{1}{4(\g_1-1)(\g_2-1)}b_W(\bs{\g})\Big|_{\g_0\rightarrow\g_0-1,\,\g_1\rightarrow\g_1-1}
i_{[d,\,\Delta_1-1,\,\Delta_2-1,\,\Delta_3]}. 
\]

\subsubsection{4-point function}

At 4-points, we find
\begin{align}
b_W(\bs{\g}) &= \frac{1}{16}(-\g_0-\g_1-\g_2+\g_3+\g_4)
(-\g_0-\g_1-\g_2-\g_3+\g_4)\nn\\&\qquad\times  
(-\g_0-\g_1-\g_2+\g_3-\g_4)
(-\g_0-\g_1-\g_2-\g_3-\g_4)
\end{align}
and hence
\begin{align}
B_W(\t,\tb) &= (\tb_1+\tb_2+\t_3+\t_4) (\tb_1+\tb_2+\tb_3+\t_4)
 (\tb_1+\tb_2+\t_3+\tb_4)
  (\tb_1+\tb_2+\tb_3+\tb_4).
\end{align}
Once again, to find $W_{12}^{--}$ we must factorise
\[
W_{12}^{--}\bar{\p}_1\mathcal{I}_{\bs{\g}}  = \p_2 B_W(\t,\tb)\mathcal{I}_{\bs{\g}}.
\]
As a first step, we expand
\[
B_W(\t,\tb) = Q_0(\tb_1+\tb_2)+Q_3\t_3\tb_3+Q_4\t_4\tb_4
\]
where the coefficients 
\begin{align}\label{WQco}
Q_0 &=(u_3 + u_4 + \tb_1+\tb_2) \Big(2 (v_3 + v_4) + (u_3 + \tb_1+\tb_2) (u_4 + \tb_1+\tb_2)\Big),\nn\\
Q_3 &= (u_3+u_4)u_4+v_3-v_4,\nn\\
Q_4 &= (u_3+u_4)u_3-v_3+v_4,
\end{align}
and 
\[
u_k=\t_k+\tb_k,\qquad v_k=\t_k\tb_k, \qquad k=3,4.
\]
Now, since all coefficients are independent of  $\t_2$, 
\begin{align}
\p_2B_W(\t,\tb)\mathcal{I}_{\bs{\g}} &=\Big[
Q_0(\bar{x}_1\p_2\bar{\p}_1
+ \bar{x}_2 \p_2\bar{\p_2})+Q_3 x_3\bar{x}_3\p_2\bar{\p}_3\p_3
+Q_4 x_4\bar{x}_4\p_2\bar{\p}_4\p_4\Big]\mathcal{I}_{\bs{\g}} 
\\&
=\Big[\bar{x}_1 Q_0\Big|_{\tb_1\rightarrow\tb_1+1}\p_2
+\bar{x}_2  Q_0\Big|_{\tb_2\rightarrow\tb_2+1}\p_1
\nn\\&\qquad
+
x_3 \bar{x}_3 Q_3\Big|_{\t_3\rightarrow\t_3+1,\,\tb_3\rightarrow\tb_3+1}\p_2\p_1
+x_4 \bar{x}_4 Q_4\Big|_{\t_4\rightarrow\t_4+1,\,\tb_4\rightarrow\tb_4+1}\p_2\p_1\Big]\bar{\p}_1\mathcal{I}_{\bs{\g}} \nn
\end{align}
where in the second line we used the toric equations \eqref{toriccont}.
We thus have
\begin{align}
W_{12}^{--}&=(\bar{x}_1 \p_2 +\bar{x}_2\p_1)Q_0\Big|_{\tb_1\rightarrow\tb_1+1}
\nn\\&\qquad
+\p_1\p_2
\Big(x_3 \bar{x}_3 Q_3\Big|_{\t_3\rightarrow\t_3+1,\,\tb_3\rightarrow\tb_3+1}
+x_4 \bar{x}_4 Q_4\Big|_{\t_4\rightarrow\t_4+1,\,\tb_4\rightarrow\tb_4+1}\Big)
\end{align}
where in the first line we used the fact that $\tb_1$ and $\tb_2$ enter $Q_0$ only in the combination $\tb_1+\tb_2$ and so the replacement $\tb_1\rightarrow \tb_1+1$ produces the same result as $\tb_2\rightarrow \tb_2+1$ allowing us to combine the two $Q_0$ terms.  We have in addition moved $Q_0$, $Q_3$ and $Q_4$ to the right (noting that all coefficients are independent of $\t_1$ and $\t_2$)  so as to be able to use the Euler equations for $\mathcal{I}_{\bs{\g}}$ to project to the physical hypersurface.   For this, we set  $\bar{x}_k\rightarrow 1$ and $\tb_k\rightarrow \t_k-\g_k$ inside all $Q_k$ coefficients giving 
\begin{align}\label{QWformula}
(W_{12}^{--})_{\mathrm{ph}} &= (\p_1+\p_2)Q_0\Big|_{\tb_k\rightarrow\t_k-\g_k+\delta_{k,1}}
\\&\quad
+\p_1\p_2
\Big(x_3  Q_3\Big|_{\t_3\rightarrow\t_3+1,\,\tb_k\rightarrow\t_k-\g_k+\delta_{k,3}}
+x_4 Q_4\Big|_{\t_4\rightarrow\t_4+1,\,\tb_k\rightarrow\t_k-\g_k+\delta_{k,4}}\Big).\nn
\end{align}
In all the replacements here, $\tb_k$ stands for any index $k=1,\ldots, 4$.
Evaluating this formula explicitly using the coefficients in \eqref{WQco}, we find
\begin{align}\label{4ptWv1}
&(W_{12}^{--})_{\mathrm{ph}} =
(\p_1+\p_2)(1-\g_0-\t_1-\t_2)\Big(2\t_3(\t_3-\g_3)+2\t_4(\t_4-\g_4) \\ &\qquad +
(1+2\t_3-\g_3+\t_1-\g_1+\t_2-\g_2)(1+2\t_4-\g_4+\t_1-\g_1+\t_2-\g_2)\Big)\nn\\&
\qquad
+\p_1\p_2\Big[x_3\Big((2+2\t_3-\g_3+2\t_4-\g_4)(2\t_4-\g_4)+(1+\t_3)(1+\t_3-\g_3)-\t_4(\t_4-\g_4)\Big)\nn\\&\qquad
+x_4\Big((2+2\t_3-\g_3+2\t_4-\g_4)(2\t_3-\g_3)+(1+\t_4)(1+\t_4-\g_4)-\t_3(\t_3-\g_3)\Big)\Big]\nn
\end{align}
where for the 4-point function $\g_0=d$ from \eqref{contparams}.

Alternatively, we can use the DWI \eqref{DWIcont}
projected to the physical hypersurface, 
\[
0=\Big(\frac{1}{2}(\g_0-\g_t)+\sum_{k=1}^4\t_k\Big)\mathcal{I}_{\bs{\g}},\qquad \g_t=\sum_{k=1}^4\g_k,
\]
to eliminate the factors of $\t_1+\t_2$ on the second line of \eqref{4ptWv1}.  After further moving all factors of  $\p_1$ and $\p_2$  to the right, this gives the equivalent form
\begin{align}\label{4ptWv2}
(W_{12}^{--})_{\mathrm{ph}} &=
-(\g_0+\t_1+\t_2)\Big((\t_3+\t_4)(\t_3+\t_4-\g_3-\g_4)
\\ &\qquad\qquad\qquad \qquad\quad 
+
\frac{1}{4}(2-\g_0-\g_t+2\g_3)(2-\g_0-\g_t+2\g_4 )\Big)(\p_1+\p_2)\nn\\&\hspace{-12mm}
+\Big[x_3\Big((2+2\t_3-\g_3+2\t_4-\g_4)(2\t_4-\g_4)+(1+\t_3)(1+\t_3-\g_3)-\t_4(\t_4-\g_4)\Big)\nn\\&\hspace{-12mm}
+x_4\Big((2+2\t_3-\g_3+2\t_4-\g_4)(2\t_3-\g_3)+(1+\t_4)(1+\t_4-\g_4)-\t_3(\t_3-\g_3)\Big)\Big]\p_1\p_2.\nn
\end{align}
The action of $W_{12}^{--}$ is
\begin{align}\label{Weqn5}
W_{12}^{--}\mathcal{I}_{\bs{\g}} = b_W(\bs{\g})\Big|_{\g_0\rightarrow\g_0-1,\,\g_1\rightarrow\g_1-1}\mathcal{I}_{\bs{\g}}\Big|_{\g_1\rightarrow\g_1-1,\,\g_2\rightarrow\g_2-1}
\end{align}
where, once again, the shift on the $b$-function derives from the fact that in projecting from GKZ variables to the physical hypersurface we chose $W_{12}^{--}$ to act on the unshifted integral $\mathcal{I}_{\bs{\g}}$ requiring us to shift the $\bs{\g}$ parameters present in \eqref{Weqn2}.  Explicitly, this is
\begin{align}
 b_W(\bs{\g})\Big|_{\g_0\rightarrow\g_0-1,\,\g_1\rightarrow\g_1-1}
&=\frac{1}{16}\Big(\g_3^2+\g_4^2-(\g_0+\g_1+\g_2-2)^2\Big)^2-\frac{1}{4}\g_3^2\g_4^2.
\end{align}
Acting on the holographic contact diagram, from \eqref{icontIg} we again have
\begin{align}\label{Woncontdi}
W_{12}^{--}i_{[d;\,\Delta_1,\Delta_2,\Delta_3,\Delta_4]} = \frac{1}{4(\g_1-1)(\g_2-1)} b_W(\bs{\g})\Big|_{\g_0\rightarrow\g_0-1,\,\g_1\rightarrow\g_1-1}i_{[d;\,\Delta_1-1,\Delta_2-1,\Delta_3,\Delta_4]}. 
\end{align}
To our knowledge,  this is the first time an operator that shifts the 4-point contact diagram in this fashion has been identified.  We emphasise that the 3-point operator \eqref{curlyWdef}, when applied to 4-point contact diagrams, generates shifted contact diagrams but with derivative vertices and hence does not satisfy this requirement  \cite{Bzowski:2022rlz}.

\paragraph{Examples:}

Contact diagrams for which the Bessel functions have half-integer indices can be evaluated directly.  This yields many simple examples for which the action of $W_{12}^{--}$ can be checked.  For instance, with  $(d,\Delta_1,\Delta_2,\Delta_3,\Delta_4)=(5,3,4,3,4)$, we find
\begin{align}
i_{[5;3,4,3,4]} &= \frac{1}{p_t^3}\Big(p_1^2+p_3^2+2(p_2^2+p_4^2)+3(p_1+p_3)(p_2+p_4)+2p_1p_3+6p_2p_4\Big)
\end{align}
where $p_t=\sum_{j=1}^4p_j$, while the shifted integral with $(d,\Delta_1,\Delta_2,\Delta_3,\Delta_4)=(5,2,3,3,4)$ is
\begin{align}
i_{[5;2,3,3,4]} &= -\frac{(p_t+2p_4)}{p_1 p_t^3}.
\end{align}
Evaluating the action of $W_{12}^{--}$ in \eqref{4ptWv2} using \eqref{badnotation}, we can verify \eqref{Woncontdi}, namely
\[
W_{12}^{--}i_{[5;3,4,3,4]} =-\frac{63}{2}i_{[5;2,3,3,4]}.
\]

\subsubsection{Combinations of operators}

To round up our discussion of shift operators for contact diagrams, we have identified operators mapping
\begin{align}
&\p_i:\quad \g_0\rightarrow\g_0+1,\quad \g_i\rightarrow\g_i-1\qquad
\bar{\p}_i:\quad \g_0\rightarrow\g_0+1,\quad \g_i\rightarrow\g_i+1\nn\\
&\mathcal{C}_i:\quad \g_0\rightarrow\g_0-1,\quad\g_i\rightarrow\g_i+1\qquad
\bar{\mathcal{C}}_i:\quad \g_0\rightarrow\g_0-1,\quad\g_i\rightarrow\g_i-1\nn\\
&W_{ij}^{\sigma_i\sigma_j}:\quad  \g_i\rightarrow\g_i+\sigma_i,\quad \g_j\rightarrow\g_j+\sigma_j,\qquad \{\sigma_i,\sigma_j\}\in\pm 1.
\end{align}
Combining these allows us to construct still further shifts, for example:
\begin{align}
\mathcal{C}_i\bar{\mathcal{C}}_i:\quad \g_0\rightarrow \g_0 - 2,
\qquad
\mathcal{C}_i\bar{\partial}_i: \quad \g_i\rightarrow\g_i+2,\qquad
\bar{\mathcal{C}}_i\partial_i: \quad \g_i\rightarrow\g_i-2.
\end{align}
Acting on the 3-point function specifically,
\[
\mathcal{C}_1W_{23}^{++}:\quad \g_0\rightarrow\g_0-1,\quad \g_i\rightarrow\g_i+1\quad\forall\,\,i=1,2,3
\]
which is equivalent to shifting $d\rightarrow d-2$ while preserving all operator dimensions $\Delta_i$.

Finally, one might wonder why all these operators produce a shift of two units:  why, for example, can one not construct an operator shifting $\g_0\rightarrow \g_0+1$ only, or just $\g_1\rightarrow \g_1+1$?  The absence of such operators can be traced to the spacing of the singular hyperplanes of the contact diagram,
 specifically the term $-2m$ appearing in the singularity condition \eqref{contsing}.  As $m\in\mathbb{Z}^+$, this means that the singularities are effectively spaced by two units.  Any operator that produced a shift of a single unit would require a $b$-function containing an infinite number of factors, since
there are infinitely many finite integrals  that are only one unit away from a singular integral.  (Namely, those for which $m$ is half-integer.)  As the number of factors in the $b$-function corresponds to the order of the differential operator, there is thus no single-shift operator of finite order.
In contrast, for an operator shifting by two units, the number of finite integrals that can be mapped to singular integrals is finite, and hence the $b$-functions and shift operators are also of finite order.

\subsection{Exchange diagrams}
\label{sec:exch}

Having analysed contact diagrams, we now turn to the  $s$-channel exchange diagrams \eqref{iexch}.
Rather than  constructing an explicit GKZ representation,  here we simply note that shifts of the form
\[
 i_{[d;\,\Delta_1, \Delta_2; \,\Delta_3, \Delta_4; \,\Delta_x]}\rightarrow  i_{[d;\,\Delta_1+\sigma_1, \Delta_2+\sigma_2; \,\Delta_3, \Delta_4; \,\Delta_x]}
\]
for any $ \{\sigma_1,\sigma_2\}\in \pm 1$ can be obtained by combining the 3- and 4-point   $W_{12}^{--}$ operators given in \eqref{W3ptphys} and \eqref{4ptWv2} with the $s$-channel Casimir operator.
As with contact diagrams, it is sufficient to focus on the case $\sigma_1=\sigma_2=-1$, since all remaining operators follow by shadow conjugation according to \eqref{Wshadows}.
We emphasise however that both the original and the shifted exchange diagrams we consider have purely {\it non-derivative} vertices. Moreover, any operator and spacetime dimensions are permitted, provided we work in dimensional regularisation where necessary to avoid divergences.

For purposes of disambiguation, let us define the operator
  \begin{align}
\mathcal{W}^{--}_{12} 
&=(d+2\t_1+2\t_2)(\p_1+\p_2)-2s^2 \p_1\p_2
\end{align}
where $d$ is the boundary spacetime dimension and $\p_i=\p/\p x_i$ with $x_i=p_i^2$ as usual.
This is simply the 3-point operator $-2W_{12}^{--}$   in \eqref{curlyWdef}, but with $p_3^2$ replaced by the Mandelstam variable $s^2=(\bs{p}_1+\bs{p}_2)^2$ as appropriate for acting on $s$-channel exchange diagrams.  (The factor of $-2$ is included for consistency with the $\mathcal{W}^{--}_{12}$ defined in \cite{Baumann:2019oyu, Bzowski:2022rlz}.)
In the following, we will then use $W_{12}^{--}$ to refer exclusively to the {\it 4-point} $W_{12}^{--}$  operator given in \eqref{4ptWv2}.

As shown in \cite{Bzowski:2022rlz},  
the action of $\mathcal{W}^{--}_{12}$  on an $s$-channel exchange diagram is to produce a {\it linear combination} of a shifted exchange and a shifted contact diagram:
\begin{align}\label{Wonexch}
\mathcal{W}^{--}_{12}\,i_{[d;\,\Delta_1,\Delta_2;\Delta_3,\Delta_4;\,\Delta_x]} &= 
\mathcal{N}_{exch.}\,
i_{[d;\,\Delta_1-1,\Delta_2-1;\Delta_3,\Delta_4;\,\Delta_x]}
 \nn\\& \quad 
 +\mathcal{N}_{cont.}\, i_{[d;\,\Delta_1-1,\Delta_2-1,\Delta_3,\Delta_4]}
\end{align} 
where the coefficients\footnote{ Where the shifted exchange diagram has a pole (or double pole) in dimensional regularisation,   one (or both) of the factors on the right-hand side of \eqref{Nexch} vanish, see \cite{Bzowski:2022rlz}.  
}
\begin{align}\label{Nexch}
\mathcal{N}_{exch.} &=\Big(\frac{d}{2}-2+\g_1
+\g_2+\g_x\Big)\Big(\frac{d}{2}-2+\g_1
+\g_2-\g_x\Big)\mathcal{N}_{cont.} \\
\mathcal{N}_{cont.} &=-\frac{1}{8(\g_1-1)(\g_2-1)}
\label{Ncont}
\end{align}
where $\g_i=\Delta_i-d/2$ and  $\g_x = \Delta_x-d/2$.
Thus, in order to go from an exchange diagram to shifted exchange diagram only, the shifted contact contribution in \eqref{Wonexch} must be subtracted.  

This can be accomplished in two steps.  First, the {\it unshifted} contact diagram is obtained by acting on the original  exchange diagram with the reduced Casimir operator, 
\begin{align} 
\hat{\mathcal{C}}_{12}\,i_{[d;\,\Delta_1,\Delta_2;\Delta_3,\Delta_4;\,\Delta_x]}  = i_{[d;\,\Delta_1,\Delta_2,\Delta_3,\Delta_4]}, \label{Casonex} 
\end{align}
where
\begin{align}\label{redCas}
\hat{\mathcal{C}}_{12}&=
2s^2\Big((\t_1+1-\g_1)\p_1+(\t_2+1-\g_2)\p_2\Big)-\Big(2\t_1+2\t_2-\g_1-\g_2+\frac{d}{2}\Big)^2+\g_x^2
\end{align}
with $\t_i=x_i\p_i$.
The action of this operator  on an $s$-channel exchange  is equivalent to that of the Casimir operator plus the square of the exchanged mass \cite{Bzowski:2022rlz}.\footnote{Specifically, 
$\hat{\mathcal{C}}_{12} = \tilde{\mathcal{C}}_{12}+m_x^2$ with $\tilde{\mathcal{C}}_{12}$ as defined in (6.44) of \cite{Bzowski:2022rlz} and $m_x^2=\g_x^2-d^2/4$.}
If desired, $\hat{\mathcal{C}}_{12}$ can be shortened using the identity
\[
0=\Big((\t_1+1-\g_1)\p_1-(\t_2+1-\g_2)\p_2\Big)i_{[d;\,\Delta_1,\Delta_2,\Delta_3,\Delta_4\,x\,\Delta_x]} 
\]
which corresponds to the difference of the Bessel operators acting on legs 1 and 2, {\it i.e.,} $K_1-K_2$ where $K_i = \p_{p_i}^2 + (1-2\g_i)p_i^{-1}\p_{p_i}$.  However,  \eqref{redCas} is symmetric under $1\leftrightarrow 2$.

For the second step, we now construct the shifted contact diagram using the 4-point $W_{12}^{--}$ operator defined in \eqref{4ptWv2}.
From \eqref{Woncontdi}, this has the action 
\begin{align}
&W_{12}^{--}i_{[d;\,\Delta_1,\Delta_2,\Delta_3,\Delta_4]} =\mathcal{N}_{W} \,\mathcal{N}_{cont.}\,i_{[d;\,\Delta_1-1,\Delta_2-1,\Delta_3,\Delta_4]}
\end{align}
with $\mathcal{N}_{cont.}$ from \eqref{Ncont} and 
\begin{align}
\mathcal{N}_W=
- \frac{1}{8}
\Big[\left(\g_3^2+\g_4^2-(d+\g_1+\g_2-2)^2\right)^2-4\g_3^2\g_4^2\Big].
\end{align}

Putting everything together, we find the operator
\begin{align}
\Omega_{12}^{--} = 
\mathcal{N}_W\,\mathcal{W}_{12}^{--} - W_{12}^{--}\hat{\mathcal{C}}_{12}
\end{align}
whose action is
\begin{align}\label{OmegaAction}
\Omega^{--}_{12}\,i_{[d;\,\Delta_1,\Delta_2;\,\Delta_3,\Delta_4;\,\Delta_x]} &= \mathcal{N}_W
\mathcal{N}_{exch.}\,
i_{[d;\,\Delta_1-1,\Delta_2-1;\,\Delta_3,\Delta_4;\,\Delta_x]}.
\end{align} 
This is therefore the desired operator mapping an exchange to a shifted exchange diagram.

Written out explicitly, with $\g_t=\sum_{j=1}^4\g_j$, we have
\begin{align}\label{OmegaOp}
\Omega^{--}_{12}&= - \frac{1}{8}
\Big[\left(\g_3^2+\g_4^2-(d+\g_1+\g_2-2)^2\right)^2-4\g_3^2\g_4^2\Big]\Big((d+2\t_1+2\t_2)(\p_1+\p_2)-2s^2 \p_1\p_2\Big)\nn\\
&\quad 
-\Big[
-(d+\t_1+\t_2)\Big((\t_3+\t_4)(\t_3+\t_4-\g_3-\g_4)\nn
\\ &\qquad\qquad\qquad \qquad\qquad 
+
\frac{1}{4}(2-d-\g_t+2\g_3)(2-d-\g_t+2\g_4 )\Big)(\p_1+\p_2)\nn\\&\hspace{-2mm}
+x_3\Big((2+2\t_3-\g_3+2\t_4-\g_4)(2\t_4-\g_4)+(1+\t_3)(1+\t_3-\g_3)-\t_4(\t_4-\g_4)\Big)\p_1\p_2\nn\\&\hspace{-2mm}
+x_4\Big((2+2\t_3-\g_3+2\t_4-\g_4)(2\t_3-\g_3)+(1+\t_4)(1+\t_4-\g_4)-\t_3(\t_3-\g_3)\Big)\p_1\p_2\Big]\times\nn\\
&\hspace{-2mm}
\times \Big[2s^2\Big((\t_1+1-\g_1)\p_1+(\t_2+1-\g_2)\p_2\Big)-\Big(2\t_1+2\t_2-\g_1-\g_2+\frac{d}{2}\Big)^2+\g_x^2\Big].
\end{align}

\paragraph{Examples:}  All exchange diagrams involving fields of $\Delta=2,3$ in $d=3$ were computed recently in \cite{Bzowski:2022rlz} and are available in the associated Mathematica package {\tt HandbooK.wl}.  These results enable many tests of the operator $\Omega_{12}^{--}$ in \eqref{OmegaOp} and its shadow conjugates
\begin{align}\label{Omshadows}
\Omega_{12}^{+-}&= x_1^{1+\g_1} \Omega_{12}^{--}\,  x_1^{-\g_1},\\
\Omega_{12}^{-+}\,  &= x_2^{1+\g_2} \Omega_{12}^{--},  x_2^{-\g_2},\\
\Omega_{12}^{++}\,  &= x_1^{1+\g_1} x_2^{1+\g_2} \Omega_{12}^{--}\,  x_1^{-\g_1}x_2^{-\g_2}.
\end{align}
For this, we work in the dimensionally regulated theory with $d\rightarrow d+2\ep$ and $\Delta_i\rightarrow \Delta_i+\ep$ for all $i=1,2,3,4,x$.   This scheme has the virtue of preserving the half-integer values of all Bessel function indices  $\g_i=\Delta_i-d/2$.  
The simplest such example is 
\begin{align}
i_{[3;22;22; 2]}&=-\frac{1}{2s}\mathcal{D}^{(+)},\\
i_{[3;33;22; 2]}&= \frac{1}{2}(p_3+p_4)\Gamma(2\ep) p_T^{-2\ep}+\frac{1}{4s}(p_1^2+p_2^2-s^2)\mathcal{D}^{(+)}
\nn\\&\quad 
+\frac{1}{2}(p_1+p_2)\Big[\log\Big(\frac{l_{34+}}{p_T}\Big)+1\Big]+\frac{7}{8}(p_3+p_4)+O(\ep)
\end{align}
where 
\[
p_T = \sum_{i=1}^4 p_i,\qquad l_{ij\pm} = p_i+p_j\pm s,
\]
and 
\[
\mathcal{D}^{(+)}=\Li_2\Big(\frac{l_{12-}}{p_T}\Big)+\Li_2\Big(\frac{l_{34-}}{p_T}\Big)+\log\Big(\frac{l_{12+}}{p_T}\Big)\log\Big(\frac{l_{34+}}{p_T}\Big)-\frac{\pi^2}{6}.
\]
By direct differentiation, one then finds
\[
\Omega_{12}^{--} i_{[3;33;22;2]}= (90+261\ep+O(\ep^2)) i_{[3;22;22;2]} + O(\ep)
\]
consistent with \eqref{OmegaAction}.  Note  $\mathcal{N}_W
\mathcal{N}_{exch.}$ on the right-hand side here is expanded to order $\ep$ since $i_{[3;22;22;2]}$ has an $\ep^{-1}$ pole.  
We have performed similar checks for all other values of the $\Delta_i$ and $\Delta_x$, and for the shadow conjugated operators.

This ability to shift exchange diagrams directly to other exchange diagrams means that, instead of computing all the  diagrams individually, we can compute the easiest diagram (namely, $i_{[3;22;22;2]}$) to sufficiently high order in the regulator $\ep$, and then obtain all others by acting with $\Omega_{12}^{\sigma_1\sigma_2}$ and $\Omega_{34}^{\sigma_3\sigma_4}$.

\section{Creation operators for Feynman diagrams}\label{sec:Feynman}

In this section, we analyse various Feynman integrals presenting their GKZ representations,  their singularities, and the associated creation operators.  Many of the examples we study have  appeared in the recent works \cite{de_la_Cruz_2019, Klausen:2019hrg, Feng_2020, Chestnov:2022alh}.  Here, our focus will be the construction of  creation operators and ways to automate their computation using standard Gr{\"o}bner basis and convex hulling algorithms.

In all cases, we start with an $L$-loop scalar integral in the momentum representation
\begin{align}\label{feynprop}
I=\Big(\prod_{j=1}^{L} \int \frac{\dd^d \bs{k}_j}{(2\pi)^d}\Big)\frac{1}{P_1^{\g_1}\ldots P_N^{\g_N}},
\end{align}
where the propagators $P_i$ for $i=1,\ldots, N$ are raised to  generalised powers $\g_i$.
As shown in appendix \ref{LPrep} (see also \cite{de_la_Cruz_2019, Weinzierl:2022eaz}), the corresponding GKZ integral is 
\begin{align}\label{feyngkz}
\mathcal{I}_{\bs{\g}}=  \Big(\prod_{i=1}^N \int_0^\infty \D z_i \,z_i^{\g_i-1}\Big)\,\mathcal{D}^{-\g_0},\qquad \g_0=\frac{d}{2}
\end{align}
where the denominator $\mathcal{D}$ is formed from the Lee-Pomeransky denominator $\mathcal{G}=\mathcal{U}+\mathcal{F}$, the sum of first and second Symanzik polynomials, by replacing the coefficient of every term with an independent variable $x_k$.
The Feynman integral \eqref{feynprop} now corresponds to
\[\label{feyntogkz}
I=c_{\bs{\g}}\mathcal{I}_{\bs{\g}},\qquad c_{\bs{\g}} = \frac{(4\pi)^{-L\g_0}\Gamma\left(\g_0\right)}{\Gamma\left((L+1)\g_0-\g_t\right)\prod_{i=1}^N\Gamma(\g_i)},\qquad \g_t=\sum_{i=1}^N \g_i
\] 
with the $x_k$ restored to their physical (Lee-Pomeransky) values. 
Knowing the coefficient $c_{\bs{\g}}$ enables the action of a creation operator on the GKZ integral $\mathcal{I}_{\bs{\g}}$ to be related to its action on the  Feynman integral $I$.

\subsection{Bubble diagram}

First, we consider the 1-loop bubble integral with propagators of mass $m_1$ and $m_2$.
As a warm-up, we begin with the single-mass case $(m_1,m_2)=(0,m)$ before turning to general masses.
The fully massless case $m_1=m_2=0$ is trivial (evaluating to a simple power of the momentum) and so will be omitted.

\subsubsection{1-mass bubble}
\begin{figure}[t]
\centering
\begin{tikzpicture}
\draw [dashed, thick](0,1)circle (1);
\draw[thick,->](-1.5,1)--(-1.2,1);
\draw[thick] (-1.2,1)--(-1.0,1);
\draw[thick](1.0,1)--(1.5,1);
\draw[thick] (1,1) arc (0:-180:1);
\node[text width=0.5cm, text centered ] at (0,1.7) {$1$};
\node[text width=0.5cm, text centered ] at (0,-0.3) {$2$};
\node[text width=0.5cm, text centered ] at (-1.7,1.0) {$\bs{p}$};
\end{tikzpicture} 
\caption{The single-mass bubble integral \eqref{1massbub1}, with massless and massive propagators represented by dashed and undashed lines respectively.}
\label{onemass-bubble}
\end{figure}
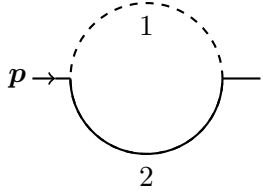
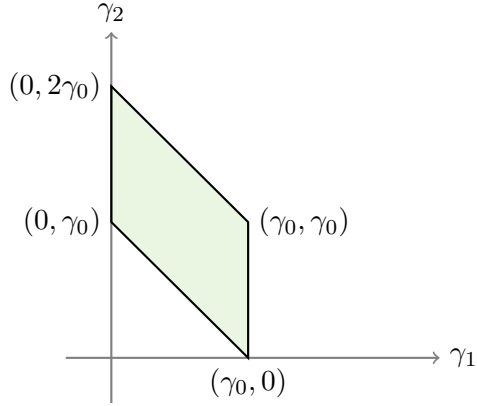
\begin{figure}[t]
\centering

\begin{tikzpicture}[scale=1.8]
\draw[->,thick,gray] (-1/3,0) -- (2.4,0); 
\draw[->,thick,gray] (0,0) -- (0,2.4); 
\draw[thick,gray] (0,0) -- (0,-1/3); 
\draw[white] (0,-1/3) -- (0,-1/3-0.08);  

\draw[thick,black,fill=YellowGreen, fill opacity=0.2] (1,0)--(0,1) -- (0,2) -- (1,1) --cycle;

     \draw node[left] at (0,1) {($0,\gamma_0$)};   
   \draw node[below] at (1,0) {($\gamma_0,0$)};
   \draw node[left] at (0,2) {($0,2\gamma_0$)};
   \draw node[right] at (1,1) {($\gamma_0,\gamma_0$)};
    \draw node[right] at (2.4,0) {$\gamma_1$};
     \draw node[above] at (0,2.4) {$\gamma_2$};
   
   \end{tikzpicture}
\caption{The rescaled Newton polytope associated to the 1-mass bubble integral \eqref{feyn3gkz}.}
\label{fig:1-mass-bubble}
\end{figure}

The single-mass bubble diagram
\begin{align}\label{1massbub1}
I= \int \frac{\dd^d \bs{k}}{(2\pi)^d}\frac{1}{\bs{k}^{2\g_1}\left((\bs{p}-\bs{k})^2+m^2\right)^{\g_2}},
\end{align}
corresponds via \eqref{feyntogkz} to the GKZ integral \cite{de_la_Cruz_2019, Klausen:2019hrg}
\begin{align}\label{feyn3gkz}
\mathcal{I}_{\bs{\g}}= \int_{\mathbb{R}_+^2} \dd z_1 \dd z_2\frac{z_1^{\g_1-1} z_2^{\g_2-1}}{(x_1z_1+x_2z_2+x_3 z_1 z_2+x_4 z_2^2)^{\g_0}}
\end{align}
evaluated on the physical hypersurface
\[\label{1bubphyshyp}
(x_1,x_2,x_3,x_4)=(1,1,p^2+m^2,m^2).
\]
In this simple case,  the GKZ integral can of course be evaluated directly,
\begin{align}\label{1massbubbleeval}
\mathcal{I}_{\bs{\g}}&=
 \frac{\Gamma(\g_1)\Gamma \left(\g_0 -\g_1\right)
 \Gamma \left(\g_1+\g_2-\g_0\right)\Gamma \left(2\g_0-\g_1-\g_2\right)}{\Gamma(\g_0)^2} \nn\\&\qquad \times m^{2(\g_0-\g_1-\g_2)}{}_2F_1\left(\g_1,\g_1+\g_2-\g_0;\g_0; -\frac{p^2}{m^2}\right),
\end{align}
enabling the action of all creation operators to be verified.
The $\mathcal{A}$-matrix  is
\begin{align}
\mathcal{A}=\left(
\begin{array}{cccc}
 1 & 1 & 1 & 1 \\
 1 & 0 & 1 & 0 \\
 0 & 1 & 1 & 2 \\
\end{array}
\right),
\label{pol-one-mass-bubble}
\end{align}
and from its kernel, we find a single toric equation 
\[
0=(\p_1\p_4-\p_2\p_3)\mathcal{I}_{\bs{\g}}.
\]
The Euler equations can be read off from the rows of the $\mathcal{A}$-matrix, 
\[\label{eul1mass}
0=(\g_0+\t_1+\t_2+\t_3+\t_4)\mathcal{I}_{\bs{\g}},\quad 0=(\g_1+\t_1+\t_3)\mathcal{I}_{\bs{\g}},\quad 0=(\g_2+\t_2+\t_3+2\t_4)\mathcal{I}_{\bs{\g}}.
\]
The (rescaled) Newton polytope derived from the column vectors of the $\mathcal{A}$-matrix is the parellelogram shown in figure \ref{fig:1-mass-bubble}. From \eqref{hypsings}, the GKZ integral is then singular for
\[\label{sing1mass}
2\g_0-\g_1-\g_2=-k_1,\quad -\g_0+\g_1+\g_2=-k_2,\quad \g_0-\g_1=-k_3,\quad \g_1=-k_4,\quad k_i\in \mathbb{Z}^+
\]
consistent with the poles of the gamma functions in  \eqref{1massbubbleeval}.

The annihilation operators $\partial_j$ send $\bs{\gamma}\rightarrow\bs{\gamma}'$ while the creation operators $\mathcal{C}_j$ send $\bs{\gamma}'\rightarrow\bs{\gamma}$
where, for each $j$, these parameters are related by  
\begin{align}\label{MlessbCshifts}
j&=1: & \gamma_0'&=\gamma_0+1,&\gamma_1'&=\gamma_1+1, & \gamma_2'&=\gamma_2, \nn\\
j&=2: & \gamma_0'&=\gamma_0+1, & \gamma_1'&=\gamma_1,&\gamma_2'&=\gamma_2+1,\nn\\
j&=3: & \gamma_0'&=\gamma_0+1, & \gamma_1'&=\gamma_1+1, & \gamma_2'&=\gamma_2+1, \nn\\
j&=4: & \gamma_0'&=\gamma_0+1,&\gamma_1'&=\gamma_1, & \gamma_2'&=\gamma_2+2.
\end{align}
Knowing the location of the singular  hyperplanes and the shifts generated by the creation operators, 
the $b$-functions can be  constructed according to \eqref{stdbfn},
\begin{align}
&b_1=\g_1(2\g_0-\g_1-\g_2),\nn\\
&b_2=(\g_0-\g_1)(2\g_0-\g_1-\g_2),\nn\\
&b_3=\g_1(-\g_0+\g_1+\g_2),\nn\\
&b_4=(\g_0-\g_1)(-\g_0+\g_1+\g_2).
\end{align}
Their zeros serve to cancel the singularities that arise whenever the action of a creation operator shifts us from a finite to a singular integral.
For example, $\mathcal{C}_4$ shifts $k_2\rightarrow k_2+1$ and $k_3\rightarrow k_3+1$ which, according to \eqref{sing1mass}, generates a singular integral when acting on finite integrals with either $k_2=-1$ or $k_3=-1$.  These singularities, however, are cancelled by the zeros of $b_4$.

Using the DWI and the Euler equations \eqref{eul1mass},  we can now re-write
\[\label{credef}
\mathcal{C}_j\p_j \mathcal{I}_{\bs{\g}} = b_j(\bs{\g}) \mathcal{I}_{\bs{\g}}=B_j(\t) \mathcal{I}_{\bs{\g}}
\]
where 
 \begin{align}
&B_1=(\t_1+\t_3)(\t_1+\t_2)=(\t_1+\t_2+\t_3)\t_1+\t_2\t_3,\nn\\
&B_2=(\t_2+\t_4)(\t_1+\t_2)=(\t_1+\t_2+\t_4)\t_2+\t_1\t_4,\nn\\
&B_3=(\t_1+\t_3)(\t_3+\t_4)=(\t_1+\t_3+\t_4)\t_3+\t_1\t_4,\nn\\
&B_4=(\t_2+\t_4)(\t_3+\t_4)=(\t_4+\t_2+\t_3)\t_4+\t_2\t_3.
\end{align}
By inspection, every term in $B_j$ either contains an explicit factor of $\p_j$ already  through $\t_j$, or else such a factor can be introduced using the toric equations.  In $B_1$ and $B_4$, for instance, we replace $\t_2\t_3=x_2x_3\p_2\p_3\rightarrow x_2x_3\p_1\p_4$.   This enables the $B_j$ to be factored (modulo the toric equations) in the form \eqref{credef} yielding the creation operators 
\begin{align}\label{bubcre}
\mathcal{C}_1 &=x_1(1+\t_1+\t_2+\t_3)+x_2x_3\p_4,\nn\\
\mathcal{C}_2 &=x_2(1+\t_1+\t_2+\t_4)+x_1x_4\p_3,\nn\\
\mathcal{C}_3 &=x_3(1+\t_1+\t_3+\t_4)+x_1x_4\p_2,\nn\\
\mathcal{C}_4 &=x_4(1+\t_2+\t_3+\t_4)+x_2x_3\p_1.
\end{align}
These creation operators act on the full GKZ integral \eqref{feyn3gkz}.
To obtain their counterparts acting on the  Feynman integral \eqref{1massbub1}, we must project to the physical hypersurface \eqref{1bubphyshyp}.  
Given the form of the operators \eqref{bubcre}, it is useful to first simplify using the DWI to
\begin{align}\label{bubcre2}
\mathcal{C}_1 &=x_1(1-\g_0-\t_4)+x_2x_3\p_4,\nn\\
\mathcal{C}_2 &=x_2(1-\g_0-\t_3)+x_1x_4\p_3,\nn\\
\mathcal{C}_3 &=x_3(1-\g_0-\t_2)+x_1x_4\p_2,\nn\\
\mathcal{C}_4 &=x_4(1-\g_0-\t_1)+x_2x_3\p_1.
\end{align}
Next, as all factors of $x_j$ are placed to the left of all derivatives, we set
\[\label{proj}
(x_1,x_2,x_3,x_4)\rightarrow (1,1,m^2+p^2,m^2)
\]
and replace all derivatives lying in directions off this hypersurface (namely $\p_1$ and $\p_2$)  with those lying along the hypersurface.  This can be accomplished using the Euler equations \eqref{eul1mass}  projected according to \eqref{proj}, namely
\[
\p_1\rightarrow -\g_1-(m^2+p^2)\p_3,\qquad
\p_2\rightarrow -\g_2-(m^2+p^2)\p_3-2m^2 \p_4.
\]
In addition, we use the chain rule 
with $p^2=x_3-x_4$ and $m^2=x_4$ 
to replace 
\[
\p_3  = \p_{p^2},\qquad \p_4 = -\p_{p^2}+\p_{m^2}.
\]
This yields
\begin{align}
\mathcal{C}_1^{\mathrm{ph}} &=1-\g_0+p^2\p_{m^2}-\t_{p^2},\nn\\
\mathcal{C}_2^{\mathrm{ph}} &=1-\g_0-\t_{p^2},\nn\\
\mathcal{C}_3^{\mathrm{ph}} &=(1-\g_0)m^2 + (1-\g_0+\g_2)p^2+(p^2-m^2)\t_{p^2}+2p^2 \t_{m^2}\nn\\
\mathcal{C}_4^{\mathrm{ph}} &=(1-\g_0)m^2+\g_1p^2 -(p^2+m^2)\t_{p^2}.
\end{align}
From \eqref{creationeq}, the  action on the projected GKZ integral is then 
\[
\mathcal{C}_1^{\mathrm{ph}}\mathcal{I}_{\gamma_0',\gamma_1',\gamma_2'}(p^2,m^2)=-\gamma_0^{-1}b_1\mathcal{I}_{\gamma_0,\gamma_1,\gamma_2}=-\gamma_0^{-1}\g_1(2\g_0-\g_1-\g_2)\mathcal{I}_{\gamma_0,\gamma_1,\gamma_2}
\]
and similarly for the other operators.  
When acting the original Feynman integral, there is an additional  factor of $c_{\bs{\g}'}/c_{\bs{\g}}$ from \eqref{feyntogkz} we must  take into account giving
\[
\mathcal{C}_1^{\mathrm{ph}}I_{\gamma_0',\gamma_1',\gamma_2'}(p^2,m^2)=-\frac{1}{4\pi}I_{\gamma_0,\gamma_1,\gamma_2}.
\]
All these results can be checked directly using \eqref{1massbubbleeval} and the standard shift operators for the ${}_2F_1$ (see {\it e.g.,} \cite{NIST:DLMF}).

\subsubsection{Massive bubble}
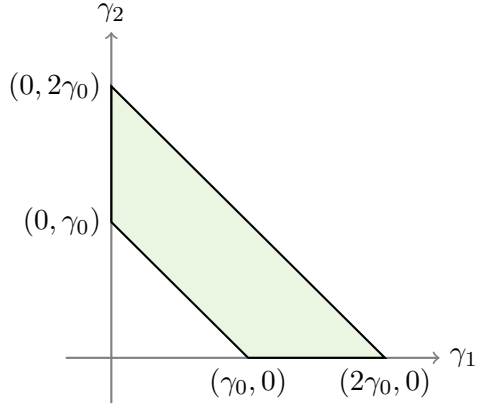
\begin{figure}[t]
\centering

\begin{tikzpicture}[scale=1.8]
\draw[->,thick,gray] (-1/3,0) -- (2.4,0); 
\draw[->,thick,gray] (0,-1/3) -- (0,2.4);

\draw[thick,black,fill=YellowGreen, fill opacity=0.2] (1,0)--(0,1) -- (0,2) -- (2,0) --cycle;

     \draw node[left] at (0,1) {($0,\gamma_0$)};   
   \draw node[below] at (1,0) {($\gamma_0,0$)};
   \draw node[left] at (0,2) {($0,2\gamma_0$)};
   \draw node[below] at (2,0) {($2\gamma_0,0$)};
    \draw node[right] at (2.4,0) {$\gamma_1$};
     \draw node[above] at (0,2.4) {$\gamma_2$};
   
   \end{tikzpicture}
\caption{The rescaled Newton polytope associated with the massive bubble GKZ integral \eqref{massbubgkz}.}
\label{fig:2-mass-bubble}
\end{figure}

Next we consider the full bubble graph with general masses $m_1$ and $m_2$,
\begin{align}\label{massbub1}
I= \int \frac{\dd^d \bs{k}}{(2\pi)^d}\frac{1}{(\bs{k}^2+m_1^2)^{\g_1}\left((\bs{p}-\bs{k})^2+m_2^2\right)^{\g_2}}.
\end{align}
The corresponding GKZ integral is
\begin{align}\label{massbubgkz}
\mathcal{I}_{\bs{\g}}=\int_{\mathbb{R}_+^2} \dd z_1 \dd z_2\frac{z_1^{\g_1-1} z_2^{\g_2-1}}{(x_1z_1+x_2z_2+x_3 z_1^2+x_4 z_2^2+x_5 z_1 z_2)^{\g_0}},
\end{align}
where $\g_0=d/2$ and the physical hypersurface is 
\[\label{physhypfullbub}
(x_1,x_2,x_3,x_4,x_5)=(1,1,m_1^2,m_2^2,m_1^2+m_2^2+p^2).
\]
\begin{figure}
\centering
\begin{tikzpicture}
\draw [thick](0,1)circle (1);
\draw[thick,->](-1.5,1)--(-1.2,1);
\draw[thick] (-1.2,1)--(-1.0,1);
\draw[thick](1.0,1)--(1.5,1);
\draw[thick] (1,1) arc (0:-180:1);
\node[text width=0.5cm, text centered ] at (0,1.7) {$1$};
\node[text width=0.5cm, text centered ] at (0,-0.3) {$2$};
\node[text width=0.5cm, text centered ] at (-1.7,1.0) {$\bs{p}$};
\end{tikzpicture} 
\caption{The massive bubble integral \eqref{massbubgkz}.}
\label{mass-bubble}
\end{figure}
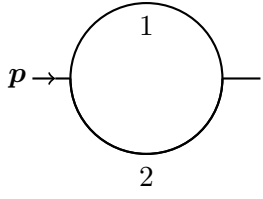
From the kernel of the $\mathcal{A}$-matrix 
\begin{align}
\mathcal{A}=\left(
\begin{array}{ccccc}
 1 & 1 & 1 & 1 & 1\\
 1 & 0 & 2 & 0 & 1\\
 0 & 1 & 0 & 2 & 1\\
\end{array}
\right)
\label{mass-bubble-mat}
\end{align}
we obtain the toric equations
\[
0=(\p_3\p_4-\p_5^2)\mathcal{I}_{\bs{\g}},\qquad 0=(\p_2\p_3-\p_1\p_5)\mathcal{I}_{\bs{\g}},\qquad 0=(\p_1\p_4-\p_2\p_5)\mathcal{I}_{\bs{\g}},
\]
while the DWI and the Euler equations can be read off from the rows:
\[
0=\Big(\g_0+\sum_{i=1}^5\t_i\Big)\mathcal{I}_{\bs{\g}},\qquad 0=(\g_1+\t_1+2\t_3+\t_5)\mathcal{I}_{\bs{\g}},\qquad 0=(\g_2+\t_2+2\t_4+\t_5)\mathcal{I}_{\bs{\g}}.
\]
The rescaled Newton polytope corresponding to this $\mathcal{A}$-matrix is the quadrilateral shown in figure \ref{fig:2-mass-bubble}. The singular hyperplanes lie parallel to and outside the facets of this polytope:
\[
\g_1=-k_1,\quad \g_2=-k_2,\quad 2\g_0-\g_1-\g_2=-k_3,\quad -\g_0+\g_1+\g_2=-k_4,\qquad k_i\in \mathbb{Z}^+.
\]

For illustration, let us now discuss the creation operator $\mathcal{C}_5$.  All others can be obtained by similar computations.  The annihilator $\partial_5$ sends $\bs{\gamma}\rightarrow\bs{\gamma}'$ where
\[
\p_5:\qquad \gamma_0'=\gamma_0+1,\qquad \gamma_1'=\gamma_1+1,\qquad \gamma_2'=\gamma_2+1,
\]
while the creation operator $\mathcal{C}_5$ acts  in the opposite direction sending $\bs{\gamma}'\rightarrow\bs{\gamma}$. 
Given this shift and the location of the singular hyperplanes, we identify  the $b$-function as
\[
b_5=\g_1\g_2(\g_0-\g_1-\g_2).
\]
Using DWI and Euler equations, this can be re-written  in terms of Euler operators as
\[
B_5=(\t_1+2\t_3+\t_5)(\t_2+2\t_4+\t_5)(\t_3+\t_4+\t_5).
\]
This expression can now be factorised as $\mathcal{C}_5\p_5$ by expanding out and using the  toric equations to replace any terms not involving $\p_5$ with equivalent terms containing this factor.
Stripping off the factor of $\p_5$ then yields  $\mathcal{C}_5$ in GKZ variables,
\begin{align}
\mathcal{C}_5&=x_5\big[2\t_3^2+2\t_4^2+8\t_3\t_4+3(\t_3+\t_4)(1+\t_5)+(1+\t_5)^2+\t_2(1+3\t_3+\t_4+\t_5)\nn\\& \qquad+\t_1(1+\t_2+\t_3+3\t_4+\t_5)
\big]
+x_2x_3\p_1(1+\t_1+2\t_3)
+x_1x_4\p_2(1+\t_2+2\t_4)\nn\\& \quad +2x_3x_4\p_5(4+\t_1+\t_2+2\t_3+2\t_4).
\end{align}
To project this operator to the physical hypersurface \eqref{physhypfullbub}, 
we first use the Euler equations to replace
\[\label{Eulrepl}
\t_1\rightarrow -\g_1-2\t_3-\t_5,\qquad
\t_2\rightarrow -\g_2-2\t_4-\t_5.
\]
The two occurrences of $\p_1$ and $\p_2$ can be dealt with similarly by writing $\p_i =(x_i)^{-1}\t_i$ for $i=1,2$ and using \eqref{Eulrepl}.  Then, setting 
$(x_1,x_2,x_3,x_4,x_5)\rightarrow (1,1,x_3,x_4,x_5)$, 
we obtain
\begin{align}
\mathcal{C}_5^{\mathrm{ph}}=&x_5\bigl[(1-\g_1)(1-\g_2)+
(1-\g_1-\g_2-\t_5)(\t_3+\t_4)
\bigr]
\nn\\&
+x_3(\g_1+2\t_3+\t_5)(\g_1-1+\t_5)
+x_4(\g_2+2\t_4+\t_5)(\g_2-1+\t_5)\nn\\&\quad
-2x_3x_4\p_5(\g_1+\g_2-4+2\t_5)
\end{align}
The remaining variables here are all physical since
\[
(x_3,x_4,x_5)=(m_1^2,m_2^2,m_1^2+m_2^2+p^2)
\]
and
\begin{align}
&\p_3=\p_{m_1^2}-\p_{p^2}, \qquad
\p_4=\p_{m_2^2}-\p_{p^2},\qquad
\p_5=\p_{p^2}.
\end{align}

\subsection{Massive triangle}\label{toricidealdisc}

Since the {\it massless} triangle integral is equivalent \cite{Bzowski:2013sza} to the 3-point contact Witten diagram studied in sections \ref{sec:3KC} and \ref{sec:3KW},  let us examine here the massive triangle integral
\begin{align}\label{masstri}
I= \int \frac{\dd^d \bs{k}}{(2\pi)^d}\frac{1}{(\bs{k}^2+m_3^2)^{\g_3}\left((\bs{k}-\bs{p}_1)^2+m_2^2\right)^{\g_2}\left((\bs{k}+\bs{p}_2)^2+m_1^2\right)^{\g_1}}.
\end{align}
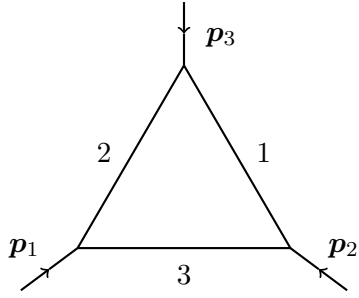
\begin{figure}[t]
\centering
\begin{tikzpicture}[scale=1.4]
   \draw[thick] (-1,0)--(1,0) -- (0,1.73); 
   \draw[,thick] (0,1.73) -- (0,1.73+0.3);
   \draw[<-,thick] (0,1.73+0.3)--   (0,1.73+0.6) ;
   \draw[thick] (0,1.73) -- (-1,0);
  \draw[thick] (-1,0)--(-1.2666,-0.2);
  \draw[<-,thick](-1.2666,-0.2)-- (-4/3-0.2,-0.4) ;
   \draw[thick] (1,0) -- (1.2666,-0.2);
   \draw[<-,thick] (1.2666,-0.2)--(4/3+0.2,-0.4) ;
   
   \node[text width=0.5cm, text centered ] at (-1.5,0) {$\bs{p}_1$};
   \node[text width=0.5cm, text centered ] at (1.5,0) {$\bs{p}_2$};
   \node[text width=0.5cm, text centered ] at (1/4+0.1,2) {$\bs{p}_3$};
   \node[text width=0.5cm, text centered ] at (-3/4,0.9) {$2$};
   \node[text width=0.5cm, text centered ] at (3/4,0.9) {$1$};
   \node[text width=0.5cm, text centered ] at (0,-1/4) {$3$};

\end{tikzpicture}
\caption{The massive triangle graph \eqref{masstri}.}
\label{fig:mass-tri}
\end{figure}
The corresponding GKZ integral according to \eqref{feyntogkz} is
\begin{align}\label{masstrigkz}
\mathcal{I}_{\bs{\g}}= \int_{\mathbb{R}_+^3} \dd z_1 \dd z_2 \dd z_3 \,z_1^{\g_1-1} z_2^{\g_2-1}z_3^{\g_3-1}\mathcal{D}^{-\g_0},
\end{align}
where 
\begin{align}\label{gkzden-masstri}
\mathcal{D}=&x_1z_1+x_2z_2+x_3z_3+x_4z_2z_3+x_5z_1z_3+x_6z_1z_2+x_7z_1^2+
x_8z_2^2+x_9z_3^2
\end{align}
with $\g_0=d/2$. The physical hypersurface is
\[\label{masstriphys}
\bs{x}= (1,1,1,p_1^2+m_2^2+m_3^2,p_2^2+m_1^2+m_3^2,
p_3^2+m_1^2+m_2^2,m_1^2,m_2^2,m_3^2)
\]
and the $\mathcal{A}$-matrix  reads
\begin{align}
\mathcal{A}=\left(
\begin{array}{ccccccccc}
 1 & 1 & 1 & 1 & 1 & 1 & 1 & 1 & 1\\
 1 & 0 & 0 & 0 & 1 & 1 & 2 & 0 & 0\\
 0 & 1 & 0 & 1 & 0 & 1 & 0 & 2 & 0\\
 0 & 0 & 1 & 1 & 1 & 0 & 0 & 0 & 2\\
\end{array}
\right).\label{masstri-mat}
\end{align}
For larger $\mathcal{A}$-matrices such as this one, it is useful to automate the calculation of creation operators using 
Gr{\"o}bner basis algorithms. 
To this end, 
in place of the five independent toric equations spanning the kernel of the $\mathcal{A}$-matrix, we will use instead the full set of 17 (non-independent) toric equations forming the toric ideal:\footnote{
These can be obtained using the Singular code \cite{singular}:
\vspace{-1mm}
\begin{alltt}
LIB "toric.lib";\\
ring r=0,(x1,x2,x3,x4,x5,x6,x7,x8,x9),dp;\\
intmat A[4][9]=1,1,1,1,1,1,1,1,1,1,0,0,0,1,1,2,0,0,0,1,0,1,0,1,0,2,0,0,0,1,1,1,0,0,0,2;\\
ideal I=toric{\_}ideal(A,"du");\\
I;
\end{alltt}
} 
\begin{align}\label{Itoric}
I_{toric}=\{&\p_2\p_6-\p_1\p_8,~\p_7\p_8-\p_6^2,~\p_1\p_6-\p_2\p_7,~\p_1\p_5-\p_3\p_7,~\p_3^2\p_9-\p_7,~\p_1\p_4-\p_3\p_6,\nn\\&~\p_4^2\p_9-\p_8,~\p_5\p_6-\p_4\p_7,~\p_4\p_6-\p_5\p_8,~\p_4\p_5\p_9-\p_6,~\p_3^2\p_8\p_9-\p_2^2,~\p_3^2\p_7\p_9-\p_1^2,\nn\\&~\p_3^2\p_6\p_9-\p_1\p_2,~\p_3\p_5\p_9-\p_1,~\p_3\p_4\p_9-\p_2,~\p_2\p_5-\p_3\p_6,~\p_2\p_4-\p_3\p_8\}.
\end{align}
Each entry here corresponds to a toric equation, for example the first is $0=(\p_2\p_6-\p_1\p_8)\mathcal{I}_{\bs{\g}}$ and similarly for the rest.
Since all the partial derivatives commute, these equations can be treated as a system of polynomial equations by mapping $\p_i$ to an ordinary commutative variable $y_i$.  As we will show below, this enables the factorisation step to be handled via ordinary commutative Gr{\"o}bner basis methods.  (For alternative constructions of creation operators  using {\it non-commutative} Gr{\"o}bner bases over the Weyl algebra, see  \cite{saito_sturmfels_takayama_1999}.)

\begin{figure}[t]
\centering
\begin{tikzpicture}[scale=2.0]
   \draw[thick,black] (0,0,2)--(2,0,0) -- (0,2,0)  ; 
   \draw[thick,black] (0,2,0) -- (0,0,2) ;
    \draw[style=dashed, color=black] (0,1,0) --(0,2,0);
    \draw[style=dashed, color=black] (0,1,0) --(0,2,0);
    \draw[style=dashed, color=black] (1,0,0) --(0,1,0)-- (0,0,1)-- (1,0,0);
     \draw[style=dashed, color=black] (1,0,0) --(2,0,0);
     \draw[style=dashed, color=black] (0,0,1) --(0,0,2);
    \draw node[right, above] at (1,0,0) {\{1\}};      
     \draw[color=black] node at (1,0,0) {$\bullet$}; 
   \draw node[right] at (0,1,0) {\{2\}};
   \draw[color=black] node at (0,1,0) {$\bullet$}; 
   \draw node[above] at (0,0,1) {\{3\}};
   \draw[color=black] node at (0,0,1) {$\bullet$}; 
   \draw node[left] at (0,1,1) {\{4\}};
   \draw[color=black] node at (0,1,1) {$\bullet$}; 
   \draw node[below] at (1,0,1) {\{5\}};
   \draw[color=black] node at (1,0,1) {$\bullet$}; 
    \draw node[right] at (1,1,0) {\{6\}};
    \draw[color=black] node at (1,1,0) {$\bullet$}; 
    \draw node[right] at (2,0,0) {\{7\}};
    \draw[color=black] node at (2,0,0) {$\bullet$}; 
    \draw node[above] at (0,2,0) {\{8\}};
    \draw[color=black] node at (0,2,0) {$\bullet$}; 
    \draw node[left] at (0,0,2) {\{9\}};
    \draw[color=black] node at (0,0,2) {$\bullet$}; 

    \draw[thick,black,fill=YellowGreen, fill opacity=0.2] (0,0,2)--(2,0,0) -- (0,2,0) --cycle;
    
   \end{tikzpicture}
   \caption{The Newton polytope associated to the denominator of the massive triangle integral \eqref{gkzden-masstri}. The label \{i\} denotes the vector defined by the $i$th column of the $\mathcal{A}$-matrix.}
\label{fig:masstri}
\end{figure}
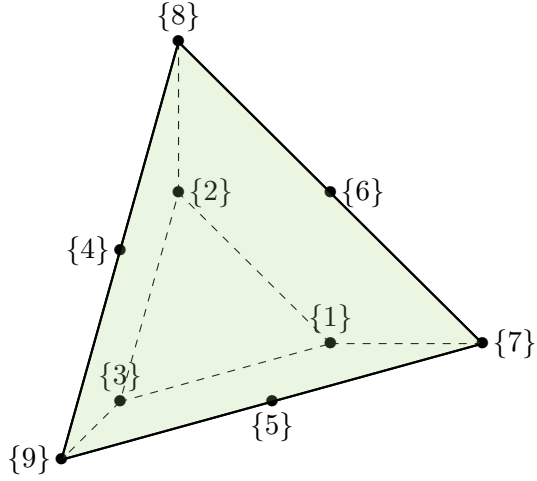

The DWI and Euler equations for this $\mathcal{A}$-matrix are
\begin{align}
&0=\Big(\g_0+\sum_{i=1}^9\t_i\Big)\mathcal{I}_{\bs{\g}}, \nn\\ &0=(\g_1+\t_1+\t_5+\t_6+2\t_7)\mathcal{I}_{\bs{\g}},\nn\\&0=(\g_2+ \t_2+\t_4+\t_6+2\t_8)\mathcal{I}_{\bs{\g}}, \nn\\
 &0=(\g_3+\t_3+\t_4+\t_5+2\t_9)\mathcal{I}_{\bs{\g}}
\end{align}
and the corresponding Newton polytope  is depicted in figure \ref{fig:masstri}.
From its facets, we obtain the  singularity conditions
\begin{align}
\begin{array}{llll}
\{23489\}:& \g_1=-k_1, &\{1379\}:& \g_2=-k_2,\\
\{1278\}:& \g_3=-k_3, &\{123\}:& \g_0-\g_1-\g_2-\g_3=+k_4,\\
\{4789\}:& \g_1+\g_2+\g_3-2\g_0=+k_5, \,\,\,\,\,& & 
\end{array}
\end{align}
where all $k_i\in \mathbb{Z}^+$.  For the facet $\{123\}$, for example, we have the outward-pointing normal $\bs{n}=(-1,-1,-1)$ which leads via \eqref{Jshift} to the spacing  of singular hyperplanes $\delta^{(J)}=1$.

Let us now compute the creation operator $\mathcal{C}_4$ which acts on the GKZ integral to shift $\bs{\g}'\rightarrow \bs{\g}$, where
\[
\g_0'=\g_0+1,\qquad \g_1'=\g_1, \qquad \g_2'=\g_2+1,\qquad \g_3'=\g_3+1.
\]
From these parameter shifts and the location of the singular hyperplanes,  the corresponding $b$-function is
\[
b_4=\g_2\g_3(\g_0-\g_1-\g_2-\g_3).
\]
Using the DWI and Euler equations, this can be re-expressed as 
\[
B_4=\Big(\sum_{i=4}^9\t_i\Big)(\t_2+\t_4+\t_6+2\t_8)(\t_3+\t_4+\t_5+2\t_9).
\]
Our goal is now to  factorise $B_4$ as $\mathcal{C}_4\p_4$ using the toric equations.  
To achieve this in an automated fashion, we decompose $B_4$ over the Gr{\"o}bner basis formed from the toric ideal \eqref{Itoric} and $\p_4$.
Treating the partial derivatives as ordinary commutative variables and computing 
 this Gr{\"o}bner basis, we obtain 
\[
\bs{g}=\{\p_4,~\p_2,~\p_8,~\p_3^2\p_7\p_9-\p_1^2,~\p_6,~\p_1\p_5-\p_3\p_7,~\p_3\p_5\p_9-\p_1,~\p_5^2\p_9-\p_7\}.
\]
Expanding out $B_4$ and rewriting all terms in the form \eqref{Bterm} so that all partial derivatives $\p_i$ lie to the right of all $x_i$, we can now decompose each term of $B_4$ in this Gr{\"o}bner basis.
This yields 
\[
B_4=\bs{Q}\cdot \bs{g}=Q_1\p_4+\sum_{i=2}^{8}Q_ig_i.
\]
where the coefficients $Q_i$ are polynomials in the $x_j$ and $\t_j$ (with $j=1,..,9$) which can be computed automatically.\footnote{
In Mathematica, for example, after writing $B_4$  in the form \eqref{Bterm} with all derivatives to the right, we replace all  $\p_i$  (both in $B_4$ and in the toric ideal) by commutative variables $y[i]$.  The code
\vspace{-1mm}
\begin{alltt}
v = \lcb y[5], y[6], y[7], y[8], y[9], y[1], y[2], y[3], y[4]\rcb;\\ 
toric = \lcb y[2] y[6] - y[1] y[8], y[6]$^2$ - y[7] y[8], y[1] y[6] - y[2] y[7], ...\rcb;\\
g = GroebnerBasis[Append[toric, y[4]], v]\\
Q = PolynomialReduce[B4, g, v][[1]]
\end{alltt}
\vspace{-1mm}
then evaluates the $Q_i$ coefficients with all derivatives $y[i]$ placed to the right.  These can then be re-expressed in terms of Euler operators by rewriting $y_i^n = x_i^{-n}\t_i(\t_i-1)\ldots(\t_i-n+1)$ leading to \eqref{Qcotri}.
}
To extract the required overall factor of $\p_4$, we now re-express those $g_i$ ($i=2,..,18$) that are not already complete toric equations (and hence zero) in terms of $\p_4$.  For example, using the third from last toric equation in \eqref{Itoric}, we can replace
\[
g_2=\p_2~\rightarrow~\p_3\p_4\p_9.
\]
In this fashion, we can replace the basis $\bs{g}$ with the equivalent basis  (modulo the toric equations)
\[
\bs{\tilde{g}}=\{\p_4,~\p_3\p_4\p_9,~\p_4^2\p_9,~0,~\p_4\p_5\p_9,~0,~0\}.
\]
All surviving terms then have an explicit factor of $\p_4$ which can be removed to obtain the creation operator 
\begin{align}
\mathcal{C}_4&=Q_1+Q_2\p_3\p_9+Q_3\p_4\p_9+Q_5\p_5\p_9,
\end{align}
where the coefficients are
\begin{align}\label{Qcotri}
Q_1&=x_4 \big[ (1+\t_4)(\t_5+\t_7+\t_9)+\t_5(\t_6+\t_7+3\t_8+\t_9)+\t_6(\t_7+\t_8+3\t_9)\nn\\&\qquad\quad+(\t_8+\t_9)(1+2\t_7+\t_8+\t_9)+4\t_8\t_9+(1+\t_4+\t_5+\t_6+\t_8+\t_9)^2\nn\\&\qquad\quad +\t_3(1+\t_4+\t_5+2\t_6+\t_7+3\t_8+\t_9)\nn\\&\qquad\quad +\t_2(1+\t_3+\t_4+2\t_5+\t_6+\t_7+\t_8+3\t_9)
\big],\nn\\
Q_2&=x_2(\t_3+\t_5+2\t_9)(\t_5+\t_6+\t_7+\t_8+\t_9),\nn\\
Q_3&=x_8(\t_3+\t_5+2\t_9)\bigl(2\t_5+3\t_6+2(1+\t_7+\t_8+\t_9)\bigr),\nn\\ 
Q_5&=x_6(\t_3+\t_5+2\t_9)(1+\t_5+\t_6+\t_7+\t_9).
\end{align}
Finally, to project  to the physical hypersurface, we use the Euler equations  to eliminate the unphysical variables $\t_1,\t_2, \t_3$ and set $x_1=x_2=x_3=1$. This yields the physical creation operator
\begin{align}
\mathcal{C}_4^{\mathrm{ph}}=&x_4\bigl[(1-\g_2)(1-\g_3)+(\t_5+\t_6+\t_7+\t_8+\t_9)(1-\g_2-\g_3-\t_4)\bigl]\nn\\
&+(\g_3+\t_4)\bigl[\p_9(\t_5+\t_6+\t_7+\t_8+\t_9)(\g_3+\g_4+\g_5+2\t_9)\nn\\&\quad -x_6\p_5(1+\t_5+\t_6+\t_7+3\t_8+\t_9)-2x_8\p_4\p_9(1+\t_5+\t_7+\t_8+\t_9)\bigr]
\end{align}
where the $x_i$ are as given in \eqref{masstriphys} and
\begin{align}
\p_4&=\p_{p_1^2}, & \p_5&=\p_{p_2^2}, & \p_6&=\p_{p_3^2},\nn\\ \p_7&=\p_{m_1^2}-\p_{p_2^2}-\p_{p_3^2}, & \p_8&=\p_{m_2^2}-\p_{p_1^2}-\p_{p_3^2}, & \p_9&=\p_{m_3^2}-\p_{p_1^2}-\p_{p_2^2}.
\end{align}
The automated approach outlined here can be  applied similarly to other examples.

\subsection{Massless on-shell box}\label{sec:hulling}

Next we  consider the massless  box integral
\begin{align}\label{box1}
I &= \int \frac{\mathrm{d}^d\bs{q}}{(2\pi)^d}\frac{1}{|\bs{q}|^{2\g_1}|\bs{q}+\bs{P}_1|^{2\g_2}|\bs{q}+\bs{P}_2|^{2\g_3}|\bs{q}+\bs{P}_3|^{2\g_4}},
\end{align}
where 
\[\bs{P}_k=\sum_{j=1}^k\bs{p}_j,\quad \mathrm{for}\quad k=1,2,3, \qquad\sum_{i=1}^4 \bs{p}_i=0.
\] 
\begin{figure}[t]
\centering
\begin{tikzpicture}
\draw[dashed,thick](-1,-1)--(-1.35,-1.35);
\draw[<-,thick,dashed](-1.35,-1.35)--(-1.7,-1.7);
\draw[dashed,thick](1,-1)--(1.35,-1.35);
\draw[<-,thick,dashed](1.35,-1.35)--(1.7,-1.7);
\draw[dashed,thick](1,1)--(1.35,1.35);
\draw[<-,thick,dashed](1.35,1.35)--(1.7,1.7);
\draw[dashed,thick](-1,1)--(-1.35,1.35);
\draw[<-,thick,dashed](-1.35,1.35)--(-1.7,1.7);
\draw[dashed, thick](-1,-1)--(-1.0,1.0)--(1.0,1.0)--(1.0,-1.0)--cycle;
\node[text width=0.5cm, text centered ] at (-1.35,0) {$1$};
\node[text width=0.5cm, text centered ] at (0,-1.3) {$2$};
\node[text width=0.5cm, text centered ] at (1.3, 0) {$3$};
\node[text width=0.5cm, text centered ] at (0,1.3) {$4$};
\node[text width=0.5cm, text centered ] at (-2,2){$\bs{p}_1$};
\node[text width=0.5cm, text centered ] at (2,2){$\bs{p}_2$};
\node[text width=0.5cm, text centered ] at (2,-2){$\bs{p}_3$};
\node[text width=0.5cm, text centered ] at (-2,-2){$\bs{p}_4$};
\end{tikzpicture} 
\caption{The on-shell massless box integral \eqref{box1}.}
\label{os-massless-box}
\end{figure}
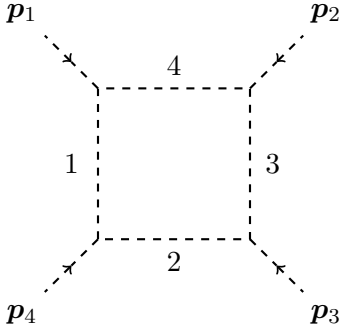
For simplicity, we will restrict to the on-shell case\footnote{ Creation operators for the off-shell box are also computable but the results are rather long.} 
where all $p_i^2=0$ for $i=1,..,4$.   
According to \eqref{feyntogkz}, the corresponding GKZ integral  is 
\begin{align}
\mathcal{I}_{\bs{\g}}=\prod_{i=1}^4\left(\int_0^\infty \dd z_i \:z_i^{\g_i-1}\right)  
(x_1z_1+x_2z_2+x_3z_3+x_4z_4 +x_5 z_1 z_3 + x_6 z_2 z_4)^{-\g_0} ,
 \label{integral-box}
\end{align}
where the physical hypersurface 
\[\label{boxphyshyp}
\bs{x}=(1,1,1,1,s^2,t^2)
\] 
with $s^2=(\bs{p}_1+\bs{p}_2)^2$ and $t^2=(\bs{p}_2+\bs{p}_3)^2$  the Mandelstam invariants. The integral can be evaluated  as a linear combination of the hypergeometric function $_3F_2$ \cite{Tarasov:2017jen}.

The $\mathcal{A}$-matrix 
\begin{align}\label{Amatboxon}
\mathcal{A}=\left(
\begin{array}{cccccc}
 1 & 1 & 1 & 1 & 1 & 1 \\
 1 & 0 & 0 & 0 & 1 & 0 \\
 0 & 1 & 0 & 0 & 0 & 1 \\
 0 & 0 & 1 & 0 & 1 & 0 \\
 0 & 0 & 0 & 1 & 0 & 1 \\
\end{array}
\right)
\end{align}
yields a single toric equation
\[\label{boxtoric}
0=(\p_1\p_3\p_6-\p_2\p_4\p_5)\mathcal{I}_{\bs{\g}},
\]
along with the DWI and Euler equations
\begin{align}\label{eulboxon}
&0=\Big(\g_0+\sum_{i=1}^6\t_i\Big)\mathcal{I}_{\bs{\g}},\quad 0=(\g_1+\t_1+\t_5)\mathcal{I}_{\bs{\g}},\quad 0=(\g_2+\t_2+\t_6)\mathcal{I}_{\bs{\g}},\nn\\& 0=(\g_3+\t_3+\t_5)\mathcal{I}_{\bs{\g}},\quad 0=(\g_4+\t_4+\t_6)\mathcal{I}_{\bs{\g}}.
\end{align}
To determine the singularities of the integral, we need to find the equations of the facets of the rescaled Newton polytope corresponding to the GKZ denominator in \eqref{integral-box}. 
As this polytope lives in four dimensions, 
it is convenient to use an automated hulling algorithm.  
Using the Mathematica package \cite{Nhull}, for example, we can enter the vertices $\bs{a}_j$ of the non-rescaled Newton polytope (where $\bs{a}_j$ is the $j$th column of the $\mathcal{A}$-matrix without the top row)  as row vectors:
\begin{verbatim} 
   verts = {{1,0,0,0},{0,1,0,0},{0,0,1,0},{0,0,0,1},{1,0,1,0},{0,1,0,1}};
\end{verbatim}
The command {\tt CHNQuickHull[verts]} then returns a list of the vertex vectors that make up the convex hull (labelled according to  the numbering specified in the input), followed by a list of the facets.  The latter are specified by the vertex vectors they contain.
Thus, in this example, we obtain
\begin{verbatim}
      {{1,2,3,4,5,6}, {{1,2,3,5}, {1,2,4,3}, {1,2,5,6}, {1,2,6,4},
            {1,3,4,5}, {1,4,6,5}, {2,3,5,6}, {2,3,6,4}, {3,4,5,6}}}
\end{verbatim}
where the first set
indicates that all six vertices belong to the convex hull, while the remainder  ($\{1,2,3,5\}$, $\{1,2,4,3\}$, {\it etc}.) list  the facets.  Here, each $\{ijkl\}$ is a co-dimension one facet containing the points $(\bs{a}_i,\bs{a}_j,\bs{a}_k,\bs{a}_l)$.

The equations for the facets of the {\it rescaled} Newton polytope with vertices $\g_0\bs{a}_i$ can now be computed through a determinant such as \eqref{fullgammaAdet}.  
For the facet $\{1,2,3,4\}$, for example, we have
\[
0=\bs{\g}\cdot\bs{N} = \det\,(\,\bs{\g}\,|\,\bs{\mathcal{A}}_1\,|\,\bs{\mathcal{A}}_2\,|\,\bs{\mathcal{A}}_3\,|\,\bs{\mathcal{A}}_4\,)=\g_0-\g_1-\g_2-\g_3-\g_4 
\]
and hence $\bs{N} = (n_0,\bs{n}) = (1,-1,-1,-1,-1)$.
The fact that $\bs{n}$ is outwards-pointing can be verified by showing $d_i^{(J)}=-\bs{\mathcal{A}}_i\cdot\bs{N}>0$ for any vertex $i=5,6$ not lying in the facet. 
 The spacing of the set of singular hyperplanes parallel to this facet is then $\delta^{(J)}=1$ using \eqref{Jshift} and \eqref{ddef},  with the singular hyperplanes themselves then following from \eqref{hypsings}.  
Automating this procedure and applying it to the other facets, the singularities for the GKZ integral \eqref{integral-box} are 
\begin{align}
&\g_i = -k_i, \quad i=1,2,3,4, & \g_1+\g_2+\g_3+\g_4-\g_0=-k_5,\nn \\
& \g_1+\g_2-\g_0=+k_6,   & \g_2+\g_3-\g_0=+k_7, \label{sing-box-un} \\ & \g_3+\g_4-\g_0=+k_8, & \g_4+\g_1-\g_0=+k_9,\nn
\end{align}
where all $k_i\in \mathbb{Z}^+$. 

We are now in a position to  compute the creation operators.  Let us choose $\mathcal{C}_1$, which acts on the GKZ integral to shift $\bs{\g}'\rightarrow \bs{\g}$ where
\[
\g_0'=\g_0+1,\qquad\qquad \g_1'=\g_1+1.
\]
From the singularities \eqref{sing-box-un}, the corresponding $b$-function is
\[
b_1=-\g_1(\g_2+\g_3-\g_0)(\g_3+\g_4-\g_0),
\]
which in terms of the Euler operators reads
\begin{align}
B_1&=(\t_1+\t_2)(\t_1+\t_4)(\t_1+\t_5)\nn\\& =
\big[ (\t_1 + \t_2) (\t_1 +\t_4) +  (\t_1 + \t_2 + \t_4) \t_5\big]\t_1+\t_2\t_4\t_5
\end{align}
Applying the toric equation \eqref{boxtoric} to the final term  now enables us to factorise $B_1$ as $\mathcal{C}_1\p_1$ giving the creation operator
\[
\mathcal{C}_1=x_1\left[(1+\t_1+\t_2)(1+\t_1+\t_4)+(1+\t_1+\t_2+\t_4)\t_5\right]+x_2x_4x_5\p_3\p_6,
\]
where we shifted the factor of $x_1$ to the left sending each $\t_1\rightarrow \t_1+1$.
Finally, to obtain the creation operator acting on the physical variables,  we use the Euler equations \eqref{eulboxon} to replace $\t_i$ for $i=1,2,3,4$ with $\t_5$ and $\t_6$ and project to  \eqref{boxphyshyp}.  This yields the operator 
\[
\mathcal{C}_1^{\mathrm{ph}}=(1-\g_1-\g_2-\t_{t^2})(1-\g_1-\g_4-\t_{t^2})+(\g_1-1)\t_{s^2}- s^2(\g_3+\t_{s^2})\p_{t^2}.
\]
Using the automated determination of the convex hull in this example,  and the  factorisation of the $b$-function via Gr{\"o}bner basis methods in the previous example,  the calculation of any creation operator can be fully automated.

\section{Discussion}
\label{sec:Discussion}

The GKZ formalism enables the construction of highly non-trivial shift operators known as creation operators.  
As we have shown, the calculation is very systematic.  First, a Feynman or Witten diagram is represented as a GKZ or $\mathcal{A}$-hypergeometric function. 
Second, the $b$-function
is identified by examining the parameter shifts produced by the creation operator in conjunction with the location of all singular hyperplanes of the integral.  
The $b$-function is a function of parameters 
that multiplies the shifted integral, and its
zeros serve physically to cancel the singularities that would otherwise arise when the creation operator maps a finite to a singular integral.
Such singularities cannot arise under the action of a finite differential operator on a finite integral.
Next, using the Euler equations and DWI, the $b$-function is expressed as a function of Euler operators 
and factorised into a product of a creation and an annihilation operator with the aid of the toric equations.
The creation operator thus extracted is then re-expressed in terms of physical variables ({\it i.e.,} the momenta and masses) 
using once again the Euler equations and DWI.

This algorithm has a number of interesting features.  
First, the parametric singularities of the integral all lie on hyperplanes parallel to the facets of the Newton polytope associated with the integral's denominator.   We derived a precise formula for the spacing of these hyperplanes in \eqref{hypsings}.
  The $b$-function therefore has a geometrical character, as originally shown by Saito in \cite{Saito_param_shift}. 
Second, the algorithm makes heavy use of 
the higher-dimensional GKZ space obtained by promoting the coefficient of every term in the Lee-Pomeransky denominator to an independent variable.    
This systematises the set of PDEs obeyed by the integral into  two distinct classes: the Euler equations and DWI, and the toric equations.  
Using the former, we can uplift to GKZ space by exchanging all dependence on the parameters $\bs{\g}$ for dependence on the additional unphysical coordinates.  Conversely, we can project back to physical variables by using the Euler equations and DWI to exchange derivatives with respect to the unphysical variables for derivatives with respect to the physical variables and dependence on the parameters $\bs{\g}$.

This last step is however a potential weakness of the algorithm.  To project a creation operator from GKZ space back to the physical hypersurface, the total number of Euler equations (including the DWI) must be equal to, or greater than, the number of unphysical coordinates.  
This enables every derivative in unphysical variables to be  replaced by an equivalent expression in purely physical variables.
For higher-loop Feynman integrals, however, the number of terms in the Lee-Pomeransky denominator, and hence the dimension of the full GKZ space, typically grows more rapidly than the number of propagators and hence Euler equations.
Thus, while a full set of creation operators can be constructed in GKZ space, in general we lack sufficient Euler equations to project back to the physical hypersurface.  For this reason, we have focused initially on 1-loop Feynman integrals. 

One possible workaround for this issue is to construct an alternative projectible GKZ system based on some representation other than the Lee-Pomeransky.
For example, for higher-loop massive sunset (aka melon or banana) diagrams, one can construct a GKZ representation based on their position-space formulation as a product of Bessel functions \cite{Klemm_2020,Zhang:2023fil}.  In this manner, these diagrams can be related to (analytic continuations of) the momentum-space contact Witten diagrams for which we have already constructed creation operators.   For more general classes of diagrams,  projectible GKZ representations can also be obtained from   Mellin-Barnes representations as shown in \cite{Feng_2020, Feng:2022ude, Zhang:2023fil}.  A further possibility might be to develop a GKZ representation starting from the Baikov representation.

Nevertheless, using the simplest formulation based on the Lee-Pomeransky representation, we have already identified 
a number of useful new shift operators.  In particular, for computations in AdS/CFT, we have found:
\begin{itemize}
\item The creation operators \eqref{3KC1} and \eqref{4KC1}, along with their permutations and shadow conjugates, connecting 3- and 4-point momentum-space contact Witten diagrams of different operator and spacetime dimensions.  These new operators are the inverse of the simple annihilators first identified in \cite{Bzowski:2013sza, Bzowski:2015yxv}.  
The corresponding creation operators for position-space contact diagrams are given in appendix \ref{sec:Dfn}.
\item The creation  operators \eqref{W3ptphys} and  \eqref{4ptWv2}, plus their permutations and shadow conjugates, relating 3- and 4-point  momentum-space contact Witten diagrams of different operator dimensions but the same spacetime dimension.  While the 3-point operator \eqref{W3ptphys}  is known  \cite{Karateev:2017jgd,Baumann:2019oyu}, the 4-point operator \eqref{4ptWv2}  is new.

\item  Using \eqref{4ptWv2}, we obtained a further new operator \eqref{OmegaOp} connecting exchange Witten diagrams of different external operator dimensions but the same spacetime dimension. 
Unlike any previous construction, this operator connects  exchange diagrams with purely {\it non-derivative vertices}.  Working in dimensional regularisation where necessary to avoid divergences \cite{Bzowski:2022rlz}, it also applies for arbitrary operator  dimensions.  
\end{itemize}
There is ample scope for building on this first application of creation operators to Witten diagrams.   In particular, our results for exchange Witten diagrams were obtained from the analysis of  contact diagrams.  It would be preferable to develop a GKZ representation for the exchange diagram directly, both in momentum and in position space.  This in turn may enable more compact expressions to be found, as well as operators acting to shift the dimension of the exchanged leg.  At present, operators achieving this latter goal are known only for a very restricted set of  external scaling dimensions \cite{Baumann:2019oyu, Bzowski:2022rlz}.  
The application of the creation operator formalism to 
cosmological correlators 
in de Sitter spacetime is also worthy of exploration.  
We hope to return to these matters 
in future.

\paragraph{Acknowledgments:}
FC thanks
the School of Maths, Statistics \& Physics for support. 
PM is supported in part by the UKRI consolidated grant ST/T000708/1.

\appendix

\section{GKZ representation of Feynman integrals}
\label{LPrep}

In this appendix we relate a generic $L$-loop Feynman integral of the form \eqref{feynprop} to the corresponding GKZ integral \eqref{feyngkz}.  Related discussions can be found in, {\it e.g.,} \cite{de_la_Cruz_2019, Weinzierl:2022eaz}. 

After exponentiating the propagators and integrating out the loop momenta, \eqref{feynprop} has  the
Schwinger parametrisation 
\[
I = (4\pi)^{-\g_0 L}\Big(\prod_{i=1}^N \frac{1}{\Gamma(\g_i)}\int_0^\infty \D t_i \, t_i^{\g_i-1}\Big)\, \mathcal{U}[t]^{-\g_0}\exp\Big(-\frac{\mathcal{F}[t]}{\mathcal{U}[t]}\Big),\qquad \g_0=\frac{d}{2},
\]
where $\mathcal{U}[t]$ and $\mathcal{F}[t]$ are the first and second Symanzik polynomials respectively, which are homogeneous  of weights $L$ and $L+1$ in the Schwinger parameters $t_i$.   
The prefactor of $(4\pi)^{-\g_0L}$ is simply  that in \eqref{feynprop} multiplied by $L$ factors of $\pi^{d/2}$ from integrating out the loop momenta.
The corresponding Feynman representation is obtained by reparametrising
\[
t_i = \sigma y_i, \qquad y_t = \sum_{i=1}^N y_i = 1
\]
and integrating out the variable $\sigma$. 
Using the Jacobian\footnote{See, {\it e.g.,} Appendix B of \cite{Bzowski:2020kfw}.}
\[
\prod_{i=1}^N \D t_i = \sigma^{N-1}\D\sigma \prod_{i=1}^N \D y_i \,\delta(1-y_t),
\]
as well as the homogeneity of the Symanzik polynomials, we find
\begin{align}
I &= (4\pi)^{-\g_0 L}\Big(\prod_{i=1}^N\frac{1}{\Gamma(\g_i)} \int_0^1 \D y_i \, y_i^{\g_i-1}\Big)\delta(1-y_t)\mathcal{U}[y]^{-\g_0}\int_0^\infty \D \sigma\, \sigma^{\g_t-\g_0L-1}\exp\Big(-\sigma\frac{\mathcal{F}[y]}{\mathcal{U}[y]}\Big)\nn\\
&= (4\pi)^{-\g_0 L}\Gamma(\g_t-\g_0L)\Big(\prod_{i=1}^N\frac{1}{\Gamma(\g_i)} \int_0^1 \D y_i \, y_i^{\g_i-1}\Big)\delta(1-y_t)\mathcal{U}[y]^{\g_t-\g_0(L+1)}\mathcal{F}[y]^{-\g_t+\g_0L}.
\end{align}
In special cases where $\g_t-\g_0(L+1)$ vanishes ({\it e.g.,} $d=2$ multi-loop sunsets with standard propagators) one can use the $\mathcal{F}$ polynomial alone to construct a GKZ representation \cite{Vanhove:2018mto}.  More generally, one can use the 
Lee-Pomeransky representation \cite{Lee:2013hzt, de_la_Cruz_2019} obtained by combining the two Symanzik polynomial factors using the Euler beta identity
\[
\mathcal{U}[y]^{-a}\mathcal{F}[y]^{a-b} = \frac{\Gamma(b)}{\Gamma(a)\Gamma(b-a)}\int_0^\infty\D s\,s^{a-1}(\mathcal{F}[y]+s\,\mathcal{U}[y])^{-b}
\]
with $a=\g_0(L+1)-\g_t$ and $b=\g_0$ giving
\begin{align}
I&=c_{\bs{\g}}\,\Big(\prod_{i=1}^N\int_0^1 \D y_i \, y_i^{\g_i-1}\Big)\delta(1-y_t)\int_0^\infty \D s\, s^{\g_0(L+1)-\g_t-1}(\mathcal{F}[y]+s\,\mathcal{U}[y])^{-\g_0}
\end{align}
where
\[
c_{\bs{\g}}=\frac{(4\pi)^{-L\g_0}\Gamma\left(\g_0\right)}{\Gamma\left((L+1)\g_0-\g_t\right)\prod_{i=1}^N\Gamma(\g_i)}.
\]
Setting $y_i = s z_i$ and using once again the homogeneity of the Symanzik polynomials, we can eliminate the $s$ integral since
\[
\int_0^\infty\frac{\D s}{s}\,\delta(1-s z_t) = \int_0^\infty \frac{\D s}{sz_t}\delta(z_t^{-1}-s) = 1
\]
after which 
\begin{align}
I &= c_{\bs{\g}}\, \Big(\prod_{i=1}^N\int_0^\infty \D z_i \, z_i^{\g_i-1}\Big)(\mathcal{F}[z]+\mathcal{U}[z])^{-\g_0}.
\end{align}
Finally, this Lee-Pomeransky representation is upgraded to the GKZ representation by replacing the coefficient of every term in the denominator $\mathcal{F}[z]+\mathcal{U}[z]$ with an independent variable $x_k$.  
For the massless triangle integral, for example, 
\[
\mathcal{U}[z] = z_1+z_2+z_3,\qquad \mathcal{F}[z]=p_1^2 z_2z_3+p_2^2 z_3z_1+p_3^2 z_1 z_2
\]
and so we replace the Lee-Pomeransky denominator
\[
\mathcal{G} = \mathcal{F}[z]+\mathcal{U}[z] = p_1^2 z_2z_3+p_2^2 z_3z_1+p_3^2 z_1 z_2+z_1+z_2+z_3
\]
with the GKZ denominator
\[
\mathcal{D} = x_1 z_2z_3+x_2 z_3z_1+x_3 z_1 z_2+x_4 z_1+ x_5 z_2+x_6 z_3.
\]
The GKZ integral 
\[
\mathcal{I}_{\bs{\g}} = \Big(\prod_{i=1}^N\int_0^\infty \D z_i \, z_i^{\g_i-1}\Big)\mathcal{D}^{-\g_0}
\]
is then related to the massless triangle integral by
\[
I = c_{\bs{\g}}\mathcal{I}_{\bs{\g}} 
\]
evaluated on the physical hypersurface
\[
x_i = p_i^2, \qquad x_{i+3}=1, \qquad i=1,2,3.
\]

\section{Creation operators for the position-space contact Witten diagram}
\label{sec:Dfn}

In position space, the $n$-point AdS contact Witten diagram 
\[
I_n = \int_0^\infty\frac{\D z}{z^{d+1}}\int\D^d\bs{x}_0\prod_{i=1}^n C_{\Delta_i}\Big(\frac{z}{z^2+x_{i0}^2}\Big)^{\Delta_i},\qquad
C_{\Delta_i} = \frac{\Gamma(\Delta_i)}{\pi^{d/2}\Gamma(\Delta_i-\frac{d}{2})},
\]
has the parametric representation\footnote{See, {\it e.g.,} equations (5.46)--(5.51) and (B.1)--(B.11) of \cite{Bzowski:2020kfw}.}
\begin{align}\label{Sym_res1}
I_n &= C_n\,\Big(\prod_{i=1}^n \int_0^\infty \D z_i\, z_i^{\Delta_i-1} \Big)\delta\big(1-\sum_{i=1}^n \kappa_i z_i\big) \Big(\sum_{i<j} z_i z_j x_{ij}^2\Big)^{-\Delta_t/2}.
\end{align}
where
\[
C_n = \frac{\pi^{d/2}}{2}\Gamma\Big(\frac{\Delta_t}{2}\Big)\Gamma\Big(\frac{\Delta_t-d}{2}\Big)\prod_{i=1}^n\frac{C_{\Delta_i}}{\Gamma(\Delta_i)}, \qquad \Delta_t = \sum_{i=1}^n\Delta_i,\qquad \bs{x}_{ij}=\bs{x}_i-\bs{x}_j.
\]
The parameters $\kappa_i\ge 0$ can be chosen arbitrarily provided they are not all zero.  For the 4-point function specifically, choosing $\kappa_i=\delta_{i4}$ and eliminating $y_4$ using the delta function leads to the GKZ representation
\begin{align}
I_4&=C_4 \mathcal{I}_{\bs{\g}},\qquad \mathcal{I}_{\bs{\g}}= \Big(\prod_{i=1}^3 \int_0^\infty \D z_i \, z_i^{\g_i-1}\Big)\mathcal{D}^{-\g_0}
 \label{Dfunction}
\end{align}
where
\[
\mathcal{D} =  x_1 z_2 z_3 + x_2 z_1 z_3+x_3 z_1 z_2+x_4 z_1 + x_5 z_2 + x_6 z_3,
\]
the parameters
\[
\g_1=\Delta_1,\qquad \g_2 =\Delta_2,\qquad
\g_3=\Delta_3, \qquad \g_0 
=\frac{1}{2}(\Delta_1+\Delta_2+\Delta_3+\Delta_4),
\]
and the GKZ variables are related to the physical coordinate separations by 
\[\label{Dfnphyshyp}
(x_1,x_2,x_3,x_4,x_5,x_6)=
(x_{23}^2,x_{13}^2,x_{12}^2,x_{14}^2,x_{24}^2,x_{34}^2).
\]
Comparing with \eqref{triD}, the position-space 4-point contact diagram, also known as the holographic $D$-function \cite{DHoker:1999kzh}, is thus equivalent to the massless triangle integral (see also \cite{Bzowski:2020kfw}).
As shown on page \pageref{tripleKex}, the massless triangle integral is itself equivalent to the triple-$K$ integral (or momentum-space 3-point contact diagram) under  affine reparametrisation of the GKZ integral. 
The creation operators for the position-space contact diagram are thus  those analysed in section \ref{sec:3KC} and \ref{sec:3KW}, except that no final projection to the physical hypersurface is required as all the GKZ variables in \eqref{Dfnphyshyp} are  physical.

Concretely, the $\mathcal{A}$-matrix \eqref{Atri} leads to the  Euler equations \eqref{triEulers} and DWI \eqref{triDWI}, and the toric equations \eqref{tritorics}. 
The Newton polytope corresponds to the right-hand panel in figure \ref{fig:tripolytope}.
From its facets we obtain the singularity conditions
\begin{align}\label{Dfun-sing}
& \g_i=-n_i,\qquad\qquad \g_0- \g_i=-m_i,\qquad\qquad i =1,2,3,\nn\\&
 \g_1+\g_2+\g_3-\g_0=-n,\qquad  2\g_0-\g_1-\g_2-\g_3=-m,
\end{align}
where  $n_i,m_i,n,m\in \mathbb{Z}^+$. 
The action of the annihilator $\p_1 = \p/\p x_{23}^2$ is to raise $\g_0$, $\g_2$ and $\g_3$ by one which corresponds to  raising $\Delta_2$ and $\Delta_3$ by one, and the 
 action of the creation operator $\mathcal{C}_1$ is the reverse of this.
The corresponding $b$-function 
\[
b_1=\g_2\g_3(\g_0-\g_1)(\g_1+\g_2+\g_3-\g_0),
\]
when re-expressed in terms of Euler operator is 
\[
B_{1}=(\t_1+\t_3+\t_5)(\t_1+\t_2+\t_6)(\t_1+\t_5+\t_6)(\t_1+\t_2+\t_3).
\]
As expected, this is simply  \eqref{3ptbigBex} under the mapping $\bar{\t}_i=\t_{i+3}$ 
since the affine reparametrisation from the $\mathcal{A}$-matrix \eqref{Atri} to \eqref{3KA} leaves the creation operators unchanged.
Expanding out 
and using the toric equations to factorise $B_1 = \mathcal{C}_1\p_1$, we recover the creation operator \eqref{3KC1gkz} in GKZ variables.  In our present variables \eqref{Dfnphyshyp}, this is 
\begin{align}
\mathcal{C}_1&
 = x_1(\t_1+1+u_2+u_3)\big((\t_1+1+u_2)(\t_1+1+u_3)+2(v_2+v_3)\big)\notag\\
&\quad +x_2x_5\p_4\bigl(
1+u_2+v_2-v_3+(u_2+u_3+2)u_3\bigr)\notag\\
&\quad +x_3x_6\p_4\bigl(1+u_3+v_3-v_2+(u_2+u_3+2)u_2\bigr)
\end{align}
where $u_i = \t_i+\t_{i+3}$ and $v_i = \t_i\t_{i+3}$.  
One likewise obtains the operator \eqref{W3ptgkz}, namely
\begin{align}
W_{12}^{--}&=(\t_4+\t_5+\t_6+\t_3)(x_4\p_2+x_5\p_1)+x_3x_6 \p_2\p_1. 
\end{align}
Both these operators can be rewritten in various equivalent forms using the DWI and Euler equations.
Their action on the position-space contact diagram follows from \eqref{affinetransfofg}, namely
\[
\mathcal{C}_1:\,\,\Delta_2\rightarrow\Delta_2-1,\quad \Delta_3\rightarrow\Delta_3-1,\qquad
W_{12}^{--}:\,\,\Delta_3\rightarrow \Delta_3+1,\quad\Delta_4\rightarrow\Delta_4-1.
\]

\section{Non-minimal b-functions}
\label{sec:nonnormal}

As we have seen, creation operators are constructed starting from a polynomial $b(\bs{\g})$ in the spectral parameters  known as the $b$-function.
In section \ref{sec:findingb}, we argued that $b(\bs{\g})$ must possess a certain {\it minimal} set of zeros, namely,  those required to cancel the singularities arising when a creation operator shifts us from a finite to a singular integral.  Notice however that this  argument does  not preclude the existence of {\it additional} zeros besides this minimal set.  
For all the Feynman and Witten diagram examples in the main text, the minimal $b$-functions were sufficient for the construction of all creation operators.  As these $b$-functions contain the fewest factors, the resulting creation operators were moreover of lowest possible order in derivatives.
Nevertheless, there are instances where the minimal $b$-function is not sufficient: a simple example, which we analyse in this appendix, is the GKZ integral \eqref{houseint}.
As we will show, additional factors must be appended to the minimal $b$-functions in order to be able to apply the toric equations and factorise 
into a product of creation and annihilation operators.  The zeros of these additional factors are all parallel to the facets of the rescaled Newton polytope, and in most (though not all) cases correspond to additional singular hyperplanes of the GKZ integral. 

Let us recall the necessary analysis of section \ref{sec:sings}.  The integral \eqref{houseint}, namely
\[\label{houseint2}
\mathcal{I}_{\bs{\gamma}} = \int_0^\infty\D z_1\int_0^\infty\D z_2 \,z_1^{\g_1-1}z_2^{\g_2-1} (x_1+x_2 z_2+x_3 z_1^2+x_4 z_1 z_2^2)^{-\gamma_0},
\]
corresponds to the $\mathcal{A}$-matrix 
\[\label{houseA2}
\mathcal{A} = \left(\begin{matrix} 1 & 1 & 1& 1\\ 0 & 0 & 2 & 1\\ 0 & 1 & 0 & 2\end{matrix}\right)
\]
with DWI and Euler equations
 \[\label{houseEul}
0=(\g_0+\t_1+\t_2+\t_3+\t_4)\mathcal{I}_{\bs{\gamma}},\quad 0=(\g_1+2\t_3+\t_4)\mathcal{I}_{\bs{\gamma}},\quad 0=(\g_2+\t_2+2\t_4)\mathcal{I}_{\bs{\gamma}},
\]
and a single toric equation
\[\label{housetor}
0=(\p_1^3\p_4^2-\p_2^4\p_3)\mathcal{I}_{\bs{\g}}.
\] 
The  singularities of this integral, derived in \eqref{housesing}, are 
\begin{align}\label{housesing2}
\gamma_1=-m_1,\quad \gamma_2= -m_2,\quad \gamma_0+\gamma_1-\gamma_2 = -m_3,\quad
4\gamma_0-2\gamma_1-\gamma_2=-3m_4,
\end{align}
for all $m_j\in\mathbb{Z}^+$.
The annihilation operators $\partial_j$ send $\bs{\gamma}\rightarrow\bs{\gamma}'$ while the creation operators $\mathcal{C}_j$ send $\bs{\gamma}'\rightarrow\bs{\gamma}$,
where for each $j$ these parameters are related by
\begin{align}
j&=1: & \gamma_0'&=\gamma_0+1,&\gamma_1'&=\gamma_1, & \gamma_2'&=\gamma_2 \nn\\
j&=2: & \gamma_0'&=\gamma_0+1, &\gamma_1'&= \gamma_1, & \gamma_2'&=\gamma_2+1\nn\\
j&=3: & \gamma_0'&=\gamma_0+1, & \gamma_1'&=\gamma_1+2,&\gamma_2'&=\gamma_2,\nn\\
j&=4: & \gamma_0'&=\gamma_0+1, & \gamma_1'&=\gamma_1+1, & \gamma_2'&=\gamma_2+2.
\end{align}
According to \eqref{stdbfn}, the minimal $b$-functions containing only the  zeros necessary to cancel the singularities produced by the action of the $\mathcal{C}_j$ are 
\begin{align}\label{housebmin}
b_1^{\mathrm{min}}&=(\g_0+\g_1-\g_2)\prod_{m_4=0}^1(4\gamma_0-2\g_1-\gamma_2+3m_4),\nn\\
b_2^{\mathrm{min}}&=\gamma_2(4\gamma_0-2\g_1-\gamma_2),\nn\\
b_3^{\mathrm{min}}&=\prod_{m_1=0}^1(\g_1+m_1)\prod_{m_3=0}^2(\g_0+\g_1-\g_2+m_3),\nn\\
b_4^{\mathrm{min}}&=\gamma_1\prod_{m_2=0}^1(\g_2+m_2).
\end{align}
For example, $\mathcal{C}_3$ shifts $m_1\rightarrow m_1+2$ and $m_3\rightarrow m_3+3$, and so the five singular integrals with $m_1=0, 1$ and $m_3=0,1,2$ in \eqref{housesing2} are accessible starting from finite integrals.  This means $b_3^{\mathrm{min}}$ has the five zeros shown, which act to cancel these singularities. 
The operator $\mathcal{C}_1$ is however a special case: this sends $m_3\rightarrow m_3+1$ and $m_4\rightarrow m_4 + 4/3$, corresponding to a non-integer $F_1^{(4)}=4/3$ in \eqref{Fdef}.  Only the singularities with $m_3=0$ and $m_4=0,1$ are then accessible starting from finite integrals for which all $m_j<0$.  (In other words, integrals for which the GKZ  representation \eqref{houseint2} converges without meromorphic continuation.)

Using the DWI and Euler equations to rewrite these  $b$-functions in terms of Euler operators, we then find
\begin{align}\label{Bjmin}
B_1^{\mathrm{min}}&=-(\t_1+3\t_3)\prod_{m_4=0}^1(4\t_1+3\t_2-3m_4),\nn\\
B_2^{\mathrm{min}}&=(\t_2+2\t_4)(4\t_1+3\t_2),\nn\\
B_3^{\mathrm{min}} &= -(2\t_3+\t_4)(2\t_3+\t_4-1)\prod_{m_3=0}^2(\t_1+3\t_3-m_3),\nn\\
B_4^{\mathrm{min}}&=-(2\t_3+\t_4)(\t_2+2\t_4)(\t_2+2\t_4-1).
\end{align}
At this point a problem appears:
to extract a creation operator requires factorising
\[\label{housefac}
B_j\mathcal{I}_{\bs{\g}} = \mathcal{C}_j\p_j\mathcal{I}_{\bs{\g}},
\]
however the only toric equation we have available for this purpose, \eqref{housetor}, is of {\it fifth} order in derivatives.  While $B_3^{\mathrm{min}}$ is indeed of fifth order, the remaining $B_j^{\mathrm{min}}$ are of at most third order.
Upon expanding out and ordering terms according to \eqref{Bterm}, we find
\begin{align}
B_1^{\mathrm{min}} &= (\ldots)\p_1 - 27 x_2^2 x_3\p_2^2\p_3,\nn\\
B_2^{\mathrm{min}} &= (\ldots)\p_2+8x_1x_4\p_1\p_4, \nn\\
B_3^{\mathrm{min}} &= (\ldots)\p_3-x_1^3x_4^2 \p_1^3\p_4^2,\nn\\
B_4^{\mathrm{min}} &= (\ldots)\p_4 -2x_2^2 x_3\p_2^2\p_3.
\end{align}
For $B_3^{\mathrm{min}}$, we obtain the necessary factorisation \eqref{housefac} upon using \eqref{housetor} allowing a successful construction of $\mathcal{C}_3$.  
For the others, the order in derivatives is too low  to apply \eqref{housetor}. 

To  find $\mathcal{C}_1$, $\mathcal{C}_2$ and $\mathcal{C}_4$, therefore, we look  for new (non-minimal) $B_j$ of the form:
\begin{align}
B_1 &= (\ldots)\p_1 +(\ldots)x_2^4 x_3\p_2^4\p_3,\nn\\
B_2 &= (\ldots)\p_2+(\ldots) x_1^3 x_4^2 \p_1^3\p_4^2, \nn\\
B_4 &= (\ldots)\p_4 +(\ldots)x_2^4 x_3\p_2^4\p_3.
\end{align}
By construction, these are all of fifth order and can be factorised into the desired form \eqref{housefac} using \eqref{housetor}.  
Since the $B_j$ must be functions of the Euler operators, and 
\[
x_i^n\p_i^n=\t_i(\t_i-1)\ldots(\t_i-n+1),
\] 
this is equivalent to seeking
\begin{align}\label{newBform}
B_1 &= (\ldots)\t_1 +(\ldots)\t_2(\t_2-1)(\t_2-2)(\t_2-3)\t_3,\nn\\
B_2 &= (\ldots)\t_2+(\ldots) \t_1(\t_1-1)(\t_1-2)\t_4(\t_4-1), 
\nn\\
B_4 &= (\ldots)\t_4 +(\ldots)\t_2(\t_2-1)(\t_2-2)(\t_2-3)\t_3.
\end{align}
For comparison, the singular hyperplanes in \eqref{housesing2}, when translated to Euler operators via \eqref{houseEul}, correspond to the zeros of
\begin{align}\label{houseEulsings}
(2\t_3+\t_4-m_1),\quad
(\t_2+2\t_4-m_2),\quad
(\t_1+3\t_3-m_3),\quad
(4\t_1+3\t_2-3m_4).
\end{align}
As the non-minimal $B_j$ in \eqref{newBform} must still contain the factors present in the minimal $B_j^{\mathrm{min}}$ in \eqref{Bjmin}, we see that for $B_1$,  and $B_4$ it suffices simply to append factors corresponding to additional singular hyperplanes:
\begin{align}\label{nonminBs}
B_1 &= -(\t_1+3\t_3)\prod_{m_4=0}^3(4\t_1+3\t_2-3m_4),\nn\\
B_4 &= -(2\t_3+\t_4)\prod_{m_2=0}^3 (\t_2+2\t_4-m_2).
\end{align}
Each of these non-minimal $B_j$ contain the factors already present in the minimal $B_j^{\mathrm{min}}$.  Moreover, they are of the form \eqref{newBform} since they correspond to performing a linear shift in $\t_j$ on each of the factors present in the second term of each $B_j$ in \eqref{newBform}.  (Equivalently, setting $\t_j$ to zero in each of the $B_j$ in \eqref{nonminBs} yields the second term of each $B_j$ in \eqref{newBform}.)  This also shows that they are of the smallest order in derivatives consistent with \eqref{newBform}. 

For $B_2$, the additional factors we must append to $B_2^{\mathrm{min}}$ are parallel to the singular hyperplanes in \eqref{houseEulsings} but have different spacing.  Explicitly, we require
\[\label{nonminB2}
B_2 = -\prod_{m_1=0}^1 (\t_2+2\t_4-2m_1)\prod_{m_2=0}^2(4\t_1+3\t_2-4m_2)
\]
so that, when expanded in $\t_2$, we obtain an expression of the form given in \eqref{newBform}.  Note this is not possible using the spacings in \eqref{houseEulsings}.\footnote{The zeros of \eqref{nonminB2}, and of the corresponding of $b_2$ in \eqref{finalbs}, {\it do} however coincide with the singular hyperplanes of the integral obtained by deleting the second column of the $\mathcal{A}$-matrix.  This removes a vertex from the Newton polytope changing the spacings of the singular hyperplanes; a procedure consistent with confining all $\t_2$ dependence in $B_2$ to the first factor in \eqref{newBform}.}

By construction,  the non-minimal $B_j$ in \eqref{nonminBs} and \eqref{nonminB2} all derive from corresponding non-minimal  $b$-functions which are polynomials in the spectral parameters,
\begin{align}\label{finalbs}
b_1&=(\g_0+\g_1-\g_2)\prod_{m_4=0}^3(4\g_0-2\g_1-\g_2+3m_4),\nn\\
b_2 &= \prod_{m_1=0}^1(\g_2+2m_1)\prod_{m_2=0}^2(4\g_0-2\g_1-\g_2+4m_2),\nn\\
b_4&=\g_1\prod_{m_2=0}^3(\g_2+m_2).
\end{align}
and all lead to valid creation operators via \eqref{housefac}.  From $B_2$, for example, we find 
\begin{align}
\mathcal{C}_2&=x_2 \Bigl[12 \t_4^2 \Big(48 \t_1^2+12 \t_1 (3 \t_2-5)+9\t_2 (\t_2-2) +5\Big)\nn\\&\qquad +4 \t_4 \Big(64 \t_1^3+48 \t_1^2 (3 \t_2-4) +4 \t_1 \Big(9 \t_2 (3 \t_2-5)+32\Big)+3 \t_2 \Big(9 \t_2(\t_2-2) +5\Big)\Big)\nn\\&\qquad +(\t_2-1) (4 \t_1+3 \t_2-5) (4 \t_1+3 \t_2-1) (4 \t_1+3 \t_2+3)\Big]+256x_1^3x_4^2\p_2^3\p_3.
\end{align}

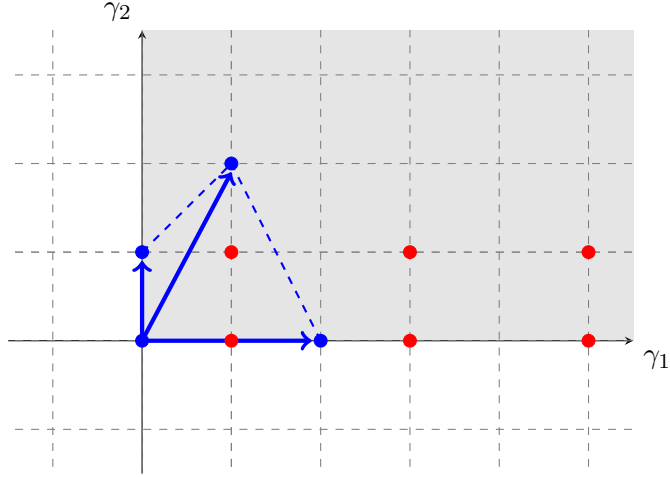
\begin{figure}[t]
\centering
\hspace{2cm}\begin{tikzpicture}[scale=1.2][>=latex]
\begin{axis}[
  axis x line=middle,
  axis y line=middle,
  xtick,
  ytick,
  xlabel={$\gamma_1$},
  ylabel={$\gamma_2$},
  xlabel style={below right},
  ylabel style={above left},
  xmin=-1.5,
  xmax=5.5,
  ymin=-1.5,
  ymax=3.5,
  axis equal image]

\addplot [mark=none,domain=0:10,draw=none,name path=A] {10};
 \addplot [mark=none,domain=0:10,draw=none,name path=B] {0};
\addplot [gray, fill opacity=0.2] fill between[of=A and B,soft clip={domain=0.01:10}];

\foreach \k in {0,...,8} {
 
\addplot [mark=none,dashed,gray,domain=-2.5:10] {\k+1};
\addplot [mark=none,dashed,gray,domain=-2.5:10] (\k+1,x);
\addplot [mark=none,dashed,gray,domain=-2.5:10] {-\k+1};
\addplot [mark=none,dashed,gray,domain=-2.5:10] (-\k+1,x);

}
\foreach \k in {0,...,8} {
\addplot[holdot,red] coordinates{(2*\k+1,0)(2*\k+1,1)};
}
\addplot [holdot,blue] coordinates{(0,0)(0,1)(1,2)(2,0)};
\draw[->, blue,ultra thick](0,0)--(0,0.9);
\draw[->,blue,ultra thick](0,0)--(1,1.9);
\draw[->,blue,ultra thick](0,0)--(1.9,0);
\draw[blue,thick, dashed](0,1)--(1,2);
\draw[blue,thick, dashed](2,0)--(1,2);

\end{axis}

\end{tikzpicture}
\caption{For purposes of illustration, a two-dimensional example of a non-normal lattice can be obtained by projecting to the $(\gamma_1,\gamma_2)$ plane.  The points $(2k+1,0)$ and $(2k+1,1)$ for $k\in\mathbb{Z}^+$ (shown in red) lie within the positive cone  $\sum_j \mathbb{R}^+\bs{a}_j$ (shaded region) spanned by the vertex vectors $\bs{a}_j$ of the Newton polytope (shown in blue). These points also belong to $\sum_j\mathbb{Z}\bs{a}_j$ since $(2k+1,0)=k(2,0)+(1,2)-2(0,1)$ and $(2k+1,1)=k(2,0)+(1,2)-(0,1)$.  However, these points cannot be expressed in the form $\sum_j \mathbb{Z}^+\bs{a}_j$ and hence the lattice generated by the $\bs{a}_j$ is non-normal.
\label{fig:non-normal}}
\end{figure}

Having solved this example, let us note that the failure of the minimal $b$-functions in \eqref{housebmin} can also be understood geometrically.   For $B_1^{\mathrm{min}}$, $B_2^{\mathrm{min}}$ and $B_4^{\mathrm{min}}$ in \eqref{Bjmin} to be factorisable as $\mathcal{C}_j\p_j$, we would need each of 
$
\p_2^2\p_3\p_1^{-1}$, $\p_1\p_4\p_2^{-1}$ and $\p_2^2\p_3\p_4^{-1}$ to be expressible as $\prod_{k=1}^N \p_k^{c_k}$ for some set of $c_k\in\mathbb{Z}^+$.  (The inverses here are purely formal: we mean that $\p_2^2\p_3 = \p_1\prod_{k=1}^N \p_k^{c_k}$, {\it etc}.)
In terms of the shifts produced by these operators on the spectral parameters $\bs{\g}=(\g_0,\g_1,\g_2)$, this is equivalent to requiring that 
\begin{align}
2\bs{\mathcal{A}}_2+\bs{\mathcal{A}}_3-\bs{\mathcal{A}}_1&=\left(\begin{matrix}2\\2\\2\end{matrix}\right),  \quad \bs{\mathcal{A}}_1+\bs{\mathcal{A}}_4-\bs{\mathcal{A}}_2= \left(\begin{matrix}1\\1\\1\end{matrix}\right),\quad  2\bs{\mathcal{A}}_2+\bs{\mathcal{A}}_3-\bs{\mathcal{A}}_4= \left(\begin{matrix}2\\1\\0\end{matrix}\right)
\end{align}
are all expressible as  $\sum_{k=1}^N c_k\bs{\mathcal{A}}_k$ for some set of $c_k\in\mathbb{Z}^+$, where $\bs{\mathcal{A}}_k$ denotes the $k$th column of the $\mathcal{A}$-matrix including the top row of ones. 
Clearly this is not possible, although these vectors do all lie in the positive cone corresponding to solutions with $c_k\in\mathbb{R}^+$ since
\begin{align}
2\bs{\mathcal{A}}_2+\bs{\mathcal{A}}_3-\bs{\mathcal{A}}_1&=2(\bs{\mathcal{A}}_1+\bs{\mathcal{A}}_4-\bs{\mathcal{A}}_2) = \frac{2}{3}(\bs{\mathcal{A}}_2+\bs{\mathcal{A}}_3+\bs{\mathcal{A}}_4),\nn\\
2\bs{\mathcal{A}}_2+\bs{\mathcal{A}}_3-\bs{\mathcal{A}}_4&=\frac{1}{2}(3\bs{\mathcal{A}}_1+\bs{\mathcal{A}}_3).
\end{align}   
 Mathematically, this is precisely the condition that lattice generated by the $\bs{\mathcal{A}}_k$ (or equivalently, the toric ideal associated with the $\mathcal{A}$-matrix) is {\it non-normal}.  Conversely, when the normality condition
\[\label{normaltoric}
\Big(\sum_k \mathbb{R}^+\bs{\mathcal{A}}_k\Big)\cap\Big(\sum_k \mathbb{Z}\bs{\mathcal{A}}_k\Big) = \Big(\sum_k \mathbb{Z}^+\bs{\mathcal{A}}_k\Big)
\]
is satisfied, it can be shown that the minimal $b$-functions \eqref{stdbfn} generate valid creation operators \cite{Saito_param_shift, Saito_restrictions}.
This condition is non-trivial as can be seen from the two-dimensional example of a non-normal lattice illustrated in figure \ref{fig:non-normal}.

\bibliographystyle{JHEP}
\bibliography{CreationOps}

\end{document}